\documentclass{iopart}
\usepackage{iopams,amsbsy,eufrak,mathrsfs,graphicx}  
\eqnobysec

\def\Journal#1#2#3#4#5{(#1) ``#2'' {#3} {\bf #4} #5} 

\def\JHEP{\em J. High Energy Phys.}
\def\CQG{\em Class. Quantum Grav.}

\def\PRD{\em Phys. Rev. D}
\def\GRG{\em Gen. Rel. Grav.}

\def\CMM{\em Commun. Math. Phys.}
\def\PR{\em Phys. Rev.}
\def\RMP{\em Rev. Mod. Phys.}

\def\JMP{\em J. Math. Phys.}

\def\PRL{\em Phys. Rev. Lett.}

\def\PLMS{\em Proc. London Math. Soc.}

\def\PRSL{\em Proc. Roy. Soc. London}

\def\G{{\bf g}}
\def\rmg{{\rm g}} 

\def\r{\mathbb R}
\def\n{\mathbb N}

\def\v{\vec{\mathbf v}}
\def\u{\vec{\mathbf u}}
\def\k{\vec{\mathbf k}}
\def\t{\mathbf t}

\def\g{\gamma}

\def\C{{\cal C}}
\def\D{{\cal D}} 
\def\NN{{\cal N}}
\def\U{{\cal U}}
\def\W{{\cal W}}

\def\Pp{{\cal P}} 

\def\l{\lambda}
\def\f{\Phi}
\def\z{\theta}
\def\X{\vec{X}}
\def\Y{\vec{Y}}

\def\xiv{\vec{{\boldsymbol\xi}}}
\def\d{\partial}
\def\fr{\frac}
\def\O{{\cal O}}

\def\less{<\!\!\!<\!}
\def\greater{>\!\!\!>\!}

\def\Uf{\mathfrak U}
\def\Bf{\mathfrak B}
\def\Cf{\mathfrak C}

\def\Ss{\mathscr{S}}
\def\Ts{\mathscr{T}}
\def\Us{\mathscr{U}}
\def\Vs{\mathscr{V}}
\def\Rs{\mathscr{R}}
\def\Xs{\mathscr{X}}

\def\Ds{\mathscr{D}}
\def\Ps{\mathscr{P}}

\def\Cs{\mathscr{C}}
\def\Is{\mathscr{I}}

\def\preca{\stackrel{1}{>}}
\def\suca{\stackrel{1}{<}}

\def\precan{\stackrel{n}{>}}
\def\sucan{\stackrel{n}{<}}
\def\precamenos{\stackrel{n-1}{>}}
\def\sucamenos{\stackrel{n-1}{<}}

\def\precu{\mbox{\raisebox{-1.5ex}{$\stackrel{
\mbox{\large$<$}}{\mbox{\tiny $\Us$}}$}}}
\def\sucu{\mbox{\raisebox{-1.5ex}{$\stackrel{
\mbox{\large$>$}}{\mbox{\tiny $\Us$}}$}}}
\def\lessu{\mbox{\raisebox{-1.5ex}{$\stackrel{
\mbox{\large$\less$}}{\mbox{\tiny$\Us$}}$}}}
\def\greatu{\mbox{\raisebox{-1.5ex}{$\stackrel{
\mbox{\large$\greater$}}{\mbox{\tiny$\Us$}}$}}}

\def\precv{\mbox{\raisebox{-1.5ex}{$\stackrel{
\mbox{\large$<$}}{\mbox{\tiny$\Vs$}}$}}}
\def\sucv{\mbox{\raisebox{-1.5ex}{$\stackrel{
\mbox{\large$>$}}{\mbox{\tiny$\Vs$}}$}}}
\def\lessv{\mbox{\raisebox{-1.5ex}{$\stackrel{
\mbox{\large$\less$}}{\mbox{\tiny$\Vs$}}$}}}
\def\greatv{\mbox{\raisebox{-1.5ex}{$\stackrel
{\mbox{\large$\greater$}}{\mbox{\tiny$\Vs$}}$}}}

\def\darrow{\stackrel{\mbox{\tiny$\boldsymbol\nearrow$}}
{\mbox{\tiny$\boldsymbol\searrow$}}}
\def\darrowu{\mbox{\raisebox{-1.5ex}{$\stackrel{\darrow}{\mbox{\tiny$\Us$}\
}$}}}

\def\garrow{\stackrel{\mbox{\tiny$\boldsymbol\nwarrow$}}
{\mbox{\tiny$\boldsymbol\swarrow$}}}
\def\garrowu{\mbox{\raisebox{-1.5ex}{$\stackrel{\garrow}{\mbox{\
\tiny$\Us$}}$}}}

\def\lie{\pounds_{\xiv}}

\def\J{\stackrel{\!\!\infty}{J^+}}

\def\be{\begin{equation}}
\def\ee{\end{equation}}
\def\bea{\begin{eqnarray}}
\def\eea{\end{eqnarray}}
\def\bnr{\begin{eqnarray*}}
\def\enr{\end{eqnarray*}}

\newtheorem{axiom}{Axiom}[section]
\newtheorem{defi}{Definition}[section]
\newtheorem{theo}{Theorem}[section]

\newtheorem{prop}{Proposition}[section]

\begin{document}

\topical[Causal structures and causal boundaries]{Causal structures and
causal 
boundaries}

\author{Alfonso Garc\'{\i}a-Parrado and Jos\'e M. M. Senovilla}

\address{Departamento de F\'{\i}sica Te\'orica, Universidad del Pa\'is
Vasco. 
Apartado 644, 48080 Bilbao, Spain.}

\begin{abstract}
We give an up-to-date perspective with a general overview of the theory of 
causal properties, the derived causal structures, their 
classification and applications, and 
the definition and construction of causal boundaries and of causal 
symmetries, mostly for 
Lorentzian manifolds but also in more abstract settings.
\end{abstract}




%\maketitle

\tableofcontents

\section{Introduction}
Causality is the relation between causes and their effects, or between
regularly correlated phenomena. Causes are things that bring about 
results, actions or conditions. No doubt, therefore, 
causality has been a major theme of concern for all branches of 
philosophy and science for centuries. Perhaps the first philosophical 
statement about causality is due to the presocratic atomists Leukippus and 
Democritus: ``nothing happens without the influence of a cause; 
everything occurs causally and by need''. This intuitive view was a 
matter of continuous controversy, eventually cleverly criticized by Hume, 
later followed by Kant, on the grounds that probably our {\em belief} 
that an event follows from a cause may be simply a prejudice due to an 
association of ideas founded on a large number of experiences where 
similar things happened in the same {\em order}. In summary, an 
incomplete induction. 

The task of science and particularly Physics, however, is establishing or 
unveiling relations between phenomena, with a main goal: {\em 
predictability} of repetitions or new phenomena. Nevertheless, when we 
state that, for instance, an increase on the pressure of a gas reduces its 
volume, we could equally state that a decrease of the volume increases the 
pressure. The ambiguity of the couple cause-effect is even greater 
when we reverse the sense of time. And this is one of the {\em key} 
points. The whole idea of causality must be founded upon the basic 
{\em a priori} that there is an orientation of time, a time-arrow. 
This defines the future (and the past), at least on our neighbourhood and
momentarily.

Consequently, all branches of Physics, old and new, classical or 
modern, have a causality theory lying underneath. This is specially 
the case after the unification process initiated by Maxwell with 
Electromagnetism and partly culminated by Einstein with the theories 
of Relativity, where time and space are intimately inter-related. 
Soon after, Minkowski was able to unite these two concepts in one 
single entity: {\em spacetime}, a generalization of the classical 
three-dimensional Euclidean space by adding an axis of time. A 
decisive difference with Euclidean space arises, though: time has a 
different status, it mingles with space but keeping its identity. It 
has, so to speak, a ``different sign''. This immediately leads to the 
most basic and fundamental causal object, the {\em cone} which embraces 
the time-axis, leaving all space axes outside. And upon this basic 
fact, by naming the two halves of the 
cone (future and past), we may erect a whole theory of causality and 
causal structure for spacetimes.

Four-dimensional spacetimes (see definition \ref{spacetime})
are the basic arena in General and Special
Relativity and their avatars, and almost any other theory trying to
incorporate the gravitational field or the finite speed of 
propagation of signals will have a (possibly $n$-dimensional with 
$n\geq 4$) spacetime at its base. The maximum speed of propagation is 
represented in this picture by the angle of the cone at each point. 
And this obviously determines the points that may affect, or be 
influenced by, other points.
In Physics, spacetimes represent the Universe, or that part of it,
 we live in or 
want to model. Thus, the notion of spacetime is pregnant 
with ``causal'' concepts, that is, with an inherent causality theory. 
Causality theory has played a very important role in the development of 
General Relativity.
It is fundamental for all global formalisms, for the theory of radiation, 
asymptotics, initial value formulation, mathematical developments, singularity 
theorems, and many more.
There are books dealing primarily with causality theory
(\cite{BEE,HO}) and many 
standard books about General Relativity contain at least a chapter devoted
to causality \cite{FF,WTM,Wald}.
Recently, modern approaches to ``quantum gravity'' have borrowed concepts 
widely used in causality theory such as the causal boundary (AdS/CFT
correspondence \cite{HOROWITZ}) or 
abstract causal spaces (quantum causal sets \cite{SORKIN}).

Roughly speaking the causal structure of a spacetime is determined at 
three stages: primarily, by the mentioned cones---called null cones---, 
at each event. This is the {\em algebraic} level. Secondly, by the 
the connectivity properties concerning nearby points, which is 
essentially determined by the null cone and the local differential 
structure through geodesic arcs. This is the {\em local} stage. 
And thirdly, by the connectivity at large, that is, between any possible pair of
points, be they close or not. This connection must be achieved by 
sequences of locally causally related events, usually by causal curves 
---representing the paths of physical small objects. This is the {\em 
global} level. These three stages taken together determine the causal 
structure of the spacetime. In short, the causal structure involves the sort of
relations, properties, and constructions arising between events, 
or defined on tensor objects, which depend essentially on the existence 
of the null cones, that is to say, on the existence of a {\em sole} time 
in front of all spatial dimensions. Unfortunately, a precise definition of 
``causal structure'' is in general lacking 
in the literature, as it is more or less taken for granted in the 
formalisms employed by each author. One of the 
aims of this review is to provide an appropriate, useful and rigorous
definition of ``causal structure'' of sufficient generality. 

From a mathematical viewpoint, a spacetime is a 
{\em Lorentzian manifold}: an $n$-dimensional semi-Riemannian
manifold $(V,\G)$ where at each point $x\in V$ the metric tensor 
$\G|_x$, which gives a local notion of distances and time intervals, has 
Lorentzian signature \cite{ONEILL}, the axis with the different sign 
($n>2$) indicating ``time'' (cf. definition \ref{spacetime} for more details).
Even though a great deal of research has been performed
in four-dimensional Lorentzian manifolds for obvious reasons,
almost all the results do not depend on the manifold dimension and we 
will always work in arbitrary dimension $n$. This is also adapted to more 
recent advances such as String Theory, Supergravity, et 
cetera. 
The field equations or physical conditions that the metric
tensor $\G$ must satisfy in order to lead to an acceptable representation
of an actual spacetime are in principle outside the scope of this review.
There is an entire branch of Mathematics called Lorentzian geometry whose
subject is
the study of Lorentzian manifolds and it encompasses topics (such as
causality theory) 
which traditionally have been studied by relativists. 
 While the study of proper Riemannian manifolds (the metric tensor is positive 
definite at each point of the manifold so there is no time and no 
causality) presents a status in which important questions
such as the presence and study of singularities, geodesic connectivity,
existence of minimizing geodesics, splittings,
or the completion of the manifold are ruled by powerful theorems, 
the matter is radically different in Lorentzian geometry
where equivalent or analogous results require a case-by-case discussion with no
general rule, and sometimes the answers to the ``same'' mathematical 
problems are entirely different. 
This makes of Lorentzian geometry a more difficult topic where 
there are still open questions regarded as bearing the 
same degree of importance as those already solved in proper Riemannian 
geometry. 
Perhaps the main advantage of proper Riemannian geometry in this regard
is the existence of a well-defined notion of distance between points,     
a feature absent in Lorentzian geometry where only a pseudo-distance
can be defined \cite{BEE}. As a result many new possibilities 
arise in Lorentzian geometry, see e.g. \cite{BEE,FF,ONEILL,Wald}. 
Among these new possibilities, as we will largely discuss in this 
review, those dealing with 
the {\em causal structure} and the {\em causal completion} of spacetimes 
are specially interesting mathematically and physically.

From both these perspectives, a {\em generic causal structure} 
collects all information about Lorentzian manifolds not
related to the particular geometrical form, or causal characteristics, 
of any particular spacetime. That is to say, the properties which 
totally depend on the existence of any Lorentzian metric $\G$ on the 
manifold, and are absent if $\G$ is removed. 
A {\em particular causal structure} will then be 
a class of Lorentzian manifolds carrying equivalent {\em causal}
properties.
Since any good causal property must be {\em conformally
invariant} (because the null cones remain intact under conformal 
transformations)
many authors have assumed implicitly that the causal structure is fully
determined by the conformal class of metrics: all metrics 
proportional to each other with a positive proportionality factor. 
However, recent studies (see subsections \ref{causal-preservation}, 
\ref{chains} and 
\cite{CAUSAL}) indicate that this view may be too restrictive, and that 
non-conformally related metrics can belong to the same causality class 
in a well-defined way. 
Of course, 
these more general causal classes, while not ``internally conformal'',
{\em are} conformally invariant! This opens a wide new world 
concerning causality which is still to be explored in detail.
Related to this is the question of the classification of ``exact 
solutions'', i.e. particular spacetimes satisfying the field 
equations of a given theory and thus of physical interest, 
in terms of their causal structure.
For instance, nowadays there is a large number of known exact solutions
to Einstein's field equations in four dimensions (the standard
reference is \cite{MCCALLUM}). Some of them have been analyzed from a 
global point of view \cite{FF,PENROSE,Wald,BEE}, but this study of global
causal properties has not been performed in a majority of cases.

Though most of the theory of causal structures has been carried out 
in the framework of Lorentzian manifolds, the definition of generic 
causal structures showed the possibility of a different line
of thought by making an abstraction of the existence of the metric. This is 
possible because well-defined binary relations can be set between points 
of a Lorentzian manifold
according to their connectivity by causal curves. As these
binary relations
fulfill very precise properties, one can devise abstract sets, called {\em causal
spaces}, with no metric ---nor even 
differentiable structure--- but possessing binary relations
whose properties resemble those present in
spacetimes
\cite{KRONHEIMER-PENROSE,CARTER,WOODHOUSE}. These more abstract sets 
are also considered in this review and an account of the
most popular {\em causal spaces} is given in section \ref{abstract}. 
Probably, most if not all of the results presented in this review for 
spacetimes extend in one way or another to abstract causal spaces. 
Sometimes this has been made explicit in the review, but not always. 
As a matter of fact, for example,
the new results of sections \ref{causal-preservation} and \ref{chains}
can be extended to abstract causal spaces
and in this sense the word ``causal structure'' also acquires a precise
meaning for them. 

The important issue of the {\em correspondence} between the mathematical model 
employed to describe the spacetime (be it a Lorentzian manifold or an 
abstract causal space) and the real things that we see or the
experimental measures we may perform has also been explored in the 
literature \cite{EHLERS}, and we give a brief account in subsection
\ref{physics}. 

An important branch of causality theory with many ramifications is 
the theory concerning {\em causal boundaries}. These are essentially 
sets of ``ideal points'' which, when added appropriately to a particular 
spacetime, make it {\em complete} in a definite sense. The idea is to
attach a boundary to the spacetime under study keeping the causal 
properties and bringing infinity to finite values of judiciously chosen 
coordinate systems. Also singularities are to be described as 
boundary points.
Causal completions stemmed from the Lorentzian analog of the conformal
compactification procedure which can be performed on proper Riemannian 
manifolds. In the Lorentzian case we can still 
carry out conformal compactifications, or other sensible completions,
but the boundary 
points representing infinity can usually be classified into 
further subclasses \cite{PENROSE-CONFORMAL,PENROSE-DIAGRAM}, and 
there is no {\em unequivocal} procedure to achieve the causal 
completions. Unfortunately, the conformal compactification is not always  
practical nor satisfactory, because 
either it is too difficult to achieve in {\em simple} cases, or 
one is interested in completing 
the spacetime without utilizing external elements,  
but employing only elements of the spacetime itself. This is precisely 
the idea behind the 
Geroch, Kronheimer and Penrose construction, (cf. subsection 
\ref{GeKrPe}), the most sophisticated and successful approach to 
causal boundaries historically. There is however
no unique proposal to accomplish this and by now a wide range 
of techniques and procedures to construct a ``causal boundary''
for a spacetime are available, each with its own advantages and 
disadvantages. This is an important line of research, with many relevant 
applications. Just to mention a few:
(i) the definition of asymptotically simple spacetimes and 
{\em asymptotic flatness}, see e.g.
\cite{FF,PENROSE,PENROSE-CONFORMAL,PENROSE-DIAGRAM,STEWART,Wald},
has revealed a wealth of interesting
properties of the conformal boundary permitting explicit 
computations of the gravitational power radiated to infinity and the 
construction of conserved quantities \cite{PENROSE-RSC,FRAUENDIENER,STEWART};
(ii) the study of the global characteristics, and global causal 
properties of spacetimes. For instance, it is very easy to distinguish 
globally hyperbolic spacetimes (see definition \ref{SER}) 
from the rest by studying properties 
of the boundary; and (iii) more recently, causal boundaries
have received new interest due to the presence of the causal 
boundary concept in the Maldacena conjecture \cite{MALDACENA,HOROWITZ}.
We review all new and classical matters concerning causal boundaries
in section \ref{causal-boundary}.

Having reached this point, it should be clear that studies 
related to the causal structure and the causal 
boundary in any of their varieties has been
a high priority among relativists and differential geometers. One can 
also convince oneself by taking a glimpse at the long list of references---and 
the references therein!
Why then a topical review on causal structures and causal boundaries? 
The answer is, in our opinion, twofold: this seems to be a convenient time
for 
an up-to-date revision and compilation of the ideas sketched in the previous 
paragraphs, pointing out those that are already old-fashioned
or obsolete, and 
those with a promising future. And this should be complemented with a 
compilation of the new achievements, and the new possibilities opened
in the field, which have been taking place in recent years. 
Examples of these are the developments of (realizations of) 
semigroups and monoids \cite{HILGERT,SEMIGROUP,NEEB}, the potentialities 
of causal symmetries and causal preserving vector fields
\cite{LETTER,CAUSAL-SYMMETRY}, new results 
about the splitting of Lorentzian manifolds \cite{BERNAL2,GALLOWAY}, 
new advances 
\cite{MAROLFROSS,MAROLF-ROSS,MAROLF-ROSS2,HUBENY,HARRIS1,HARRIS2,CAUSAL} 
and renewed interest \cite{MALDACENA,HOROWITZ} in the construction of causal 
boundaries, and others. It is also helpful
to establish links between ideas presented in different places and under 
various motivations and a review like this could help achieve 
this goal. The review is intended to provide an introduction with
enough basic
information to make the article interesting and informative for
non-specialist scientists. It can also serve as a reference source for 
those researchers interested in deepening their knowledge on any of the
topics presented herein. Moreover, we have really tried to bring to 
physicists attention some useful mathematical tools
available in journals and books mainly addressed to
mathematicians, and conversely, to inform mathematicians of many 
efforts and results which may be unknown to them.

In writing this review we were confronted with the decision as to what
topics should be covered while keeping the length to a reasonable size. 
The guiding principle
was to achieve a good compromise between historical relevance, 
impact on other research, applicability, and future perspectives of the 
reviewed results. Perhaps one of the topics with a greater
influence is 
Penrose's idea of attaching a boundary to a given spacetime and this is
why a major part 
of the review is devoted to the implications and generalizations of this
fruitful idea. Other relevant topics treated are the precise 
definition of causal structure,
the axiomatic study of causality theory, and the classification of
spacetimes according 
to their global causal properties as there are a number of recent
investigations which 
have profited from these classic topics in causality. 
A full account of the matters covered together 
with the outline of the review is given in subsection \ref{plan}.     

Bearing in mind the above ideas, we set no time limits on the papers or
books surveyed and 
only their compliance with the chosen topics was taken into account. 
In general we do not mean to be exhaustive in 
our exposition of each paper, rather, we give
overviews in such a way that the reader may get a general idea and
extract enough information 
to decide if the reference is worth looking at.
Thus, proofs of the results are in general omitted. 
Nor are examples included in general; although there are a 
few exceptions.

The work of a large number of authors is described in the review.
As is an unfortunate custom among us, scientists, each author often follows
his/her own notation and conventions. Hence it was a bit challenging to write a 
readable text while keeping faithful to the contents of the references.
For the sake of clarity, we have maintained a consistent standard notation 
throughout, and we have adapted the original notation in the 
references to ours in order to provide a unified clear treatment of 
the subjects. 

%% 
 % The number of existing references in each of the topics covered is very
 % huge and they
 % were chosen following the above guidelines.
 % We are aware that there is a personal measure here and  
 % references which perhaps would deserve to be quoted or even studied might
 % have been left aside.
 % We may have missed other references which we would have surely quoted had
 % we 
 % have notice of them during the writing of the review. In any case we 
 % apologize to the authors of such papers and we invite them to let us know
 % about 
 % their work or to send us their comments and suggestions prior to the
 % publishing of this review.
 % (poner esta \'ultime frase s\'olo en la versi\'on electr\'onica) 
 %%

\subsection{Plan of the paper with a brief description of contents}
\label{plan}
In this review there are sections which may be instructive for young
researchers, Ph.D. students, and non-specialist scientists interested
in the subject, as they contain introductions and enough basic
information to make them informative for a general readership.  Other
parts of the review give an account of recent results which due to
their novelty might arouse the scientific curiosity of all interested readers. 
Generally speaking, a non-specialist may be interested in reading
sections \ref{sec:essentials}, \ref{causal-characterization} and, if
willing to learn the basics about causal boundaries, also parts of
section \ref{causal-boundary} (for instance subsections
\ref{conformal-boundary}, \ref{GeKrPe} and
$\S$\ref{conformal-diagram}.)  A mathematically oriented researcher
may be interested in the more abstract constructions of section
\ref{abstract}, in the theory of semigroups and monoids and the causal
symmetries of section \ref{symmetry}, or in the topological results of
section \ref{sec:CandT}.  A theoretical physicist would probably like
to skip section \ref{abstract}, but then he/she is advised to read
subsection \ref{physics}.  On the other hand, sections
\ref{sec:essentials} (except subsection \ref{causal-tensors}),
\ref{abstract} and subsection \ref{hierarchy} may be skipped by
experts in the field, as we have tried to use a standard notation.  If
a reader wishes to pay attention only to the causal boundaries for
spacetimes, then he/she can go directly to section
\ref{causal-boundary} and consult, if necessary, other parts of the
review to which we refer there.  Finally, the review contains
up-to-date information, recent advances and a handful of new
applications which are not well known, not even to some experts. 
These new lines of research can be found in subsections
\ref{causal-tensors}, $\S$\ref{isocausal}, \ref{chains},
\ref{splitting}, $\S$\ref{m-r}, $\S$\ref{harris},
$\S$\ref{causal-relationship}, $\S$\ref{causal-diagram},
$\S$\ref{ppwaves} and the whole section \ref{symmetry}.  Next, we give
a brief outline of the contents of each section in order to help
readers to choose according to their own interests.

Section \ref{sec:essentials} is a basic summary of Lorentzian
causality in which all the standard concepts such as null cone, causal
curve and basic sets used in the causal analysis are introduced, and
we set the notation followed in the review.  Despite this being a 
basically standard section, the
relatively novel concept of future (and past) tensor, which is
relevant for the entire theory of causality and permits to carry over 
the classical null cone structure to all tensor bundles 
is presented here in subsection \ref{causal-tensors}.

Section \ref{abstract} contains an account of four axiomatic
approaches to sets possessing {\em causal relations}, trying to
reproduce at least part of the basic causal building blocks present in
a Lorentzian manifold.  These are the Kronheimer-Penrose causal
spaces, Carter's etiological spaces, the Ehlers-Pirani-Schild
axioms for a physical spacetime, later improved by Woodhouse, and the
{\em causal set} model for a quantum spacetime.  The relationship
between these approaches, their common features and differences are
analyzed. Apart from obvious historical reasons, we have decided to put 
section \ref{sec:essentials} before section \ref{abstract} because
this is very instructive, and also because almost all the abstract
concepts introduced in section \ref{abstract}, which do not need
any smooth underlying structure, were in fact inspired by the
Lorentzian geometry ideas.

Section \ref{causal-characterization} accounts for different
procedures to classify Lorentzian manifolds and abstract
causal/etiological spaces.  The standard hierarchy of causality
conditions is reviewed and briefly commented.  An improved recent
scheme to classify Lorentzian manifolds based on setting causal
mappings between them is reproduced here.  In this framework, the
ultimate abstract definition of causal structure for a differentiable
manifold is highlighted, and its implications investigated. 
A correct version of a much sought
after but frequently misrepresented folk theorem, which was unproven
until recently in its precise form, stating the local equivalence of
all Lorentzian manifolds from the causal point of view is given here. 
The definition and potentialities of causal chains of Lorentzian
structures on a given manifold are also recalled.

The interplay between topology and causality is the subject of section
\ref{sec:CandT}.  Classical topics such as possible topologies on
spaces of causal curves or spaces of Lorentzian metrics are explained
here in a succinct manner.  Remarkable recent results on splittings of
globally hyperbolic Lorentzian manifolds are included in this section.

Section \ref{causal-boundary} is the longest of the review and is
devoted to an exhaustive account of the major attempts at finding
``the boundary'' of a Lorentzian manifold.  This section could very
well deserve a full topical review on its own.  The contents include
all classical approaches explained in a concise form, but new
important and promising advances have also been described, such as the
intriguing new ideas by Marolf and Ross, the proof by Harris of the
universality of the classical future boundary using categories, or the
new definition of causal boundary and causal completions using causal
mappings.  A basic introduction to the Penrose diagrams with a
comparison to their newly defined generalization called causal
diagrams is also provided.

In section \ref{symmetry} we discuss new concepts concerning the role
of continuous transformations preserving the causal properties of
spacetimes.  This is a new line of research which incorporates some
recent mathematical advances concerning realizations of semigroups and
general cone structures.  The transformations are studied from a
finite viewpoint first (causal symmetries) and also from the
infinitesimal perspective by defining their generators
(causally-preserving vector fields).  The set of all such
transformations is no longer a group, but a monoid which is in turn a
subset of a more general algebraic structure known as semigroup. 
Actions of semigroups on manifolds have been studied in mathematics
and a brief account of this work is given in order to bring it to the
knowledge of mathematical physicists.

Finally section \ref{future} suggests new possible avenues of
research, possible applications, and some interesting open problems
which would be desirable to solve.

\subsection{Conventions and Notation}
\label{notation}
An $n$-dimensional differentiable manifold $V$ (sometimes $M$)
will be the basic arena 
in this review (except in sections \ref{abstract} and 
some other related places). Thus, $V$ is endowed with a 
{\em differentiable structure}
\cite{BEE,FF,ONEILL,Wald}, that is to say, given any two
local charts $V_1$, $V_2\in\D$ of a given atlas $\D$ with 
$V_1\cap V_2\neq\varnothing$ the induced diffeomorphism between the 
corresponding open sets of $\r^n$ is of class $C^k$, $k\in\n$ ($k$ 
times differentiable with continuity). Two atlases $\D_1$, $\D_2$ are said 
to be $C^k$-compatible if $\D_1\cup\D_2$ provides another $C^k$ differentiable 
structure for $V$. We will further assume that $V$ is Hausdorff, 
oriented and connected. 

One can define at any point $x\in V$ 
the tangent space $T_x(V)$, the cotangent space
$T^*_x(V)$ and by means of the tensor product ``$\otimes$'' of these
vector spaces the space of $r$-contravariant $s$-covariant tensors
$T^r_s|_{x}(V)$ \cite{BEE,FF,ONEILL,Wald}.  
They give rise to the bundles $T(V)$, $T^*(V)$ and
$T^r_s(V)$ respectively.  Boldface letters will be used to denote
elements of any of the mentioned spaces (and also for the sections
of the bundles), their distinction being usually obvious
contextually. Indices on $V$ are occasionally used and represented by lowercase Latin letters.  
The push-forward and pull-back of a map $\Phi$ are denoted by $\Phi'$ 
and $\Phi^*$ respectively. 

In this review we will be mostly concerned with a particular type of
pseudo-Riemannian (also called semi-Riemannian) 
manifolds \cite{ONEILL} which are manifolds where a
{\em metric tensor} $\G$ is defined: $\G$ is a non-degenerate
symmetric (at least) $C^1$ section 
of the bundle $T^0_2(V)$, so that
in local coordinates det($\rmg_{ab}(x)$)$\neq 0$,
$\rmg_{ab}(x)=\rmg_{ba}(x)$, $\forall x\in V$ and the signature of
$\G$ is constant on $V$. In any semi-Riemannian manifold there is a 
canonical isomorphism between $T_{x}(V)$ and $T^{*}_{x}(V)$ which is 
induced by $\G$ and thus indices on tensors can be ``raised and 
lowered'' adequately.
When the signature of $\G$ has a definite sign the
manifold is called proper Riemannian or simply Riemannian. However, 
generic causal properties and the intuitive concepts of `time flow', 
`future and past', et cetera, can only be defined if the signature is 
Lorentzian, so that one of the dimensions (time) has a different status with 
respect to the rest (space). This allows for two possible (equivalent) 
choices, $(+,-,\dots,-)$ or $(-,+,\dots,+)$, depending on whether 
a vector pointing on the time direction is chosen to have positive or negative 
length. Such pseudo-Riemannian manifolds are called {\em Lorentzian manifolds} 
\cite{BEE,FF,ONEILL}. Our convention will be the first one, a choice 
determined by the particular goals we have in mind (causal properties),
so that any $\v\in T_x(V)$ will be called timelike, null or
spacelike if $\G|_x(\v,\v)$ is respectively greater than, equal to or
less than zero. Non-spacelike vectors are commonly treated together and
then also simply called {\em causal} vectors.  
An important point is that such a Lorentzian metric $\G$ cannot be defined 
on manifolds with arbitrary topology (this is
a significative difference with the case of a proper Riemannian metric which can be
defined on every differentiable manifold). We will come back to this
important point in section \ref{sec:CandT}.  The set of all Lorentzian
metrics on a given differentiable manifold $V$ will be denoted by
Lor($V$).

The interior, exterior, closure and boundary of a set $\zeta$ are denoted by
int$\zeta$, ext$\zeta$, $\overline{\zeta}$ and $\partial \zeta$, respectively, 
and the inclusion, union and intersection of sets are written as
$\subset$, $\cup$ and $\cap$.  We will also need some basic
terminology concerning binary relations in set theory.  A binary
relation $R$ on a set $X$ is a subset of the Cartesian product
$X\times X$.  For any $(x,y)\in R$ we will write $xRy$.  A relation
$R'$ is larger than $R$ if $R\subseteq R'$.  The restriction of $R$ to
a subset $Y\subset X$ is the new relation $R\cap(Y\times Y)$ denoted
by $R_Y$.  The inverse relation $R^{-1}$ of $R$ is another binary
relation defined as
$$
R^{-1}=\{(x,y)\in X\times X: (y,x)\in R\}.
$$ 
In particular, the relation is called
symmetric if $R=R^{-1}$, while it is said antisymmetric if $xRy$ and $yRx$ imply 
necessarily that $x=y$. Other relevant cases are reflexive relations
($xRx$), anti-reflexive relations ($(x,x)\not\in R$), and transitive
relations ($xRy$, $yRz$ $\Rightarrow$ $xRz$). A relation which is 
reflexive and transitive is called a preorder, and antisymmetric
preorders are called partial orders, while symmetric preorders are 
also called equivalence relations.  A relation $R$
orders linearly the set $Y$ if for any pair $x,y\in Y$ either $yRx$ or
$xRy$.  The reflexive relation $R$ will be called horismotic if for
any finite sequence $\{x_i\}_{1\leq i\leq m}$ such that $x_iRx_{i+1}$
$\forall i \neq m$ then
\begin{enumerate}
\item $x_1Rx_m$ $\Rightarrow$ $x_hRx_k$, $\forall h,k$ with $1\leq h\leq k\leq m$.
\item $x_mRx_1$ $\Rightarrow$ $x_h=x_k$, $\forall h,k$ with $1\leq h\leq 
k\leq m$.
\end{enumerate}

\section{Essentials of causality in Lorentzian Geometry}
\label{sec:essentials}
We start by reviewing the basic concepts and
causal properties of Lorentzian manifolds. As we will see in section
\ref{abstract} some of these key concepts can be generalized to sets
which are not Lorentzian manifolds (in some cases they are not even
differentiable manifolds), usually keeping the same notation and 
terminology.  We will give brief definitions of most of
the concepts involved in order to fix the nomenclature followed in this
review.

A Lorentzian manifold admitting a global nonvanishing timelike vector
field $\xiv$ is said to be time orientable. Such manifolds constitute the
main 
underlying structure for most parts of Physics through the following
definition.
\begin{defi}[Spacetime]
Any oriented, connected Hausdorff $C^{\infty}$ Lorentzian manifold 
with a time orientation and a $C^1$ metric $\G$ is called a spacetime. 
The points of a spacetime are called events.
\label{spacetime}
\end{defi}
The conditions imposed in this definition may vary slightly in the
literature
(compare for instance the definitions of \cite{BEE,FF,PENROSE}). 
In particular time orientability
is not always required, and the Hausdorff requirement may be too restrictive 
in some situations (see subsection \ref{classic} for details). The question of 
when time orientability is feasible will be briefly discussed in subsection 
\ref{classic}. The relation between orientability of the manifold, and space 
or/and time orientability is discussed for instance in \cite{Gernew,FF}.
Nevertheless, the existence of a consistent time 
orientation is crucial for the global causal structure and therefore 
we are forced to assume it here.
Moreover any non time-orientable Lorentzian manifold 
has a double-cover which can be time oriented, hence we can always perform
our study in this double covering. The condition of paracompactness is
added in many references, but this is redundant because, as it was
shown by Spivak \cite{SPIVAK},
a pseudo-Riemannian manifold is necessarily paracompact. 

An important point concerning definition \ref{spacetime}
is that the $C^1$ condition for the metric tensor 
cannot be improved to a larger $C^k$ with $k>1$ if one wishes to describe 
situations with different matter 
regions (such as the interior and exterior of stars, or shock waves) 
where these regions must be properly {\em matched}. The discontinuities 
of the matter content variables ---its energy density, for instance--- 
arise usually, via appropriate field equations (e.g. Einstein's field 
equations \cite{FF,MCCALLUM} or any of their relatives), through 
discontinuities on the second derivatives of the metric tensor, which 
must thus be allowed. Unfortunately, this poses enormous problems 
concerning causality structure, specially because the basic {\em purely
local} causal properties of the spacetime, which are fundamental for 
the construction of the whole theory of causality, might not hold in 
general $C^1$ metrics. A longer description of this problem can be 
found in \cite{SINGULARITY,WOOLGAR,DGS}, and will be briefly
considered later in subsections \ref{local}, \ref{carter-study} and
\ref{classic}. In order to avoid this annoying 
problem though ---despite it being completely fundamental!---, 
we will implicitly assume for most of this review that $\G$ is at 
least of class $C^2$.

In what remains of this section we follow the logical steps which allow us
to build 
a sensible notion of causality: first we consider the direct 
{\em algebraic} implications that the existence of the Lorentzian metric
$\G$ 
has at every single point $x\in V$ (subsection \ref{causal-tensors}); 
then we go a step further and 
construct the {\em local} causal structure of the spacetime, that is, 
at appropriate local neighbourhoods of any point (subsection 
\ref{local})); finally, we explore 
the causal relations between non-locally related points (subsection 
\ref{causal}), which 
require the study of causal connectivity properties, that is, using causal 
curves, and leads to the definition of the fundamental sets used in 
causality theory.

\subsection{The null cone. Causal tensors}
\label{causal-tensors}
Our first step towards the study of general properties of Lorentzian
manifolds 
concerns the classification that the Lorentzian metric $\G$ induces on 
the tangent bundle. As is obvious, the condition of time-orientability 
incorporated in definition \ref{spacetime} implies
that any {\em causal} vector $\v\in T_x(V)$ can be classified as 
either future-directed or past-directed according to whether
$\G|_x(\xiv,\v)>0$ 
or $<0$ (recall that two causal vectors can be orthogonal 
with respect to $\G$ only if they are null and proportional; 
besides, without 
loss of generality we have implicitly assumed that the future is 
defined by the arrow of $\xiv$.) This 
provides $T_{x}(V)$ with a two-sheeted cone, the {\em null cone}, 
which is the most basic causal object. Future-directed timelike 
(respectively null) vectors lie 
inside the ``upper'' part of the null cone (resp.\ on the upper cone
itself), 
and similarly the past-directed causal vectors on the lower part of 
the cone.

Surprisingly, the extension of this classification to higher rank tensors
has 
only been formulated very recently in \cite{PI}. Probably, the main 
difficulty was that there is no simple relation between the length 
$\G(\t,\t)=t_{a\dots b}t^{a\dots b}$ of a 
tensor $\t\in T_{s}|_{x}(V)$ and its ``causal character''. 
Nevertheless, one can use an equivalent definition of 
causal vector which can be translated to all tensors, namely, a 
vector $\v\in T_{x}(V)$ is future directed if and only if $\G(\v,\u)\geq
0$ for 
all future-directed vectors $\u\in T_{x}(V)$. Hence, future tensors 
can be defined as follows \cite{PI}.
\begin{defi}[Causal tensors]
A tensor $\t\in T_s|_{x}(V)$ is said to be future (respectively past) if 
$\t(\u_1,\dots,\u_s)\geq 0$ (resp.$\leq 0$) for all future-directed 
vectors $\u_1,\dots,\u_s\in T_x(V)$. A causal tensor is a tensor which is 
either future or past. 
\label{CAUS-TENSR}
\end{defi}
Observe that $\t$ is a future tensor if and only if $-\t$ is a past 
tensor. This definition extends straightforwardly to tensor fields and
the bundles $T^r_s(V)$. It can be easily seen that
the set of all causal tensors has an algebraic structure of a graded
algebra 
of cones generalizing the null cone.

Several useful characterizations of causal tensors, as well as 
applications and their basic properties, were given in \cite{PI} and 
subsequently improved and enlarged in \cite{CAUSAL, LETTER, 
CAUSAL-SYMMETRY,thesis}. Definition \ref{CAUS-TENSR} can be equivalently 
stated \cite{PI} by
(i) just demanding that $\t(\k_1,\dots,\k_r)\geq 0$ for all
future-directed
{\em null} vectors $\k_1,\dots,\k_r$ or (ii) requiring that 
$\t(\u_1,\dots,\u_r)> 0$ for all future-directed {\em timelike} vectors 
$\u_1,\dots,\u_r$. Simple criteria to ascertain when a 
given tensor is causal are given in 
\cite{SUP,PI,CAUSAL,CAUSAL-SYMMETRY}.

Of course, the condition of future tensor was known long ago in 
General Relativity for the case of symmetric rank-2 covariant 
tensors, but with another name and purpose: 
the future property is what was usually called the ``dominant 
energy condition'' for the energy-momentum tensor. This was a 
condition to be demanded upon energy-momentum tensors likely to 
describe a physically acceptable matter content. Thus, the 
property of a future tensor is sometimes referred to as the ``dominant 
property'', see \cite{SUP,PI,thesis} and references therein. 

A more important property of causal tensors (specially in the case of
rank-2 
tensors) is their relation to maps that preserve the null cone. This 
was proved in \cite{PI} and shows the deep connection of causal 
tensors with the elementary causal structure of the Lorentzian manifold. 
This connection and the properties and applications derivable thereof 
were exploited in \cite{CAUSAL} to generalize the notion of causal 
structure, and in \cite{LETTER,CAUSAL-SYMMETRY} to look for finite and 
infinitesimal transformations which preserve the causal structure, or 
part of it; see also \cite{thesis} for a self-contained full exposition.

We will come back to these novel matters later on in sections 
\ref{causal-characterization} and \ref{symmetry}, specially in subsections 
\ref{causal-preservation}, \ref{chains}, \ref{causal-symmetry} and 
\ref{CPVF} .

\subsection{Local causality}
\label{local}
The second step in the erection of the causal program is the 
extension to local neighbourhoods. To that end we need the classical 
ideas on local Riemannian normal coordinates and normal  
neighbourhoods, and the definition of causal curves.

Recall that a piecewise $C^k$ curve with $k\in\n$, $k\geq 1$,
is a set of $C^k$ maps (called arcs)
$\l_j:I_j\rightarrow V$, $j\in\n$ where $\{I_j\}$ is a countable set of
open intervals of
the real line such that the set $\bigcup_j\overline{\l_j(I_j)}$ is a
continuous curve in $V$.
\begin{defi}[Causal and timelike curves]
A piecewise $C^k$ curve $\g\subset V$ is said to be future-directed
timelike (resp.\ null, causal) if its tangent vector at all points $x$ 
where it is well defined is a timelike (resp.\ null,
causal) future-directed 
vector of $T_x(V)$, and furthermore the causal orientations of all 
the arcs $\l_j(I_j)$ are consistent. 
\label{timelike-curve}
\end{defi}
Note that the tangent vector to a piecewise $C^k$ curve is well
defined at every point except possibly at the intersections 
$\overline{\l_j(I_j)}\cap\overline{\l_{j+1}(I_{j+1})}$, which are called
corners. At 
these corners there may be two different tangent vectors, and the 
consistency condition included in definition \ref{timelike-curve} 
requires simply that these two vectors point into the same causal 
orientation (the future, say) at every corner. Obviously, causal 
curves cannot change their time orientation within any of its 
arcs due to the differentiability of the $\l_j$, and thus a 
piecewise $C^k$ causal curve is future directed if all the ``pairs of
tangent 
vectors'' at their corners are future pointing.

For global causality one needs to consider limits of sequences 
of differentiable curves, which of course do not need to be piecewise
differentiable curves themselves. This will be briefly considered 
in subsection \ref{causal}.

%% 
 % {\bf Remark 2.} For time oriented causal curves we choose the tangent
 % vector $\d/\d t$ as pointing towards the future (i.  e. $\l_j'(\d/\d
 % t)$ is a future-directed vector at all the points of the curve where
 % this equation makes sense).  This amounts to saying that the curve
 % goes towards the future for increasing values of the parameter $t$.
 %%

The {\em exponential map} (see \cite{BEE,FF,ONEILL,SINGULARITY} for
details) 
from an open neighbourhood ${\cal O}$ of
$\vec{0} \in T_x(V)$ into a neighbourhood of $x\in V$ maps a given
$\vec{v}\in {\cal O}$ into the point that reaches the geodesic
starting at $x$ with tangent vector $\v$ a unit of affine length away 
from $x$ provided this is defined. Given that geodesics
depend continuously on the initial conditions $x$ and $\vec{v}$, by
choosing adequate neighbourhoods this exponential map is a 
homeomorphism, and actually a diffeomorphism if $\G$ is $C^2$. This 
exponential map allows to define the classical Riemannian normal
coordinates
\cite{BEE,Ei,SINGULARITY} at a neighbourhood of $x$. 
Any such neighbourhood is called a {\em normal neighbourhood} of $x$ 
and they can be chosen to be convex \cite{Whi,Fr}.
The change to normal coordinates must be at least of class $C^1$ to keep a
differentiable atlas, and consequently
the geodesics must depend differentially on the initial conditions. 
As mentioned before, this will {\em not} happen 
in physical situations requiring the matching of two different regions
with different matter contents across a common boundary.
It is well-known that under such circumstances
there exists a local coordinate system, called admissible by 
Lichnerowicz \cite{Lich},
in which the metric is $C^1$ piecewise $C^2$, see
e.g. \cite{Lich,MS}. Thus,
in these situations there is no guarantee that the normal coordinate
neighbourhoods are differentiable at the matching hypersurface. This is
{\it crucial} for causality and therefore to all theories and matters 
which rely on it; see e.\ g.\ subsection 6.1 in \cite{SINGULARITY} for a 
detailed list of troubles arising in relation with singularity 
theorems due to this problem. In general, there are many
statements concerning causality theory which have only been proven under
the
restrictive assumption of a $C^2$ metric, some notable
exceptions can be consulted in
\cite{CARTER,Cla4,FF,WOOLGAR,DGS}. An example of these difficulties 
will be commented at the end of $\S$\ref{etio}, and some 
results obtained in \cite{WOOLGAR,DGS} in subsection \ref{classic}.
Having mentioned these fundamental 
difficulties---which are usually ignored or dismissed without 
mention---, and in order to avoid complicated subtleties, we will 
consider, when needed, and for the rest of the review unless otherwise
stated, 
that the metric is $C^2$, keeping in mind that this problem should be
eventually 
considered and solved.

One of the most important results in causality theory concerns the 
local causal properties of spacetime and states that the causality 
within normal neighbourhoods of any $x\in V$ is {\em analogous} to that of 
flat spacetime. The exact meaning here of the word ``analogous'' is 
very important: it has usually be interpreted as synonymous of 
``equivalent''. As we will largely discuss in subsection 
\ref{causal-preservation}, this equivalence can actually be
proven rigorously {\em whenever an appropriate definition of causal 
equivalence} is provided, and this proper definition {\em cannot} be 
the usual and simple local conformal relation between spacetimes. 
This latter part was explicitly showed by Kronheimer and Penrose in 
\cite{KRONHEIMER-PENROSE}, while the solution to the problem, 
selecting the right definition for causal equivalence, was only very 
recently given in \cite{CAUSAL}.
We postpone the discussion of these matters to section 
\ref{causal-characterization},
and here we only want to give the basic facts supporting the 
``analogy'' between the causality in local neighbourhoods and that of 
flat spacetime. To be precise, let us define the future light cone 
(resp.\ its interior) of $x\in V$ as the image of the future null cone
(resp.\ its interior) in ${\cal O}\subseteq T_x(V)$ by the exponential
map.
Observe that the light cone is only
defined on a normal neighbourhood of $x$. 
Then, one can prove a 
fundamental proposition (see \cite{BEE,FF,KRIELE,SINGULARITY}). The
following 
is its strongest version we are aware of in terms of 
differentiability \cite{SINGULARITY} (see also \cite{CARTER}).
\begin{prop}
Any continuous piecewise $C^1$ future-directed causal curve starting at 
$x\in V$ and
entirely contained in a normal neighbourhood of $x$ lies completely on
the future light cone of $x$ if and only if it is a null geodesic from
$x$, and
is completely contained in the interior of the future light cone of $x$
after the point at which it fails to be a null geodesic.
\label{local-causal}
\end{prop}
Observe that a non-differentiable curve composed by arcs of null
geodesics is {\it not} a null geodesic. Thus, any future-directed curve
constituted by an arc of null geodesic from $x$ to $y$ followed by
another null geodesic at the corner $y$ lies on the future light cone up
to
$y$ and enters into its interior from $y$ on. Notice also that any 
future-directed causal curve which is not a null geodesic at $x$
immediately 
enters and remains in the interior of the light cone with vertex at 
$x$, in particular all timelike curves from $x$ are completely contained
in this
interior.

\subsection{Global causality}
\label{causal}
The third step towards the study of general causal properties of
Lorentzian
manifolds is to show how this structure can be used to define
certain global objects or properties on the manifold. We are specially 
interested in the possible relations arising
between ``far apart'' points of the manifold because they can be
generalized to more abstract sets other than differentiable Lorentzian
manifolds.

The nice simple local causal structure shown in Proposition 
\ref{local-causal} does not hold globally in general, unless the 
spacetime is extremely well-behaved (such as for instance in flat 
spacetime), as we will see. Thus, all non-simple results concerning 
causality arise only as global aspects of the spacetimes, and
one needs to 
investigate and try to control the properties of Lorentzian 
manifolds globally. This is somehow frustrating, and paradoxical, if 
we have in mind physical theories such as General Relativity, which 
are {\em local} in essence, as the differential equations determining 
the metric tensor field are obviously local. In such physical 
theories, in order to prove very important results, such as the 
singularity theorems, or the properties of horizons, black holes, and 
so on, the causality theory is absolutely necessary. But this is a 
{\em global} set of properties, while the theory, as already said, is 
only {\em local}. The traditional solution to solve this dichotomy has 
been the use of {\em extensions}, so that if a local solution was 
found one tries to extend it beyond its domain of validity as many 
times as needed until the resulting spacetime is inextensible 
\cite{FF,SINGULARITY}. The problem with this strategy is that 
extensions are not unique, nor they can be determined by 
physical motivations or mathematical properties, see 
\cite{SINGULARITY} for a detailed discussion. In other words, 
extensions are arbitrary. And therefore, the global solutions needed 
to ascertain the causal properties of a spacetime rely on a very weak 
foundation.

Nevertheless, if one assumes that the global Lorentzian manifold is known,
or 
given by any means, the causality analysis can be pursued without 
problems. This is an interesting task on its own, as we can learn 
the different possible causal structures compatible with spacetimes, 
and the possible difficulties, or surprises, that may arise. 

The basic objects permitting the global analysis are 
causal curves, as they connect points in $V$ which do not have to lie 
on the same normal neighbourhood.

\subsubsection{Continuous causal curves on Lorentzian manifolds.}
The simplest causal curves are the piecewise $C^k$ future- or
past-directed
timelike and causal curves introduced in definition \ref{timelike-curve}.    
Curves will usually be denoted in this review with the letter $\g$ ---or
$\g(t)$ if we want to make explicit the parametrization of the curve. 
The piecewise $C^k$ condition imposed in definition \ref{timelike-curve}
is needed in order to be able to define a tangent vector at the points
of each arc $\l_j(I_j)$ of the curve. However, it is possible and 
indeed necessary to generalize the
previous definition in order to include curves which are only
continuous see \cite{BEE,FF,KRIELE,ONEILL,PENROSE}.
\begin{defi}[Continuous causal curves]
A continuous curve $\g\subset V$ is said to be causal and 
future-directed if for every point $x\in\g$ 
there exists a convex normal neighbourhood $\NN_x$ of $x$ 
such that for every pair $y,z\in\g\cap\NN_x$ ($y=\g(t_{1})$, 
$z=\g(t_{2})$ with $t_{2}>t_{1}$)
there is a $C^1$ future-directed causal arc
contained in $\NN_x$ from $y$ to $z$. 
\label{continuous-curve}
\end{defi} 
It is easily shown that continuous causal curves are in fact Lipschitz 
and thus differentiable almost everywhere \cite{PENROSE, KRIELE}.

Other standard concepts dealing with causal curves are presented next.
\begin{defi}[Inextensible curves]
A point $x$ is said to be a future (resp.\ past) endpoint of a continuous 
future (resp.\ past) directed causal
curve $\g(t)$ if for every neighbourhood $\U_x$ of $x$ there exist a value 
$t_0$ such that $\g(t)\subset\U_x$ for all $t>t_0$ (resp.\ $t<t_0$). 
Causal curves with no future (past) endpoints will be called future 
(past) endless. A curve $\g$ is inextensible if there is no curve $\g'$ 
containing $\g$ as a proper subset.
\label{endpoint}
\end{defi}
Clearly inextensible causal curves have no endpoints. 
The set of future-directed causal curves with
a past endpoint $a$ and a future endpoint $b$ will be denoted by
$C(a,b)$.  This set is a topological space under the $C^0$ topology
defined next.
\begin{defi}[$C^0$ topology]
The collection of sets of curves $\O(\U)=\{\g\in 
C(a,b):\g\subset\U\}$ where $\U\subset V$ is an open set,
constitute a basis for a topology
in $C(a,b)$ called the $C^0$ topology.
\label{c-0}
\end{defi}
It is possible to define convergence of curve sequences contained in
$C(a,b)$ using this topology.  Other different notions of convergence
on $C(a,b)$ (or more general sets of curves) can be defined and
studied although we will not pursue this matter further in this review
(see \cite{BEE,FF,KRIELE,PENROSE}).
\subsubsection{Basic sets used in causality theory.}
\label{J+}
We are now prepared to define the fundamental
binary relations between the points of a Lorentzian manifold
according to whether they can be joined by timelike, null
or causal curves (or none of them).
\begin{defi}[Causal relations]
Let $p,q\in V$: 
\begin{itemize}
\item $p$ is chronologically related with $q$,
written $p<\!\!<\!q$, if $C(p,q)$ contains timelike curves;  
\item $p$ is causally related with $q$, written $p<q$, 
if $C(p,q)$ is not empty;  
\item the relation $p\rightarrow q$ means that $p<q$ but not $p\less q$.
\end{itemize}
\label{causal-relations}
\end{defi}
These relations are standard in causality theory and they can be found
in many textbooks, e.\ g.\ \cite{BEE,FF,KRIELE,ONEILL,PENROSE,Wald}. 
Excellent surveys for non-experts are also \cite{Gernew,GERHOR}. 
We summarize next their basic properties as they will be needed later.
\begin{prop}
For a Lorentzian manifold $V$ the binary relations ``$\less$'' and
``$<$'' fulfill the following basic properties
\begin{enumerate}
\item $<$ is reflexive.
\item $<$ and $\less$ are transitive.
\item $p<q$ and $q\less z$ $\Rightarrow$ $p\less z$. 
$p\less q$ and $q<z$ $\Rightarrow$ $p\less z$.
\end{enumerate}
\label{properties-relations}
\end{prop}
As we will see in the next section these properties can be abstracted
to more general sets which do not need even to be 
topological spaces.  Using them the sets $I^{\pm}(p)$
(chronological future (+) and past (-) of $p$), $J^{\pm}(p)$ (causal 
future of and past $p$) and $E^{\pm}(p)$ (future and past horismos of
$p$) are
defined as (from now on we only give the definitions for future
objects assuming the obvious generalization for their past
counterparts)
$$
I^{+}(p)=\{x\in V: p\less x\},\ \ J^{+}(p)=\{x\in V:p<x\},\ \ 
E^+(p)=J^+(p)-I^+(p)\, ,
$$ 
from which we can construct $I^+(\U)$, $J^+(\U)$ and  $E^+(\U)$ 
for an arbitrary set $\U\subset V$
$$
I^+(\U)=\bigcup_{p\in\U}I^+(p),\ \ J^+(\U)=\bigcup_{p\in\U}J^+(p),\ \ 
E^{+}(\U)=J^+(\U)-I^+(\U).
$$ 
There are some variants of these definitions in which an auxiliary set 
$\W$ is employed
$$
I^+(p,\W)\equiv\{x\in V:\ \mbox{$p$ and $x$ can be joined by a timelike 
curve contained in $\W$}\},
$$
and similar definitions for $J^+(p,\W)$, $E^+(p,\W)$ and $I^+(\U,\W)$,
$J^+(\U,\W)$, $E^+(\U,\W)$.  These sets have well known topological
properties which again can be found in e.\ g.\ 
\cite{FF,PENROSE,BEE,ONEILL,Wald,KRIELE,SINGULARITY}:
\begin{enumerate}
\item $I^+(\U)$ is always open.
\item $I^+(\overline{\U})=I^+(\U)$.
\item $\overline{I^+(\U)}=\{x\in V:I^+(x)\subseteq I^+(\U)\}$=
$\overline{J^+(\U)}$.
\item $\d J^+(\U)=\d I^+(\U)=
\{x\in V: x\not\in I^+(\U)\ \mbox{and}\ I^+(x)\subseteq I^+(\U)\}$. 
\item int$(J^+(\U))=I^+(\U)$. 
\end{enumerate}
Another type of subset, playing an important role in one of
the most important constructions of {\em causal boundary}, is that of 
future (or past) set which can be
defined using the chronological sets. Recall that
a set $F$ is {\em achronal} if $F\cap I^+(F)=\varnothing$ or in other 
words no pair of points in $F$ can be joined by a timelike future-directed
curve.
\begin{defi}[Future sets, achronal boundary]
A set $F\subset V$ is called a future set if $I^+(F)\subseteq F$.  The
boundary $\d F$ of a future set $F$ is called an achronal boundary.
\label{futureset}
\end{defi}
Note that the terminology ``achronal boundary'' is standard but may be 
misleading as there are boundaries of non-future sets which are achronal 
\cite{PENROSE}, and this is why some authors have changed or omitted 
this terminology \cite{KRIELE,ONEILL,SINGULARITY}.
An interesting property of achronal boundaries is the next (see e.\ g.\
\cite{FF,KRIELE,ONEILL,PENROSE,SINGULARITY} for a proof.)
\begin{prop}
Any achronal boundary is a closed achronal Lipschitz hypersurface.
\label{lipschitz}
\end{prop} 
The set $\d J^+(\U)=\d I^+(\U)$ is an achronal boundary because it is the 
boundary of the future set $I^+(\U)$. This achronal boundary 
satisfies the following fundamental property 
\cite{FF,PENROSE,SINGULARITY,Wald} 
\begin{prop}
The set $\d J^+(\U)-\overline{\U}$ is formed by a disjoint union of 
null geodesics called null generators, their future endpoints which may 
be empty and an 
acausal set. Furthermore if a null generator has a past endpoint it must 
lie on $\overline{\U}$. 
\label{null-generator}
\end{prop}
The result that null geodesics starting at a point $p$ may meet other such null 
geodesics (formation of caustics) and the fact that null geodesics from $p$ in 
general leave the boundary $\d J^+(p)$ and enter into $I^+(p)$ are the typical 
features which destroy the naive picture we have of causality based on local, 
or flat spacetime, considerations.

The definitions stated before are focused on how points of a
Lorentzian manifold influence each other.  However, we are sometimes
interested in those points of $V$ influenced solely by a given
region of the Lorentzian manifold.  This is taken care of in
the following definition of future Cauchy development, which is
again standard  \cite{GEROCH-SPLITTING}. 
The future Cauchy development $D^+(\U)$ of a set $\U\subset V$ 
is the set of points of $V$ which can be
influenced exclusively by points of $\U$. More precisely
\begin{defi}[Cauchy development]
$x\in D^+(\U)$ if and only if every past-endless past-directed
causal curve containing $x$ intersects $\U$.
\label{development}
\end{defi}
Future and past Cauchy horizons $H^{\pm}(\U)$ of $\U$ are defined
as the future and past boundaries of the Cauchy developments of $\U$. 
They are formed by the points in 
the closure $\overline{D^+(\U)}$ which 
cannot be joined by a future-directed timelike curve with any other point
of 
$D^+(\U)$:
$$
H^+(\U)=\overline{D^+(\U)}-I^{-}(D^+(\U))\, .
$$
The total domain of dependence and the total Cauchy horizon of $\U$ are 
defined respectively as $D(\U)=D^+(\U)\cup D^-(\U)$, 
$H(\U)=H^+(\U)\cup H^-(\U)$. When $\U$ is closed and achronal, 
some general properties 
of these sets are \cite{BEE,FF,KRIELE,ONEILL,PENROSE,SINGULARITY}
\begin{enumerate}
\item int$D^{+}(\U)=I^+(\U)\cap I^-(D^+(\U))$.
\item $H^+(\U)$ is an achronal set.
\item $\overline{D^+(\U)}=\{x\in V:\mbox{every endless timelike past 
directed curve from $x$ meets $\U$}\}$.
\item $I^+[H^+(\U)]=I^+(\U)-\overline{D^+(\U)}$.
\item $\d D^+(\U)=H^+(\U)\cup\U$.
\end{enumerate}
A particularly good account of 
them can be found in the fundamental paper \cite{GEROCH-SPLITTING}.

\section{Causality in abstract settings}
\label{abstract}
In the previous section we have reviewed the main concepts used in the
causal analysis of Lorentzian manifolds.  In these manifolds,
causality stems from the peculiar properties that a metric of Lorentzian
signature provides, which ultimately allows to classify vectors, 
tensors, fields, sets, and curves as future, past or neither of the two.
This causal character for 
curves led to set certain relations between points of the manifold. 
In this section we follow the opposite way and ask ourselves which are
the properties of manifolds or sets where binary relations resembling
those obtained between points of a Lorentzian manifold ({\em cf.}
proposition \ref{properties-relations}) are present.  This approach could 
be termed as axiomatic, as we are trying to isolate what is genuine of 
causality regardless of
the existence of a Lorentzian metric. This axiomatic viewpoint was
not studied systematically in the literature until the 1960's. 
Perhaps the two most important and illuminating investigations dealing
with 
this subject can be found in the work by Kronheimer and Penrose
\cite{KRONHEIMER-PENROSE}, where the
properties presented in proposition \ref{properties-relations} are
axiomatized, 
and in Carter's thorough analysis \cite{CARTER} where
a further generalization is achieved. 
These two papers are the main subject of this section and their
contents will be discussed in some detail.

\subsection{The Kronheimer-Penrose causal spaces}
\label{causalspace}
The paper by Kronheimer and Penrose \cite{KRONHEIMER-PENROSE} is the
first general study of {\em causal spaces}.  Roughly speaking these
spaces are sets in which binary relations $\less$, $<$ and
$\rightarrow$ with properties similar to those presented in definition
\ref{causal-relations} have been set.  The precise definition as is
given in the Kronheimer-Penrose paper is presented next.
\begin{defi}[Causal space]
The quadruple $(X,<,\less,\rightarrow)$ is called a {\em causal
space} if $X$ is a set and $<$, $\less$, $\rightarrow$ are three
binary relations on $X$ satisfying for each $x,y,z\in X$ the following
conditions
\begin{enumerate}
\item $x< x$.
\item if $x< y$ and $y< z$, then $x< z$.
\item if $x< y$ and $y< x$ then $x=y$.
\item not $x\less x$.
\item if $x\less y$ then $x< y$.
\item if $x< y$ and $y\less z$ then $x\less z$.
\item if $x\less y$ and $y< z$ then $x\less z$.
\item $x\rightarrow y$ if and only if $x< y$ and not $x\less y$.
\end{enumerate}
The relations $<$, $\less$ and $\rightarrow$ are called respectively 
causality, chronology and horismos.
\label{causal-space}
\end{defi}
As we can see, this definition is fully inspired in the Lorentzian causal 
structure outlined in section \ref{sec:essentials}. However, a very 
important remark is that points (iii) and (iv) have been added, and 
they do {\em not} necessarily hold in Lorentzian manifolds. This is 
connected to the existence of closed timelike or causal loops, which 
are perfectly permitted in generic Lorentzian manifolds, though they may
be 
seen as undesirable, unphysical or even paradoxical. Kronheimer and
Penrose 
want to avoid these ``ill-behaved'' causal spaces from the start. It 
must be borne in mind, however, that there are spacetimes violating 
(iii) and/or (iv). Having said this, any Lorentzian manifold 
satisfying (iii) and (iv) (these are called chronological and causal
spacetimes, see definition \ref{SER} below) is a causal space as we deduce
from
proposition \ref{properties-relations}. However, a causal space is a
rather more general structure as $X$ does not need to be even a
topological space, although if a topology is present in $X$ its
interplay with some relevant sets constructed from the causal and
chronological relations was also considered in \cite{KRONHEIMER-PENROSE} 
as we will discuss in section \ref{sec:CandT}.

Immediate properties arising from definition \ref{causal-space} are
$\rightarrow\, \subset\, <$, $\less\, \subset\, <$, $<$ is 
a partial order,
$\less$ is anti-reflexive and transitive and $\rightarrow$ is horismotic. 
Another direct consequence is that for any three points $x$, $y$, $z$
in a causal space $X$ such that $x< y< z$ and $x\rightarrow z$
then $x\rightarrow y\rightarrow z$.

From the relations of definition \ref{causal-space} we may construct
their inverses getting a new quadruple which turns out to be a new
causal space called dual causal space.  In the case of a Lorentzian
manifold it corresponds to the interchange of $<$, $\less$ and
$\rightarrow$ by $>$, $\greater$ and $\leftarrow$ being these relations
defined using past-directed causal curves instead of future-directed
ones.  Any subset $Y$ of a causal space can be transformed into a
causal space by just considering the binary relations $<_Y$,
$\less_Y$ and $\rightarrow_Y$.  In this case $Y$ is a causal subspace
of $X$.

The existence of these relations in a set $X$ is enough to define the
generalizations of the chronological future, causal future and future
horismos for any element $p\in X$
$$
I^+(p)\equiv\{x\in X:p\less x\},\ J^+(p)\equiv\{x\in X:p< x\},\ 
E^+(p)\equiv J^+(p)-I^+(p)\, .
$$
%($C^+$ is the notation for $H^+$ in Kronheimer-Penrose's paper).
From them the chronological, causal and future horismos of any subset
of the causal space $X$ are defined straightforwardly. Using the 
standard definition of chain for any given binary relation one can 
then define {\em chronological chains} and {\em causal chains} which are
in a sense the generalization of timelike and causal curves, respectively.

Other interesting objects are   
$$
[p,q]=\{z\in X:p< z< q\},\ \ <x,y>=\{z\in X: x\less z\less y\}
$$
which can be seen equivalent to $J^+(p)\cap J^-(q)$ and to $I^+(x)\cap
I^-(y)$.
The relation $x||y$ means that neither
$x< y$ nor $y< x$ and a set $S\subset X$ such that $x||y$ for 
all $x,y\in S$ is said to be {\em acausal} (notice that in 
\cite{KRONHEIMER-PENROSE} such a set is called ``spacelike'', but 
this would enter in conflict with the usual definition of spacelike 
subsets in spacetimes, as there are spacelike sets which are not 
acausal.)

As we see a great deal of the basic notions and sets introduced in
causality theory are present in abstract causal spaces. 
Kronheimer and Penrose study further different aspects of causal spaces
some
of which are detailed next.
\begin{defi}[Regular causal spaces]
A causal space is called regular if for any four different 
points $x_1$, $x_2$, $y_1$, $y_2$ such that $x_i\rightarrow y_j$,
$i,j=1,2$ 
then $x_1||x_2$ if and only if $y_1||y_2$.
\label{regular}
\end{defi}
Causal spaces can be further classified according to the properties of
the chronology $\less$.
\begin{defi}
The causal space $X$ is said to be 
\begin{enumerate}

\item future reflecting if $I^{-}(x)\subset I^{-}(y)$ whenever
$I^+(x)\supset I^{+}(y)$.  $X$ is reflecting if it is future and past
reflecting.

\item weakly distinguishing if both $I^{+}(x)=I^+(y)$, $I^-(x)=I^-(y)$
entail $x=y$.  Future distinguishing if $I^+(x)=I^+(y)$ implies
$x=y$, and similarly for past distinguishing.

\item full if the following two conditions and their dual versions hold: 
\begin{itemize}
    \item $\forall x\in X$, $\exists p\in X$ such that $p\less x$; 
    \item $\forall p_1,p_2\less x$ $\exists q$ such that $p_1\less q$, 
    $p_2\less q$ and $q\less x$.
\end{itemize}    
\end{enumerate}
\end{defi}
Note that these three properties refer exclusively to the relation 
$\less$, and thus we can talk about reflection, distinguishing and 
fullness of any anti-reflexive and transitive binary relation $\less$.
Point {\em (iii)} is always satisfied in a spacetime due to the 
fundamental proposition \ref{local-causal} and the openness of the sets 
$I^+$. Regarding {\em (i)} and {\em (ii)}, we will actually meet them
again
in definition \ref{SER} as they are basics steps in the standard causal 
hierarchic classification of Lorentzian manifolds.

A causal space $X$ can be endowed with a natural topology in view of
the properties of chronological futures and past of manifolds.

\begin{defi}[Alexandrov topology]
The Alexandrov topology for a causal space $X$ is the coarsest
topology in which the sets $I^+(x)$ and $I^-(x)$ are open for every
element $x\in X$.
\label{alexandrov}
\end{defi}
Observe that the sets $I^+(x)\cap I^-(y)$ are open for all $x,y\in X$.
We will provide further details about this and other topologies in
section \ref{natural}.

The next important topic addressed in \cite{KRONHEIMER-PENROSE} 
is the study of the necessary restrictions to be imposed on a set equipped 
with at least one binary relation with the properties of either 
$<$, $\less$ or $\rightarrow$ such that the remaining necessary 
binary relations can be added in order to get a causal space. 
The notation is fixed when we have a set $X$ and two binary
relations (say $\rightarrow$ and $<$) specified on it as follows:
\begin{defi}
For the triad $(X,\rightarrow,<)$ it is said that $\rightarrow$ is
{\em horismos compatible} (resp.  regularly compatible) with the
causality $<$ if there exists a relation $\less$ such that the
quadruple $(X,<,\less,\rightarrow)$ is a (regular) causal space. 
The set 
$$
\{\rightarrow\subset X\times X:\, \exists\less\ \mbox{such that}\  
(X,\rightarrow,<,\less)\ \mbox{is a (regular) causal space}\}
$$
is denoted by $\{(reg.)hor|cau <\}$
\label{hor-compatible}
\end{defi}
The definitions of 
$\{(reg.)hor|chr\less\}$, $\{(reg.)chr|cau <\}$ et cetera, are similar.

The construction of causal spaces from spaces with a single relation
(a chronology, a causality or a horismos) can then be attacked 
separately \cite{KRONHEIMER-PENROSE}.
\begin{defi}[Construction from the horismos]
Let $\rightarrow$ be a horismotic relation on a set $X$. 
Then, two further relations 
$<^{\Uf}$ and $\less^{\Uf}$ can be defined on $X$ as follows
\begin{enumerate}
\item $x<^{\Uf}y$ $\Leftrightarrow$ there exists a finite sequence 
$(u_i)_{1\leq i\leq n}$ satisfying
$$
x=u_1\rightarrow u_2\rightarrow\dots\rightarrow u_n=y,
$$
\item $x\less^{\Uf}y$ $\Leftrightarrow$ $x<^{\Uf}y$ and not $x\rightarrow
y$.
\end{enumerate}
\end{defi}
It is not difficult to check that
$(X,<^{\Uf},\less^{\!\!\!\Uf},\rightarrow)$ is a causal space. 
Therefore $<^{\Uf}\in\{cau|hor \rightarrow\}$ and
$\less^{\!\!\Uf}\in\{chr|hor\rightarrow\}$.  Moreover for any
$<\, \in\{cau|hor\rightarrow\}$ the relations $x<^{\Uf} y$
imply that in fact $x< y$ and the same for $\less$ and
$\less^{\Uf}$, hence $<^{\Uf}$ and $\less^{\Uf}$ can
be defined in an alternative and more transparent way
\be
<^{\Uf}=\cap\{cau|hor\rightarrow\},\ \ 
\less^{\Uf}=\cap\{chr|hor\rightarrow\}.
\label{intersection}
\ee 
\begin{defi}
A causal space $(X,<,\less,\rightarrow)$
is a $\Uf$-space if the equalities $<=<^{\Uf}$ and $\less=\less^{\Uf}$ 
hold.
\label{U-space}
\end{defi}
The construction from the chronology relation is as follows
\begin{defi}[Construction from the chronology]
Consider a set $X$ endowed with an anti-reflexive and transitive 
binary relation $\less$ . Two further relations 
$<^{\Bf}$ and $\rightarrow^{\Bf}$ on $X$ can then be defined as follows 
\bnr
x<^{\Bf}y\Leftrightarrow I^+(x)\supset I^+(y)\ 
\mbox{and}\ I^{-}(x)\subset I^{-}(y)\\
x\rightarrow^{\Bf} y\Leftrightarrow x<^{\Bf}y\ 
\mbox{and not}\ x\less y. 
\enr
\label{Bf}
\end{defi}
For any $<\in\{cau|chr\less\}$ we have that $x< y\, 
\Rightarrow\, x<^{\Bf} y$.  Conversely, if $x<^{\Bf} y$
there always exists $<\in\{cau|chr\less\}$ such that $x< y$.  Hence
\be
<^{\Bf}=\cup\{cau|chr\less\},\ \ \rightarrow^{\Bf}=\cup\{hor|chr\less\}
\label{union}
\ee
In this case the quadruple $(X,<^{\Bf},\less^{\Bf},\rightarrow^{\Bf})$ 
does not always satisfy {\em (iii)}
of definition \ref{causal-space}. In fact, 
$(X,<^{\Bf},\less^{\Bf},\rightarrow^{\Bf})$ is a causal space 
if and only if $\less$ is weakly distinguishing.

\begin{defi}
A causal space $(X,<,\less,\rightarrow)$ is called a $\Bf$-space if 
$<=<^{\Bf}$ and $\rightarrow=\rightarrow^{\Bf}$.
\end{defi}
From the above considerations we deduce that any $\Bf$-space must be
future distinguishing.  Furthermore a sufficient condition for the
causal space $(X,<,\less,\rightarrow)$ to be a $\Bf$-space is that
$x< y$ whenever $x<^{\Bf} y$.

Finally the construction of a causal space from a space with a single
causal relation $<$ can be achieved as follows.
\begin{defi}[Construction from the causality]
Let $<$ be a partial order on a set $X$. Define two new binary
relations $\rightarrow^{\Cf}$, $\less^{\Cf}$ on $X$ by
\begin{enumerate}
\item $x\rightarrow^{\Cf}y$ $\Leftrightarrow$ $x< y$ and 
$<$ linearly orders every $[u,v]\subset[x,y]$.
\item $x\less^{\Cf}y$ $\Leftrightarrow$ $x< y$ and not $x\rightarrow^{\Cf}
y$. 
\end{enumerate}
Then, $(X,<,\less^{\Cf},\rightarrow^{\Cf})$ is a causal space.
\label{Cf}
\end{defi}
In this case, the counterpart 
of (\ref{intersection}) and (\ref{union}) is 
\be
\rightarrow^{\Cf}=\cup\{(reg)hor|cau <\},\ \ 
\less^{\Cf}=\cap\{(reg)chr|cau <\}
\label{mix}
\ee
$\Cf$-spaces are defined similarly to $\Uf$-spaces and $\Bf$-spaces. 
$\Cf$-spaces do not need to be regular. A sufficient condition for a
regular 
causal space 
$(X,<,\less,\rightarrow)$ to be a $\Cf$-space is the inclusion 
$\rightarrow^{\Cf}\, \subseteq\, \rightarrow$, which is equivalent to
$\less^{\Cf}\, \subseteq\, \less$.

\subsubsection{Connectivity properties}
Kronheimer and Penrose also considered the possible representation 
that the intuitive ideas of ``null geodesic'' or ``null arc'' 
could have in their abstract setting. The main definition is
\begin{defi}[Girders, hypergirders and proper beams]
Let $G\equiv (g_i)_{1\leq i\leq N}$ be a finite chain of $N\geq 3$ points
ordered 
by the relation $<$ (i.e. $g_i< g_{i+1}$, $\forall i<N$). 
$G$ is called a girder if $g_i\rightarrow g_{i+2}$.
A nonempty subset $H\subset X$ is called a hypergirder if $\forall x,y\in
H$ 
there exists a girder $G\subset H$ containing $x$ and $y$. A maximal 
hypergirder is called a proper beam.
\label{girder}
\end{defi}
The definition of girder implies that in fact $g_i\rightarrow g_{i+1}$ 
so the elements of $G$ can be arranged according 
to the following diagram 
\bnr
\dots g_{i-2}\longrightarrow g_i\longrightarrow g_{i+2} \dots\\
\ \ \ \ \ \ \ \searrow\ \nearrow\ \ \ \searrow\ \ \nearrow\\
\ \ \ \ \dots g_{i-1}\longrightarrow g_{i+1}\dots    
\enr
As we see, a hypergirder is the analogous to a null 
geodesic arc on Lorentzian manifolds, while proper beams are the
generalization of inextensible null geodesics. Actually, using of Zorn's
lemma (or equivalently the axiom of choice)
one can prove that any hypergirder is a subset of a proper beam. 

Two points $x$ and $y$ are said proximate if they belong to some 
girder, or in other words, if there exists a girder $x\rightarrow 
z\rightarrow y$.  Then, the following result holds
true for regular causal spaces.
\begin{theo}
Each pair of proximate points belong to a single proper beam if and
only if the underlying causal space is regular.
\label{proximate}
\end{theo} 
To include certain pathological cases proper beams are generalized to
beams.  These are sets which are either proper beams or just consist
of two points $\{a,b\}$ ordered by the horismos and not contained in
any hypergirder.  This generalization allows us to formulate the
following result
\begin{theo}
In any causal space any non-trivial set linearly ordered by the
horismos is a subset of some beam.
\label{beam-subset}
\end{theo}
To end this subsection giving a flavour of the power of this abstract 
construction and the relation to standard or intuitive theorems and 
results on Lorentzian manifolds, we present a theorem
involving beams which generalizes the
well known decomposition of $\d J^+(S)$ for
spacetimes in null generators, their past endpoints and its edge 
(compare with the spacetime versions in e.\ g.\ \cite{BEE,FF,PENROSE}).
\begin{theo}
Let $A$ be any acausal subset of a regular causal space $X$ and
construct 
$$
A_0=\{x\in A: x< z\ \mbox{for some $z\in E^+(A)$}\},\ \ S=E^+(A)-(A-A_0). 
$$  
The set $B\cap S$ is called a generator of $S$ if $B$ is a beam
intersecting $S$ more than once.  In this case
\begin{enumerate}
\item $S$ is the union of its generators.
\item Each generator has a first point in $A_0$.
\item The set of last points of the generators, if not empty, is 
acausal.
\item Any point which is common to two different generators is either the
first point or the last point of both.
\end{enumerate}
\end{theo}
A generalization of the Kronheimer-Penrose construction with 
applications to quantum physics can be found in \cite{SZABO,SZABO2}.

\subsection{Carter's study of causal spaces}
\label{carter-study}
Carter's paper \cite{CARTER} takes a stride forward and gives a more
accurate analysis of the causal binary relations. Almost all the
definitions and results in \cite{CARTER} are established for a 
{\em space-time manifold}
but the author cares to point out which of these results can or cannot be
generalized to arbitrary sets. This is done by the introduction of 
{\em etiological spaces}\footnote{Carter proposed the name etiology
for the branch of topology devoted to abstract causal spaces.
Etiology is the branch of knowledge concerned with causes.}, see
subsection 
\ref{etio}. A basic difference with \cite{KRONHEIMER-PENROSE} is that 
axioms (iii) and (iv) in definition \ref{causal-space} are not assumed 
from the beginning, which is more adapted to what actually happens in 
General Relativity and general spacetimes. This allows for a thorough 
analysis of the questions of causality violation (``vice'') and 
causal good behaviour (``virtue'') in general causal spaces---we will
further
treat these matters in section \ref{causal-characterization}. 
Carter's paper is very detailed
and contains a large number of concepts which makes it, together 
with \cite{KRONHEIMER-PENROSE}, one of the
most complete references dealing with abstract causal theory.
Of course, here we cannot cover all the material presented
in \cite{CARTER} but we hope to convey the most important ideas.

To start with, the basic definition we adopted in definition
\ref{spacetime} for 
spacetime is more restrictive than that used in \cite{CARTER}, where
a spacetime is taken to be just a connected $C^1$ $n$-dimensional 
differentiable manifold $M$ on which a continuous 
oriented null-cone structure is defined.
An oriented null-cone structure is a continuous linear mapping of the
$n$-dimensional solid Euclidean half cone defined in Cartesian
coordinates $\{x^1,\dots,x^n\}$ by
$$
x^n\geq\sqrt{\sum_{r=1}^{n-1}(x^r)^2},
$$ 
into each fibre of the tangent bundle. 
Put another way we are providing each point of the manifold with a cone so
we could also term this structure as a ``Lorentzian cone field'' on the
manifold $M$.  The image of the Euclidean half cone on each tangent
space is the future null cone whereas
the image of the opposite half of the Euclidean cone is the past null
cone. Particular cases of space-time manifolds in this sense
are the so-called {\em conformal structures} 
(see \cite{KRIELE,SEGAL}), while a generalization is given by the 
{\em conal manifolds}
in which the Euclidean cone is replaced by a 
closed convex pointed cone $C\subset V$ where $V$ is an 
$n$-dimensional vector space \cite{SEGAL,LAWSON} 
(see also $\S$\ref{Lie-subsemigroup}).

Once the null-cone structure is given, definitions of concepts such as
timelike, null or spacelike vectors and future oriented continuous
curves proceed along obvious lines so all these ideas will be
taken for granted (\cite{CARTER} is systematic in the definition of
these and other related concepts such as spacelike or timelike
submanifolds, inextensible and maximal subsets, etc), as they are 
equivalent in an obvious sense to those we have already given. 
Here we are more interested in the study of binary relations arising 
from the causality in much the same way as we 
have done in the previous subsection.

\begin{defi}[Qualified causality and chronology relations]
Let $\Ss$ and $\Ts$ be two subsets of $M$ and consider an auxiliary
set $\Us$ also in $M$.  We will say that $\Ts$ lies in the causal
future of $\Ss$ with respect to $\Us$, denoted as
$\Ss$\raisebox{-1.5ex}{$\stackrel{\mbox{\large$<$}}{\mbox{\tiny$\Us$}}$}$\Ts$,
if for every point $x\in\Ts$ there is a past-directed causal continuous
semi-arc contained entirely within $\Us$ and intersecting some point
of $\Ss$. Similarly, $\Ts\sucu\Ss$ if for any $x\in\Ts$ there is a
future directed causal continuous semi-arc contained entirely within
$\Us$ and intersecting some point of $\Ss$.  The relations
$\Ss\lessu\Ts$ and $\Ts\greatu\Ss$ are defined similarly.
\label{carter-relations}
\end{defi}

The attribute ``qualified'' here means that the
relations are defined through the reference to a subset
$\Us\subset M$.  Unqualified relations are those in which $\Us$ is
the manifold $M$ itself. In this last case the
subscript will be dropped from the binary causal relations.

Observe also that in definition \ref{carter-relations} 
causal and chronological relations are defined
between subsets of $M$ as opposed to the case previously 
considered in Kronheimer and Penrose's work where
only causal relations between points were defined. In the notation of 
section \ref{sec:essentials}, and for Lorentzian manifolds,
the previous sets can be defined as follows
$$
\Ss\raisebox{-1.5ex}{$\stackrel{\mbox{\large$<$}}{\mbox{\tiny$\Us$}}$}\Ts
\Longleftrightarrow \Ts\subset J^+(\Ss ,\Us), \,\,\,\,
\Ss\lessu\Ts \Longleftrightarrow \Ts\subset I^+(\Ss ,\Us),
$$
and so on. Therefore, it should be noticed that according to definition 
\ref{carter-relations},
$\Ss\precu\Ts$ and $\Ts\sucu\Ss$ have in general different meanings
and the same happens with $\Ss\lessu\Ts$ and $\Ts\greatu\Ss$ (this is
a difference with the Kronheimer and Penrose causal relations $<$,
$\less$ and $\rightarrow$).  Only in the case in which the sets $\Ss$
and $\Ts$ consist of a single point can the statements 
$\{x\}$\precu$\{y\}$
and $\{x\}\lessu \{y\}$ be read from right to left.  In fact in this last
case
the unqualified relations are just the causal relations of Kronheimer
and Penrose. 

The notions of causal and chronological future can be given
straightforwardly 
once the previous relations have been set. Let us simply remark that 
they received specific notation in \cite{CARTER} 
$$
\lessu\ \Ss)\equiv\{x\in M:\Ss\lessu\ x \}=I^+(\Ss,\Us),\hspace{5mm}
%(\Ss\greatu\equiv\{x\in M:x\ \greatu\Ss\}\\
\precu\ \Ss)\equiv\{x\in M:\Ss\precu\ x \}=J^+(\Ss,\Us),
%\ (\Ss\sucu\equiv\{x\in M:x\ \sucu\Ss\},
$$
and similarly the qualified {\em horismoidal future} of $\Ss$ was 
written as
$$
\darrowu\Ss)=E^+(\Ss,\Us).
%\equiv\precu\Ss)-\lessu\Ss),\ 
%(\Ss\garrowu =E^-(\Ss,\Us).
%\equiv(\Ss\sucu-(\Ss\greatu.
$$
This notation does not seem to have been used ever since. 
One can define, however, a new binary relation called {\em horismos
relation}.
\begin{defi}[Qualified horismos relation]
For any two subsets $\Ss,\Ts\subset  M$, the qualified future and past
horismos relation with respect to $\Us\subset  M$ are defined by
$$
\Ss\darrowu\Ts\Leftrightarrow\Ts\subset E^+(\Ss,\Us),\ \ \ \
\Ts\garrowu\Ss\Leftrightarrow\Ts\subset E^-(\Ss,\Us).
$$
\end{defi}

Carter also used new notation for the straightforward generalization of 
the Cauchy developments, namely
$$
[\Ss]\sucu =D^+(\Ss,\Us),\ \ \ 
[\Ss]\greatu =\overline{D^+(\Ss,\Us)},\ \ \
\precu[\Ss]=D^-(\Ss,\Us),\ \ \
\lessu[\Ss]=\overline{D^-(\Ss,\Us)}
$$
which allow to define Cauchy causality relations between subsets of 
$ M$ whenever they are qualified on an {\em open} $\Us$ (or on 
timelike proper submanifolds \cite{CARTER}).
\begin{defi}[Qualified Cauchy causality relations]
Let $\Us$ be an open subset of $ M$.  For any two sets
$\Ss, \Ts\subset M$ we say that $\Ts$ lies in the future causal
Cauchy development of $\Ss$ with respect to $\Us$ if 
$\Ts\subset D^+(\Ss,\Us)$. This is denoted by $\Ss\ ]\sucu\Ts$. 
The relation $\Ts\precu[\Ss$ 
is defined analogously using $D^-(\Ss,\Us)$. Finally
$$
\Ss\ ]\greatu\Ts \Longleftrightarrow \Ts\subset 
\overline{D^+(\Ss,\Us)}\, , \hspace{5mm}
\Ts\lessu\ [\ \Ss \Longleftrightarrow \Ts\subset 
\overline{D^-(\Ss,\Us)} \, .
$$
\end{defi}

The main properties of all these binary relations between the subsets 
of $ M$ are summarized in the following proposition \cite{CARTER} 
\begin{prop}[Basic properties of Carter's causal relations]
The following statements are true for any subsets $\Us,\Vs\subset M$
(here, the qualifying subset $\Us$ must be 
taken as either open or a timelike submanifold
whenever this is necessary):
\begin{enumerate}
\item 
$$
\Us\subset\Vs\Rightarrow\left\{
\begin{array}{cc}
\precu\subset\precv, & \lessu\subset\lessv\\
\sucu\subset\sucv, & \greatu\subset\greatv
\end{array}\right.
$$
\item 
$$
\lessu\subset\precu,\ \ \ ]\sucu \subset\, ]\greatu\subset\precu,\ \ \
\greatu\subset\sucu ,\ \ \ \precu[\, \subset \lessu [ \, \subset 
\sucu \, .
$$
\item The relations $\precu,\ \sucu$ are reflexive {\em only}
for sets $\Ss\subset \Us$. The corresponding unqualified relations are 
thus reflexive.
\item The relations $]\sucu,\ \precu[,\ \ ]\greatu , 
\lessu\ [$ are all reflexive.
\item All relations $\lessu, \greatu, \precu, \sucu, 
]\sucu,\ \precu[,\ \ ]\greatu , \lessu\ [$ are transitive.
\end{enumerate}
\label{basic-carter}
\end{prop}

\subsubsection{Etiological spaces.}
\label{etio}
Even though most of \cite{CARTER} assumes a manifold structure, 
Carter also considered the question of how one can use the previous 
binary relations in order to get an axiomatic definition of {\em
etiological} 
space, which is a generalization of Kronheimer-Penrose's causal space. 
\begin{defi}[Etiological space]
A topological space $\Xs$ endowed with two binary relations $<$
and $\less$ is an etiological space if the following axioms are
fulfilled
\begin{description}
\item[Axiom 1.] $\forall x\in\Xs$, $x< x$.
\item[Axiom 2.] Both $<$ and $\less$ are transitive.
\item[Axiom 3.] $\less\subset<$ in the sense of binary relations.
\item[Axiom 4.] If $x\less x$ for some $x\in\Xs$ $\Rightarrow$ $\exists
y\in\Xs$, $y\neq x$: $x\less y$, $y\less x$.
\item[Axiom 5.] The topology of $\Xs$ is a refinement of the Alexandrov 
topology constructed from $\less$.
\end{description} 
\label{etiological}
\end{defi}
The great advantage of this definition is that {\em any} space-time
manifold 
$ M$ (in the sense used in this section, that is, with a null-cone
structure, 
hence also for those complying with the standard definition
\ref{spacetime})
is an etiological space. This is clear by identifying the binary 
relations $<$ and $\less$ with the unqualified relations between 
subsets of definition \ref{carter-relations} in an obvious sense, that is
to say, the 
meaning of $x< x$ is simply $\{x\}< \{x\}$, and analogously 
$x\less x\, \Leftrightarrow\, \{x\}\less \{x\}$.
%% 
 % If one wishes to work with the qualified relations instead the subsets
 % of $ M$ employed to define these relations cannot be arbitrary as
 % they must be part of an {\em etiological infrastructure} (qualified
 % relations in this context are as in definition
 % \ref{carter-relations}).
 % \begin{defi}
 % The qualified relations $\precu$, $\lessu$ with respect to the subset
 % $\Us\subset M$ form an etiological infrastructure if (i) they do not
 % relate points outside $\Us$; (ii) $\precu$ and $\sucu$ fulfills the
 % first four axioms of definition \ref{etiological}; (iii) if $\Us$ is
 % open then its induced topology is a refinement of the Alexandrov
 % topology defined from $\lessu$.
 % \label{infrastructure}
 % \end{defi}
 %%

Clearly an etiological space is more general than a causal
space. To start with, the former is defined in terms of just two binary 
relations, instead of three. Moreover definition \ref{etiological} relies 
on a less number of axioms than definition \ref{causal-space}. It is
easily 
checked that points {\em (i), (ii), (v)} in definition \ref{causal-space} 
are covered by the first three axioms of etiological spaces. However, 
axiom 4 is more general than the corresponding points {\em (iii)} 
and {\em (iv)} in definition \ref{causal-space}. So the definition of 
etiological space is also more general in this sense. It might seem,
however, that Kronheimer-Penrose's definition is less restrictive in 
another sense, as it does not require the background set $X$ to be 
a topological space. Nevertheless, as is manifest from definition 
\ref{alexandrov}, any causal space is always a topological space with 
the Alexandrov topology, and thus this requirement for the 
etiological space is not restrictive at all (provided, of course, 
that the chosen topology for $X$ is the Alexandrov topology.)

Etiological spaces include all possible Lorentzian manifolds, 
including those which violate points {\em (iii)} and/or {\em (iv)} of 
the definition \ref{causal-space} of causal space. As mentioned 
before, the failure of these two points may seem not very satisfactory 
from a physical point of view, as there is a general consensus that any
model 
of the real physical world should actually comply them. Nevertheless, 
and given that such ``vicious'' spaces do appear in the theory of 
General Relativity (e.g., the G\"odel universe \cite{GODEL}, see section 
\ref{causal-characterization} and \cite{KRIELE}), and there is still some 
controversy as to whether or not they might be valid in some extreme
regime, 
it seems more appropriate from a scientific point of view not to exclude
them 
{\em by axiom}, as was argued also in \cite{CARTER}. In this way, at 
least we can study them and draw conclusions about their predictions 
and their absurdity, if this is so. Carter's paper contains an 
interesting discussion and a
classification of space-time manifolds according to the
``virtuousness'' of the causal relations present in $ M$ (of course
these results can be translated to etiological spaces). Other 
relevant references are \cite{FF,KRIELE} and references therein. We will
discuss these issues in section \ref{causal-characterization} as we
believe they have a closer relationship with the contents discussed
there.

In any case, Carter's paper thoroughly discusses which further axioms must 
be added to definition \ref{etiological} in order to get a causal space 
in the sense of Kronheimer and Penrose. These are essentially two
\begin{description}
\item[\small Causality principle.] There are no $x$, $y$, $x\neq y$ such
that 
$x< y$ and $y< x$.
\item[\small Strong transitivity.]
For any $x, y, z\in \Xs$ we have 
$$
x< y,\ y\less z\Rightarrow x\less z,\ \
x\less y,\ y< z\Rightarrow x\less z \, .
$$
\end{description}
The first of these takes care of the previous discussion, as then 
points {\em (iii)} and {\em (iv)} of definition \ref{causal-space} 
hold. This is clear for point {\em (iii)} which is exactly the 
causality principle. Point {\em (iv)} is {\em then} a consequence of 
this and axiom 4. As a matter of fact, provided that the causality 
principle is assumed, one could substitute axiom 4 in definition 
\ref{etiological} by the following
\begin{description}
\item[\small Chronology principle.] There is no element $x\in\Xs$ such
that 
$x\less x$.
\end{description}
Concerning the second added axiom, the strong transitivity principle, 
it was (erroneously) argued in \cite{KRONHEIMER-PENROSE} that this is a
property 
satisfied by all spacetimes. As pointed out by Carter \cite{CARTER}, 
this is true {\em only} for sufficiently differentiable structures. 
We meet once more the question of the differentiability of the 
metric (or of the null-cone structure), that we have already 
discussed briefly in subsection \ref{local}. It will never be 
sufficiently 
stressed the importance of these differentiability matters (at least from
a 
mathematical viewpoint), as many ``trivial'' or ``obvious'' results, 
which are taken for granted, do not actually hold or have not been proved.

All spacetimes with a $C^2$ metric, or with a differentiable 
and not merely continuous null-cone structure, satisfy the strong 
transitivity principle. In summary, all spacetimes are etiological,
all causal spaces in the sense of Kronheimer-Penrose  are etiological 
too (provided the Alexandrov topology is used), and all etiological 
Lorentzian manifolds with a $C^2$ metric and satisfying 
the causality principle are causal spaces. It should be stressed, 
however, that the causality principle does not remove all possible 
causal pathologies \cite{CARTER,FF,KRIELE,SINGULARITY}, as will be
discussed in 
section \ref{causal-characterization}.

For further details and developments concerning causal and etiological 
spaces, consult \cite{SZEKERES}.

In both the Kronheimer-Penrose and Carter constructions the notions of
causal and etiological space are based on
abstractions of certain simple relations which can
be set between points and subsets of a spacetime due to the presence of a
Lorentzian metric.  However it is still not clear how one can go the
other way round, namely, starting from the axiomatic relations present
in a causal or etiological space, and
trying to reconstruct the Lorentzian metric, or a class of Lorentzian 
metrics, compatible with the given causal binary relations.  A number of
investigations have tackled this question and we will give an account
of them in section \ref{causal-characterization}.

\subsection{Physically inspired axiomatic approaches}
\label{physics}
References \cite{KRONHEIMER-PENROSE} and \cite{CARTER} adopt a set of
axioms 
abstracted
from the mathematical properties of Lorentzian manifolds and from them
they derive general results.  There is however, another way to proceed
which is trying to motivate the introduction of Lorentzian manifolds
as models of spacetimes right from objects and concepts which can be
touched and experimented upon, which exist in every day's common 
life, or which seem to be intuitively incontrovertible (taken as axioms)
\cite{EHLERS,WOODHOUSE,SCHROTER,SCHROTER2}. 
This is the idea behind the approach
taken by Ehlers, Pirani and Schild \cite{EHLERS}, improved by 
Woodhouse \cite{WOODHOUSE}, as well as in all approaches related to 
quantum physics, where the starting point is often a {\em discrete} 
set to be smoothed out by some procedure in order to produce the 
effective continuous spacetime that we see. We consider these two 
lines in the following subsections.

\subsubsection{Ehlers-Pirani-Schild-Woodhouse axiomatic construction.}
\label{epsw}
In \cite{EHLERS} the authors
took as primitive concepts ``particles'', ``light rays'' and ``events'' in
a set $M$ and imposing certain axioms they showed that $M$ must be a
four dimensional Lorentzian manifold\footnote{Only the case $n=4$ is considered in 
this \S\ref{epsw}.}. 
An interesting and powerful generalization of this construction was
performed by
Woodhouse in \cite{WOODHOUSE}. An obvious difference between the two 
approaches is that Woodhouse speaks of ``light signals'' 
instead of ``light rays''. More importantly, Woodhouse's approach is 
more ambitious, for he pays a deeper attention to the chronological 
and causal relations of events in the space-time, inasmuch as he 
proves how to endow the set of events with the structure of a causal space 
in the sense of definition \ref{causal-space}, and he also derives 
the topology from the axioms.

We outline next the essentials
of both constructions by mixing them in an appropriate way.
The axioms presented herein do not necessarily
correspond to those of \cite{EHLERS,WOODHOUSE} nor do they follow the same
order
as in those papers, rather we have produced a self-consistent version 
which takes benefit of the primordial construction performed in 
\cite{WOODHOUSE} first, and then follows along the lines of the 
original paper \cite{EHLERS}.
\begin{axiom}
The physical spacetime is represented by a set $M$ whose elements are
called {\em events} endowed with a particular subset $\Ps$ 
 of the power set of $M$, ${\cal P}(M)$. The elements $P,Q,\dots\in\Ps$ 
are called {\em particles}, each of which has the 
structure of a $C^0$ one-dimensional manifold homeomorphic to $\r$.
There is at least one particle through each event.
\label{uno}
\end{axiom}
The homeomorphism of $P,Q,\dots$ with $\r$ provides each particle 
with an orientation and thereby with an antireflexive and transitive 
binary relation denoted by $\less |_{P}$. By combining then these 
relations one can define the chronology relation $\less$ between the
points $x,y\in M$ if there is a finite sequence of 
$\{\less|_{P_{i}}\}_{i=1,\dots ,m}$ such that $x\in P_{1}$ and $y\in 
P_{m}$. In order to comply with point {\em (iv)} in definition 
\ref{causal-space} we then assume
\begin{axiom}
The orientation of all particles can be chosen in such a way that for 
all $x\in M$, not $x\less x$.
\end{axiom}
Once we have the chronology $\less$, we can endow $M$ with the 
Alexandrov topology of definition \ref{alexandrov}. Thus we have a 
topological space. Furthermore, it is possible to use definition 
\ref{Bf} in order to define an {\em almost} causal space (see the comments
that 
follow that definition) in the sense of Kronheimer and Penrose. To 
actually get a causal space we need a third axiom
\begin{axiom}
Let $x,y\in M$. If $I^+(y)\subset I^+(z),\, \forall z\in I^{-}(x)$ and 
$I^+(x)\subset I^+(z),\, \forall z\in I^{-}(y)$ then $x=y$.
\end{axiom}
From this axiom one can prove \cite{WOODHOUSE} that $\less$ is future 
and past distinguishing so that, according to the remark after 
equation \ref{union}, $(M,\less,<,\rightarrow)$ is a causal space.

At this stage, and in order to incorporate the properties of light 
propagation, another axiom is needed \cite{WOODHOUSE}
\begin{axiom}
Every $x\in M$ has a neighbourhood $N_{x}$ which is future and past 
reflecting.
\end{axiom}
Then, {\em local} ``light signals'' from $p\in N_{x}$ to $q\in N_{x}$
can be defined without ambiguity as equivalent to the relation 
$p\rightarrow_{N_{x}} q$.

Let $P$ be a particle and, for each $p\in P$, take one of the past and 
future reflecting neighbourhoods $U_{p}$. As $U_{p}$ belongs to the
Alexandrov 
topology we can always choose $U_{p}=I^-(q)\cap I^+(r)$, $q,r\in P$. 
Define a neighbourhood $U_{P}$ of $P$ as the union of all such $U_{p}$ for 
$p\in P$. Then, for $e\in U_{P}$ there is a $p\in P$ such that $e\in 
U_{p}$, and we can define the mappings $f^{\pm}: e \rightarrow\, P$ 
by 
\bnr
f^+(e)=v\in P\,\,  \mbox{such that} \,\, e\rightarrow_{U_{p}} v \, ,\\
f^-(e)=u\in P\,\,  \mbox{such that} \,\, u\rightarrow_{U_{p}} e \, ,
\enr
(see figure). The functions $f^{\pm}$ are called the {\em messages}. 
In \cite{EHLERS}, the definition of {\em echos} is also given: the 
mapping from $f^-(e)\in P$ to $f^+(e)\in P$.
%% 
 % Particles can interchange information by means of messages and echos. 
 % A message is a light ray sent from the point $p$ of the particle $P$
 % towards a point $q$ of another particle $Q$.  If this light ray
 % bounces at $q$ and travels back intersecting $P$ at the point $p'$
 % then we call it echo.  From this we see that messages and echos are
 % mappings between $p$ and $q$, $p'$ respectively.
 %%

Consider now a point $e\in M$ and two nearby particles $P$, $P'$.  In
a physical spacetime $e$ can be determined by means of two echos with
the points $u\in P$ and $u'\in P'$ as the respective domain points and
$v\in P$, $v'\in P'$ as the image points.  These points determine
four real numbers which
will change if we change the point $e$ but keep $P$ and $P'$ fixed. 
We have thus constructed a bijective map $x_{PP'}:U\subset
M\rightarrow \r^4$ in a neighbourhood $U$ of the event $e$ as shown in 
the picture (throughout this Review null rays are represented by 
lines forming $45^o$ degrees with the horizontal plane).

\begin{center}
\includegraphics[width=.4\textwidth]{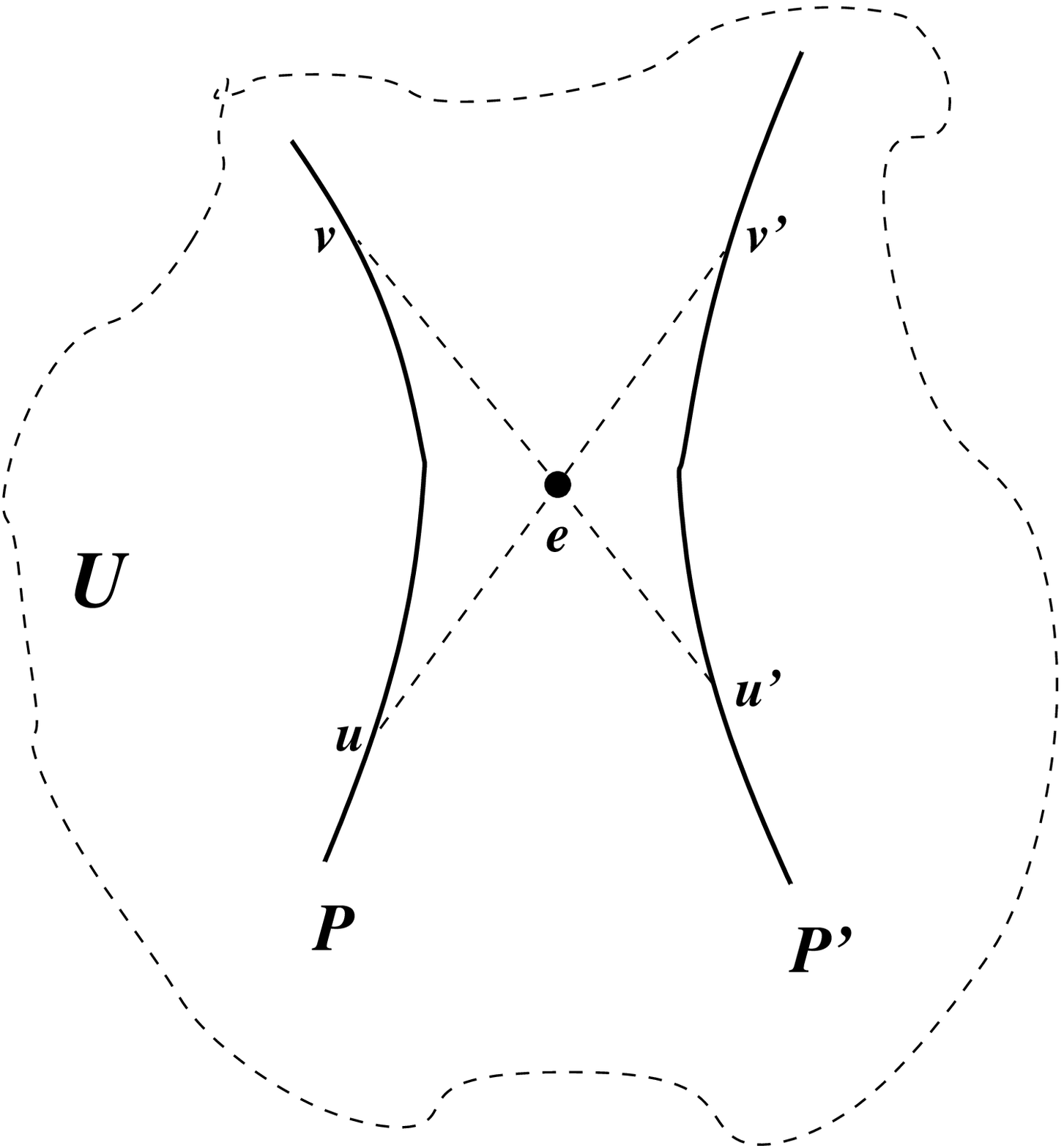}
\end{center}
\begin{axiom}
The set of maps $x_{PP'}|_{U}$ is a smooth atlas for $M$.  The
coordinate charts defined by this atlas are called radar coordinates.
\label{atlas}
\end{axiom} 

In \cite{WOODHOUSE} the differentiability of the previous functions 
$f^{\pm}$ and $x_{PP'}$ are derived from the previous axioms, while 
in \cite{EHLERS} this is simply assumed as in the previous axiom
\ref{atlas}. 
The word smooth in this context means ``as smooth as the message 
functions'', that is to say, as smooth as the manifold would look 
if we performed physical experiments with light signals and particles.
%% 
 % \begin{axiom}
 % Messages and echos are smooth mappings.  In addition to this, echos
 % are invertible.
 % \end{axiom}
 % 
 % {\bf Remark.} The above reasoning can be adapted to higher dimensional
 % spacetimes in the obvious way.
 %%
Axiom \ref{atlas} tells us that $M$ is a smooth differentiable
manifold of dimension {\em four} (this 4-dimensionality is built in in the 
above construction and cannot be avoided), and as proved before $M$ has 
also the structure of a causal space.  To proceed further in the
construction
something more about how light rays, and not merely light signals in local 
neighbourhoods, propagate on $M$ must be said. From now on we slightly
depart 
from the construction in \cite{WOODHOUSE} using the original 
ideas in \cite{EHLERS}, which are more transparent.

As before, for any event $p$ choose a particle $P\ni p$ and a
neighbourhood $U_{P}$ 
of $P$ such that any event $e\in U_{P}-P$ defines the two points 
$f^{\pm}(e)\in P$. Choose a coordinate $t$ for $P$ (that is to say, a 
parametrization on $P$) with $t(p)=0$. 
\begin{axiom}
The function $\rmg:e\rightarrow t(f^{+}(e))t(f^{-}(e))$ is of class $C^2$ 
on $U_{P}$.
\label{light-1}
\end{axiom}
At this point, light rays can be defined as an obvious 
``not-bending'' extension of local signals \cite{WOODHOUSE}. The 
following axiom is then needed in \cite{EHLERS}.
\begin{axiom}
The set of nonvanishing vectors on $T_p(M)$ which are tangent to light 
signals consist of two connected components.
\label{light-2}
\end{axiom}
Using axioms \ref{light-1} and \ref{light-2} the authors of 
\cite{EHLERS} manage to
show that, in a local coordinate system $\{x^1,x^2,x^3,x^4\}$, the
derivatives $\rmg_{ab}\equiv\rmg,_{ab}|_p$ of the above defined
function $\rmg$ form a tensor at $p$.  Moreover light ray directions
at $T_p(M)$ are characterized as those vectors $T^a$ such that
$\rmg_{ab}T^aT^b=0$ and the signature of the tensor $\rmg_{ab}$ is
Lorentzian at every event of $M$.  Nonetheless this only determines
the metric up to a positive conformal factor because a rescaling of
the coordinate $t$ of the particle used in axiom \ref{light-1} leads
to a metric $\rmg'_{ab}$ conformal to $\rmg_{ab}$. 
Therefore from the axioms laid down so far we conclude that $M$ has a
{\em conformal structure} $\Cs$ such that the null vectors are the
tangent vectors of light rays. Recall that a conformal structure is
an equivalence class of conformally related metrics through a strictly
positive conformal factor. The authors also show that light rays are
in fact $\Cs$-null geodesics (null geodesics of any metric of the
conformal
structure). As remarked in \cite{WOODHOUSE}, this 
proves as a by-product that light rays {\em are} smooth (this
smoothness was an axiom in \cite{EHLERS}.)

An important observation is in order here.  Axioms \ref{uno} to
\ref{light-2} do not guarantee that the conformal structure $\Cs$
is unique because a definite family of
particles must be chosen to carry out the procedure
described in the preceding paragraphs. It might well happen that
changes in these families would lead to another different conformal
structure $\Cs'$.

Nevertheless, if the family of particles (and thereby, the light 
signals) are fixed once and for all, then the conformal structure is 
claimed to be unique in \cite{EHLERS}. To that end, 
the authors appeal to a result (theorem \ref{hawking-th})
attributed to  Hawking \cite{HAWKING} (see also 
the related results in theorem \ref{phi-phiminusone} and in subsections 
\ref{causal-preservation} and \ref{natural}): any bijective map 
$\phi:(M,\Cs)\rightarrow(\bar{M},\bar{\Cs})$ between the $C^3$
strongly causal (see definition \ref{SER} below) manifolds $M$, $\bar{M}$ 
with $C^2$ conformal structures
$\Cs$ and $\bar{\Cs}$ such that both $\phi$ and $\phi^{-1}$ preserve
the causal relations is a $C^3$ conformal diffeomorphism. 
It is known due for 
instance to Proposition \ref{local-causal}, that all Lorentzian manifolds
are 
locally strongly causal. Thus, if we apply locally
the mentioned result to the structures $(M,\Ds,\Cs)$ and
$(M,\bar{\Ds},\bar{\Cs})$ ($\Ds$ and $\bar{\Ds}$ stand for the
differential structures of $M$) and the identity map $M\rightarrow M$
we can conclude that both structures are diffeomorphic and 
conformally related. 

By the way, observe once more the necessity of 
having a $C^2$ metric in order to obtain definite and precise results. 

Further axioms are needed to obtain the projective or metric structure
from the
conformal one \cite{EHLERS}. Here, given that we are only 
interested in the causal structure and this has already been obtained, 
we will just mention the two main axioms without further comments,
(the ideas behind both of them are more or less intuitive).
\begin{axiom}
For any event $p\in M$ and a $\Cs$-timelike direction on $T_{p}(M)$ there
is one and only one particle $P$ passing through $p$ with such
direction.
\label{curve-direction}
\end{axiom}
\begin{axiom}[Law of inertia]
For each event $p\in M$ there exists a local coordinate system
$\{x^1,x^2,x^3,x^4\}$ on a neighbourhood of $p$ such that any particle
through $p$ can be parametrized by $x^a(u)$ and
$$
\left.\fr{d^2x^a}{du^2}\right|_p=0.
$$
\label{free-falling}
\end{axiom}
An interesting
general study of manifolds in which only a projective structure is present
can be found in \cite{EHLERS-PROJECTIVE}.

\subsubsection{Quantum approaches. Causal sets.}
\label{quantum}
%\label{discrete}
An idea which has emerged from many approaches to gravity
quantization, and which seems to be an inherent feature to the sought
theory 
of ``quantum gravity" \cite{GARAY},
is that the spacetime cannot be continuous down to a
scale known as Planck's scale.  This scale is defined naturally by the
combination of three basic constants of nature, namely, Newton's
constant $G$, the speed of light $c$ and Planck's constant $\hbar$ in a
quantity with dimension of length called Planck's length and given by
$$
l_p=\fr{G^{1/2}\hbar^{1/2}}{c^{3/2}}\approx 1.61\times 10^{-35}m\, .
$$
Similarly we can combine these constants to obtain natural scales of
time and energy (Planck's time and Planck's energy).  These are
the scales upon which a quantum theory of the spacetime is
expected to take over the classical differentiable picture.
Thus, the theory of ``quantum gravity'' has flourished in the last 
decades as one of the most active branches of theoretical physics in
an attempt to merge successfully Quantum Physics and General
Relativity. A popular alternative is the use of (super-) String Theory,
where 
in particular the Maldacena conjecture has important applications 
of the theory of causal boundaries, see section \ref{causal-boundary}. 
The whole quantum issue would require a long ``topical review'' by 
itself, and we cannot treat any of the important problems arising 
there in a fair manner here. Thus, we have limited ourselves to 
present a very brief summary of main lines and a list of references 
which, hopefully, will be useful to the readers interested in this 
matter. Excellent recent reviews on this subject are for instance
\cite{ASHT,HOROWITZ} and references therein.

As mentioned before, a generalization of the Kronheimer-Penrose 
construction with applications to quantum physics can be found in 
\cite{SZABO,SZABO2}. However, the most fruitful line of research 
deals with the idea of a discrete spacetime, which is perhaps the next 
obvious stage after having discovered that matter and energy are
discontinuous or 
quantized. Despite this being a logical step, only in the nineties
explicit models of discrete spacetimes, and of ``quantum geometries'',
were constructed, see 
\cite{BG,RAINER2,RAINER,KEYL,KEYL2,MSMOLIN,MARKOPOULOU},
although early attempts can be found in the literature
\cite{SNYDER,HILL,FINKELSTEIN,PENROSE-NETWORK}.  
There are currently two main approaches to this subject.  
The first one is the causal set
approach started in \cite{SORKIN} and briefly discussed in following
paragraphs and the second one is loop quantum gravity.
For a relation between the two approaches see \cite{MSMOLIN2,SORK}
and references therein. Loop quantum
gravity manages to quantize Einstein's theory in a background
independent way and one of its most remarkable predictions is that
spacetime is formed at the quantum level by a discrete structure
called spin network.  This is a very active field of research and
there are already excellent reviews about this subject
\cite{ASHT,ASHTEKAR,ROVELLI}; as the subject falls outside the scope 
of this review we will not add anything else about this theory here. 

The foundations of the causal set approach to quantum gravity were laid
in \cite{SORKIN}. For a pedagogical introduction, see \cite{REID}.
As already mentioned, the basic idea of this approach
is to regard the would-be spacetime as a discrete set at small scales. 
Furthermore, only the relation $<$ between its elements (called also 
points) is kept at such scales. 
The specific definition of causal set as stated in \cite{SORKIN} is
reproduced here.
\begin{defi}[Causal set]
A causal set $C$ is any finite set partially ordered by a binary relation
$<$ 
with the additional condition that $<$ is
noncircular, i.e., there are no elements $x,y\in C$, $x\neq y$ such that
$x<y<x$.
\label{causal-set}
\end{defi} 
A partially ordered set is commonly called a {\em poset}.
This definition has common points with definition \ref{causal-space}
of Kronheimer and Penrose, and also with definition 
\ref{etiological} supplemented with the causality principle. 
Thus, only causal spacetimes (see definition \ref{SER} below)
could eventually be described by averaging causal sets. Actually this
restriction is even tighter, because
there is also an important difference with definitions \ref{causal-space}
and 
\ref{etiological}: the absence of binary relations of the sort of
$\less$ and $\rightarrow$. This can be put in correspondence with the 
construction of a causal space from the causality of definition 
\ref{Cf}. As discussed briefly there, and also in
$\S$\ref{causal-preservation},
the relation $<$ alone suffices to characterize the causal
properties of the ``averaged'' spacetime only if this is
distinguishing. In other words, a built-in constraint of the causal 
set approach is that spacetimes whose microscopic
behaviour is likely to be mimicked 
by causal sets must be at least distinguishing. In 
our opinion, a successful quantization of gravity ---if this exists---,
should pose no restrictions on the resulting background spacetime we are
trying
to quantize, however unphysical this background might seem to be.
Nevertheless, these issues were not discussed in \cite{SORKIN}.

A fundamental requirement of any discrete theory is the prescription
to be followed in order to get the picture of a smooth Lorentzian
manifold. In short, how can a spacetime be recovered from a causal set
$C$. 
To achieve this a proposal is given in
\cite{SORKIN}: construct an embedding
$f:C\rightarrow V$, where $(V,\G)$ is an $n$-dimensional
Lorentzian manifold with metric
tensor $\G$ complying with the following properties
\begin{enumerate}
\item $f(x)\in J^-(f(y)) \, \Longleftrightarrow \, x<y$.
\item The embedded points are uniformly spread on $V$ with unit length. 
This means that units are chosen in such a way that the numerical 
value of the integral of 
the volume element $n$-form canonically defined by $\G$ is proportional to
the number of elements of $C$ contained in the region of 
integration to the power $n$.
\item The characteristic length $\l$ of the continuous geometry (the
length
over which $\rmg_{ab}$ vary appreciably) is much larger than the
mean spacing between embedded points.
\end{enumerate}   
Any embedding $f$ meeting the above properties is called a {\em
faithful} embedding. In principle a faithful embedding in a Lorentzian
manifold $(V,\G)$ does not need to exist for a given causal set but
if it does it should be essentially unique.  This means that given a
pair of faithful embeddings $f_1:C\rightarrow (V_1,\G_1)$,
$f_2:C\rightarrow (V_2,\G_2)$ there should exist a diffeomorphism
$h:V_1\rightarrow V_2$ such that $f_2=h\circ f_1$.  In \cite{SORKIN} 
it is argued
that a causal set for which no faithful embedding in $(V,\G)$ exists,
could 
nonetheless be faithfully embedded when {\em coarse grained}. 
A coarse graining $C'$ is essentially a subset of the causal set $C$ with 
the partial order $<$ inherited from $C$. 

Once a causal set is defined one needs to find its {\em
dynamics}. Such dynamics must reduce to (say) Einstein's field equations
in the
continuum limit. In \cite{SORKIN} a method is sketched to
recover the Einstein-Hilbert action from a {\em quantum} causal set. Other
models were tried in the subsequent follow-ups.
Causal set theory has been widely investigated and by now there is a 
vast literature about this subject,  
see e.\ g.\ \cite{ASH,BDGHS,DHS,HMS,ISHAM,ISHAM2,RAPTIS,REID2,REID,RS,RS2} 
and references therein.  In fact there is no single causal set theory
and plenty of variants abiding by the idea of a basic discreteness of
the spacetime have been devised. Of course we cannot do justice here
to all this work (a field outside our expertise). We would simply like 
to remark that hitherto none of these theories accounts
for a satisfactory continuum limit in which a Lorentzian manifold is
recovered.  This seems to be the crux of most of the current
approaches to ``quantum gravity'': mind-boggling theories are put
forward but the key issue of General Relativity as a limit is often quoted
as ``under current research''.

\section{Causal characterization of Lorentzian manifolds and 
their classification}
\label{causal-characterization}
Lorentzian manifolds may have rather different global causal
properties depending on their Lorentzian metrics. More 
importantly, completely smooth regular manifolds (such 
as $\r^n$) carrying analytical metrics with perfectly good local
properties 
and which look completely innocuous at first sight may have closed 
future-directed timelike loops: violation of the chronology hypothesis.
This was not fully realized until the
publication of G\"odel's famous paper \cite{GODEL} where he
constructed an example of cosmological spacetime with rotation
containing timelike loops through every point of the manifold. 
Other physically acceptable spacetimes, such as the Kerr solution 
\cite{FF,MCCALLUM}, are extensible as solutions to Einstein's vacuum 
field equations, and their maximal extensions may contain regions 
in which such causality violations arise. 
As is obvious, one would desire to rule out these curves on the 
grounds of their unphysical properties,
because they lead to violations of the ``free-will'' principle 
and to other traditional paradoxes (they would meet their own past after a
certain proper time), see the discussion in \cite{KRIELE} and 
references therein. It has been recently claimed, however, that these
pathologies arise in physically acceptable spacetimes
\cite{BONNOR,BONNOR2}.

Timelike loops are not the only
undesirable causal feature one wishes to rule out from a spacetime for
there may also be causal or null loops, or other timelike/causal curves 
which form ``almost a loop'', or get
trapped inside a compact region of the spacetime (causal
imprisonment, see \cite{MISNER,FF} for particular examples.)
This means that the causal classification of Lorentzian manifolds in a 
hierarchy which measures their causal behaviour is not so
simple as to put in a group those having closed timelike loops and
those which do not in another group. This is the ultimate reason why
several {\em causality 
conditions}, forbidding such loops or the related ``almost loops'' 
in order to achieve physically acceptable spacetimes, 
were devised in \cite{CARTER,FF,PENROSE} and references therein. Many of
these 
conditions turned out to be insufficient to rule out all causality 
pathologies, and this is why the final classification is longer 
than expected. The subject has been studied over the years and nowadays
there is a well established hierarchy of causality conditions, which can
be
found in many textbooks of General Relativity \cite{FF,WTM,Wald}. 
The different 
good properties that each condition achieves, as well as those which 
does not forbid, are now well understood. A summary of this subject is 
presented next.

\subsection{Standard hierarchy of causality conditions}
\label{hierarchy}

\begin{defi}
A Lorentzian manifold $(V,\G)$ is said to be:
\begin{itemize}
\item {\bf not totally vicious} if $I^+(x)\cap I^-(x)\neq V, \forall x\in
V$.
\item {\bf Chronological} if $x\not\in I^{+}(x)\ \forall x\in V$.
\item {\bf Causal} if $J^{+}(x)\cap J^{-}(x)=\{x\}\ \forall x\in V$.
\item {\bf Future distinguishing} if $I^+(x)=I^+(y)$ only if $x=y$ 
for $x,y\in V$. Future and past distinguishing Lorentzian manifolds 
are simply known as distinguishing.
\item {\bf Strongly causal} if $\forall x\in V$ and for every 
neighbourhood $\W_x$ of $x$ there exists another neighbourhood
$\U_x\subset\W_x$ containing $x$ such that for every future-directed
causal curve $\g$ the intersection $\g\cap\U_x$ is either empty or a 
connected set.
\item {\bf Causally stable} if there exists a function whose gradient is
timelike everywhere (called a time function).
\item {\bf Causally continuous} if it is reflecting and distinguishing.
\item {\bf Causally simple} if it is distinguishing and $J^{\pm}(x)$ 
are closed sets $\forall x\in V$.
\item {\bf Globally hyperbolic} if there exists an edgeless acausal 
hypersurface $S$ such that $D(S)=V$. $S$ is called a Cauchy hypersurface.
\end{itemize}
\label{SER}
\end{defi}    
Conditions in definition \ref{SER} are given in an increasing order of
specialization.  This means that the hierarchy is built in such a way
that spacetimes belonging to a certain class of definition \ref{SER}
contain a causal feature regarded as better than those present in the
classes lying above.  Detailed explanations (which are beyond the scope 
of this review) about these features and
the route followed to build the hierarchy can be found in e.\ g.\
\cite{FF,BEE,SINGULARITY}. Here we are more interested in the attempts of
improving this classification either by adding more classes to the
hierarchy or defining it in more abstract terms. This will be treated 
in the next subsections. However, we will present now a summary of the 
main ideas behind each condition in the hierarchy, and the relation 
between them, for the sake of clearness and completeness.

The worst behaved spacetimes are totally vicious ones. An equivalent 
characterization of them is $\exists x\in V$ such that $I^+(x)\cap 
I^-(x)=V$, from where one immediately derives that $I^{\pm}(\zeta)=V$ 
for any set $\zeta\subset V$, see also \cite{KK}. 
This terminology was put forward in 
\cite{CARTER}, where probably the first detailed abstract classification 
was given. Carter uses the name ``(almost) vicious'' for those 
spacetimes with (causal) timelike future-directed loops. Accordingly, 
he used the names ``(almost) virtuous'' for spacetimes complying with the 
(chronology) causality condition. A well-known result \cite{BW} is that 
all {\em compact} spacetimes are vicious \cite{BEE,FF,PENROSE}, see 
also \cite{KK,MATORI}. A large set of non-compact vicious spacetimes 
was described in \cite{SANCHEZ2} from a mathematical point of view.
 It has recently been claimed, see  \cite{BONNOR,BONNOR2} and references 
therein, that vicious spacetimes may arise in physically realistic 
situations, although, their conclusions have been criticized in 
\cite{MR,MR2} and this claim appears to be doubtful.

The distinguishing condition was devised to forbid these ``vices''. 
An equivalent statement for future distinction is (see \cite{SINGULARITY} 
for a proof): every neighbourhood of $x$ contains another 
neighbourhood $\U_x$ of $x$ such that every causal future-directed 
curve starting at $x$ intersects $\U_x$ in a connected set (the 
condition for past distinction is the same replacing future by past).
From here it is clear that (i) future distinction and past distinction 
are actually inequivalent and independent, and (ii)
all distinguishing spacetimes are virtuous.
Nevertheless, a spacetime can be distinguishing and still there may be 
future-directed causal curves starting nearby a point $x\in V$--- 
but {\em not} at $x$---and intersecting all neighbourhoods of $x$ in 
a disconnected set. This is the reason behind strong causality, which 
forbids these behaviours. An equivalent statement for strong causality 
is: $\forall x\in V$ there is a neighbourhood $\U_{x}$ such that for 
all $p,q\in \U_{x}$ with $p\less q$, $I^+(p)\cap I^-(q)\subset \U_{x}$. 
Trivially strong causality implies, and is stronger than, the 
distinguishing condition. Moreover, strong causality forbids the above 
mentioned pathology of imprisonment \cite{BEE,CARTER,FF}.

Carter was the first to realize that the strong causality condition 
was actually not {\em strong} enough to avoid all simple causal 
pathologies. To show this, in \cite{CARTER} a classification by means of 
new causal relations called $n$-degree causal relations was elaborated.
%% 
 % \begin{defi}
 % A set $\Ss\subset M$ with at least two points is said to be virtuous
 % with respect to $\Us\subset M$ if for every pair $x, y$ of different
 % points of $\Ss$ at most one of $x\precu y$ and $y\precu x$ is true. 
 % If the same is true for the relation $\lessu$ instead then $\Ss$ is
 % almost virtuous with respect to $\Us$.
 % \label{virtuous}
 % \end{defi}  
 % This definition ensures that no causal or timelike loops going through
 % the set $\Us$ are present in $\Ss$.  If $\Us= M$ then the space-time
 % manifold is virtuous or almost virtuous.  Space-time manifolds in
 % which causality is violated can be also classified as follows.
 % 
 % \begin{defi}
 % Under the hypotheses of definition \ref{virtuous} the set $\Ss$ is
 % said to be vicious with respect to $\Us$ if for every pair $x, y\in
 % \Ss$ both $x\lessu y$ and $y\lessu x$ hold.  If we replace $\lessu$ by
 % $\precu$ then $\Ss$ is almost vicious.
 % \label{vicious}
 % \end{defi}  
 %%
 First of all, one defines the sets
\bnr
\suca\Ss)\equiv\left\{x\in V:\left( J^+(\Ss)\cap\overline{J^-(x)}
\right)\bigcup\left(\overline{J^+(\Ss)}\cap 
J^-(x)\right)\neq\varnothing\right\},\\
(\Ss\preca\equiv\left\{x\in V:\left( J^-(\Ss)\cap\overline{J^+(x)}
\right)\bigcup\left(\overline{J^-(\Ss)}\cap 
J^+(x)\right)\neq\varnothing\right\}.
\enr
Next, new binary relations between subsets of $V$ are arranged by 
$$
\Ss\suca\Ts\Longleftrightarrow \Ts\subset\suca\Ss),\ \ \ \ \ \ \
\Ts\preca\Ss\Longleftrightarrow \Ts\subset (\Ss\preca .
$$
Of course all this is valid for sets consisting of a single point. 
These relations are used in the first step of an inductive chain
of definitions
\bnr
\fl\sucan\Ss)\equiv\left\{x\in V:
\left(\overline{\sucamenos\Ss)}\cap(x>\right)\bigcup\left(<\Ss)\cap
\overline{(x\precamenos}\right)\bigcup
\left(\bigcup_{r=1}^{n-1}\stackrel{r}{<}\Ss,x\stackrel{n-r}{>}\right)
\neq\varnothing\right\}\\
\fl(\Ss\precan\equiv\left\{x\in V:
\left(< x)\cap\overline{(\Ss\precamenos}\right)\bigcup
\left(\overline{\sucamenos x)}\cap
(\Ss>\right)\bigcup
\left(\bigcup_{r=1}^{n-1}\stackrel{r}{<}\Ss,x\stackrel{n-r}{>}\right)
\neq\varnothing\right\},
\enr
where
$$
\stackrel{r}{<}\Ss,x\stackrel{n-r}{>}
\equiv\stackrel{r}{<}\Ss)\cap(x\!\!\stackrel{n-r}{>}.
$$ 
From these relations one constructs binary relations $\sucan$,
$\precan$ between subsets of $V$ in the same fashion as above. 
These are then used to generalize the concepts of virtuous and
vicious (again, $x\stackrel{r}{<}y$ will be used for 
$\{x\}\stackrel{r}{<}\{y\}$ and so on):
\begin{defi}
A subset $\Ss\subset V$ is said to be virtuous to the $n^{th}$ degree 
($n\geq 0$) if no pair of different points $x, y\in\Ss$ satisfies 
$x\stackrel{r}{<}y$ and $y\stackrel{s}{<}x$ with 
$r+s\leq n$. And $\Ss$ is said to be sub-vicious by $n$ degrees ($n\geq
0$) 
if for any two points $x, y\in\Ss$ we have $x\stackrel{r}{<}y$, 
$y\stackrel{s}{<}x$ with $r+s\leq n$.
\label{nvirtuous} 
\end{defi}
According to this definition, the most vicious 
spacetimes are those sub-vicious by zero degree, which correspond 
to the totally vicious spacetimes of definition \ref{SER}. From here, 
going up the ladder of causal virtue, the spacetimes virtuous to the 
1$^{st}$ degree are simply the distinguishing ones, and those virtuous 
to the 2$^{nd}$ degree are the strongly causal spacetimes. 
The next degrees can be understood intuitively as follows \cite{PENROSE,FF}. 
One can have future-directed causal curves starting arbitrarily close to a 
point $p_1$, never passing again close to $p_1$ but going arbitrarily near 
$p_2$ from where new future-directed causal curves starting arbitrarily close to $p_2$ and
passing arbitrarily near $p_1$ can exist. And the same with $p_1,p_2,p_3$ and so on.
The $n^{th}$-degree virtuous spacetimes have increasing 
virtue for larger $n$, still they do not remove all causal problems. 
As 
a matter of fact, there is not only an infinite number of such 
conditions, but also the concept of ``unlimited
virtue'' as Carter called it cannot be obtained from definition
\ref{nvirtuous}.
%, not even by passing to the limit when $n\rightarrow \infty$. 
%In other words, even if we define the $\omega$-degree 
%virtuousness, with $\omega$ the first transfinite ordinal, the process 
%can be continued from $\omega$ on indefinitely, and 
The process of gaining virtue can be continued indefinitely, and all such 
spacetimes are in some sense {\em unstable} with respect to their 
virtue: small perturbations may always produce $1^{st}$-degree 
vicious spacetimes.  

This raises the question as to whether there exist
Lorentzian manifolds which are maximally virtuous in this sense, that 
is to say, so that small perturbations will not affect their 
virtuousness. We will come back to the question of maximum 
``good causal behaviour'' in subsections \ref{causal-preservation} and 
\ref{chains}. Carter's paper proposed to consider the {\em causally
simple} 
spacetimes as those with ``stable virtue'' (though the virtue may be
larger or 
improved within the class.)
As a matter of fact, there are intermediate conditions which capture 
the idea of stable virtue: the stable and the continuous causality
conditions. The first of these was introduced by Hawking in \cite{H} with
this 
precise purpose, see also \cite{SEIFERT2}. An equivalent formulation of
this 
condition is: there is a continuous timelike vector field $\v$ such 
that $\G+\v\otimes\v$ is strongly causal. This means that we can open 
up slightly the null cones and the spacetime remains causal, see for 
more details proposition \ref{causa-stable}.
%and the entire subsection \ref{Cr-topo}. 
Hawking showed in \cite{H} that causal stability defined in terms of the 
vector field $\v$ implies that there exists a {\em cosmic time} which
is a function $f:V\rightarrow\r$ increasing along every causal future
directed
curve on $V$. A complete proof of the differentiability of such cosmic
time has 
only been accomplished recently
in \cite{BERNAL2} (earlier claims were made in \cite{SEIFERT2} but 
this proof seems to have unclear points.)  
In \cite{BERNAL2} these results were carried a step forward: 
if a cosmic time exists then one can find a smooth function $f$ 
(a time function) 
with an everywhere future-directed timelike gradient $df$, see 
theorems \ref{bernal-sanchez} and \ref{bernal-splitting}. 
 The hypersurfaces $f=$const.\ are spacelike and acausal, and 
they foliate the spacetime (see also \cite{H,VYAS-JOSHI,JOSHI}). 
For the causal continuity hypothesis, see \cite{HAWKINGSACHS}.

Nevertheless, causally stable and causally continuous spacetimes may fail 
to be causally simple. This may happen if $\d J^{+}(x)\neq E^{+}(x)$ 
for some point $x$ (or its past version), whose negation may be taken as 
an equivalent definition of causal simplicity (subsection \ref{classic}). 
A remarkable example of a non-causally-simple spacetime was given by 
Penrose \cite{PENROSE-WAVE} using the so-called plane waves in 
General Relativity \cite{MCCALLUM}, see paragraph \ref{ppwaves}. 
Finally, even the foliation $\{\Sigma_{f}: f=\mbox{const.}\}$ 
constructed with a time 
function $f$ in causally simple spacetimes does not necessarily define 
Cauchy hypersurfaces, as $D(\Sigma_{f})\neq V$ can certainly happen. 
Explicit typical examples are anti de Sitter spacetime \cite{FF} and 
the maximal extension of the Reissner-Nordstr\"om solution 
\cite{CARTER3}. Therefore, globally hyperbolic 
spacetimes, which were originally introduced by Leray (see 
\cite{FF,PENROSE}) using the 
condition that $J^+(x)\cap J^-(x)$ be compact for all $x\in V$ (see 
subsection \ref{classic}), is 
regarded as the strongest causal restriction considered hitherto. See 
subsection \ref{splitting} for further details.

Interesting alternative characterizations of some of the conditions 
of definition \ref{SER} can be found in \cite{RACZ3,NATARIO,NATARIO-TOD}. In the first of these references, spacetimes satisfying some of these conditions are distinguished in terms of 
the injectivity of certain maps from the manifold $V$ onto its power set 
${\cal P}(V)$. 
\begin{prop}
A spacetime $V$ is 
\begin{enumerate}
\item causal if and only if one (or both) 
of the maps $J^{\pm}:V\rightarrow{\cal P}(V)$
is injective.
\item future (respectively past) distinguishing if and only if $I^+:V\rightarrow{\cal P}(V)$ (resp.\ $I^-:V\rightarrow{\cal P}(V)$) is injective.
\item strongly causal if and only if one (or both) of the maps 
$\downarrow I^+,\uparrow I^-:V\rightarrow{\cal P}(V)$ are injective.
%\item Causally stable if and only if one (or both) of the maps 
%$J^{\pm}:V\rightarrow{\cal P}(V)$ are injective.
\end{enumerate}
\label{injective}
\end{prop}
The sets $\downarrow I^+(U)$ and $\uparrow I^-(U)$ are known as 
the chronological common past of $I^+(U)$ and the chronological 
common future of $I^-(U)$ respectively (see definition \ref{common-pf}). In \cite{RACZ3} there is also a characterization of causally stable spacetimes by means of injective maps, but this requires a generalization of the causal futures and pasts using the $C^r$ topology on Lor$(V)$ (definition \ref{Crtopology}), or equivalently, a definition of the common future (or past) of a set within a class of metrics of the type $\G+\v\otimes\v$ mentioned in a previous paragraph. With regard to
references \cite{NATARIO,NATARIO-TOD}, they are concerned with globally hyperbolic spacetimes and the relationship between causal properties of such
spacetimes and the set of its null geodesics which is
a smooth $(2n-1)$-dimensional contact manifold. The topology 
and geometry of this manifold can be used to study causal properties 
of the spacetime. Explicit results for the case of dimension $n=3$ were given in \cite{NATARIO,NATARIO-TOD}.

\subsection{Mappings preserving causal properties}
\label{causal-preservation}
In this subsection we explore a very interesting approach for the
causal classification of Lorentzian manifolds and, more generally, 
etiological/causal spaces. We may ask ourselves
when two different such spaces can in some sense be termed as
causally equivalent, or bearing the same causal structure.
The basic idea, of course, is
to define mappings between them which in a precise sense
preserve the causal properties and regard those spaces
which can be put into correspondence by means of one of these mappings
as ``causally isomorphic'' or sharing the same causality. 
These sort of mappings have been considered many times in the 
literature, e.\ g.\ 
\cite{ZEEMAN,KRONHEIMER-PENROSE,BUDICSACHS,HAWKING,MALAMENT,VYAS,VYAS2,PARK}.
Kronheimer and Penrose introduced this subject in 
\cite{KRONHEIMER-PENROSE}, and then Budic and Sachs used an 
improvement in a paper \cite{BUDICSACHS} devoted to
generalising the construction of the causal boundary (see
$\S$\ref{budic-sachs} for more details) and hence they did not use
these mappings to classify spacetimes in terms of their causal
properties.
Although the nomenclature in these and others papers is sometimes 
conflicting with 
each other, we have tried to unify the terminology with an up-to-date 
perspective. 
\begin{defi}
Let $(Z,<,\less)$ and $(X,<,\less)$ be etiological spaces (including 
in particular the causal spaces in the sense
of Kronheimer and Penrose and the Lorentzian manifolds)
and let $\theta:X\rightarrow Z$ be a mapping. $\theta$ is said to be
\begin{itemize}
\item chronal preserving if $x\less y$ implies $\theta(x)\less\theta(y)$.
\item causal preserving if $x< y$ implies $\theta(x)< \theta(y)$. 
\item a chronal isomorphism if $\theta$ is bijective and $\theta$ and 
its inverse $\theta^{-1}$ are chronal preserving.
\item a causal isomorphism if $\theta$ is bijective and both 
$\theta,\, \theta^{-1}$ are causal preserving.
\end{itemize}
$(Z,<,\less)$ and $(X,<,\less)$ are causally isomorphic if there is a 
causal and chronal isomorphism between them.
\label{theta}
\end{defi}
Perhaps the first reference dealing with these mappings, prior to 
\cite{KRONHEIMER-PENROSE,BUDICSACHS}, is Zeeman's paper
\cite{ZEEMAN}, where the causal isomorphisms (called ``causal
automorphisms'' 
in that paper) of flat Minkowski spacetime were studied. The set of 
such causal isomorphisms clearly forms a group, called the causality group
of 
flat spacetime, and Zeeman was able to prove that
this group is generated by the orthochronous Lorentz group, the
group of translations, and what he called the dilatation group 
(multiplication of the flat metric by scalars.) This was a 
preliminary result which was soon to be generalized in several ways.
For instance, if the etiological spaces of
definition \ref{theta} are in fact Lorentzian manifolds 
then Malament \cite{MALAMENT}, elaborating on 
previous results in \cite{HAWKING}, was able to prove
the following very important theorem (see also theorem 
\ref{huang-res} and subsection \ref{natural}).
\begin{theo}
Let $(V,\G)$, $(V',\G')$ be a pair of Lorentzian manifolds and
$f:V\rightarrow V'$ a bijection. If either of the following two 
conditions hold
\begin{enumerate}
\item both $f$ and $f^{-1}$ preserve continuous timelike curves, or
\item $(V,\G)$ and $(V',\G')$ are distinguishing and $f$ is a 
chronal isomorphism,
\end{enumerate}    
then $f$ is a smooth conformal isometry.
\label{malament-result}
\end{theo}
Clearly, a bijection $f$ with the first of these properties 
is a chronal isomorphism, but the converse is not true in general.
This is why further requirements upon $V$ must be imposed. 
As Malament showed explicitly in \cite{MALAMENT}, there are examples 
of chronal isomorphisms between non past-distinguishing spacetimes 
which do not preserve continuous timelike curves. Thus, the 
distinguishing hypothesis is essential in the second point of 
theorem \ref{malament-result}. For distinguishing spacetimes, however,
a map $f:V\rightarrow V'$ preserving the chronology preserves also the 
continuous timelike curves, the
ultimate reason for this is that for distinguishing spacetimes a curve
$\g$ is timelike if and {\em only if}, $\forall$ $p,q\in\g$, $p\less q$
\cite{CAUSAL}. 
An important consequence of the above is that, for distinguishing
spacetimes, the chronological relation $\less$ determines the Lorentzian
metric up to a conformal factor because if two such Lorentzian
manifolds are {\em chronally isomorphic} then by theorem
\ref{malament-result} they must be conformally related. Once this is 
understood, the following statements proving the invariance of the
hierarchy of
causality conditions under chronal or causal isomorphisms
\cite{VYAS,VYAS2,PARK} become rather obvious.
\begin{theo}
Let $(V,\G)$ and $(V',\G')$ be chronally isomorphic spacetimes. 
Then, $(V,\G)$ is globally hyperbolic, causally simple,
causally continuous, reflecting, 
strongly causal, distinguishing, causal, chronological, or totally 
vicious, if and only if so is $(V',\G')$. 

If $(V,\G)$ and $(V',\G')$ are causally isomorphic, and one of 
them is causally stable, so is the other.
\label{obvious}
\end{theo}

From the above we realize that the true applicability of these
mappings to the classification of Lorentzian manifolds is rather
limited in the relevant cases as we can only compare spacetimes
{\em conformally related} to each other. This is too strong 
a restriction, as we are going to argue in what follows. For example, 
it prevents us from being able to give a meaning to the {\em local 
equivalence} of causal structures. This question was addressed
by Kronheimer and Penrose in their seminal paper 
\cite{KRONHEIMER-PENROSE}. They showed explicitly a negative result which
is 
often ignored: they proved, in a precise sense, that
the statement ``any Lorentzian manifold has a causal structure 
locally equivalent to that of flat spacetime'' is wrong as it
stands---despite 
being considered as obvious too many times--- if the local equivalence 
is of conformal type.

Of course, in order to give a meaning to the previous assertion 
between quotes, and to claim
that it is false, one must first of all provide a definition for
``equivalent'' causal structures.  This is precisely what was given in
\cite{KRONHEIMER-PENROSE}, where a definition which appeared to be
``natural''
at first
sight was put forward.  In \cite{KRONHEIMER-PENROSE} the authors were
only interested to discuss as to what extent a Lorentzian manifold
regarded as a causal space is {\em locally} similar, from the
causality point of view, to flat Minkowski spacetime, so the 
definition they gave was stated as follows
\begin{quote}
{\em For each point $x$ of the $n$-dimensional manifold $V$ we can
find a small open neighbourhood $U_{x}$ of $x$ together with
a homeomorphism $h$ of $U_{x}$ onto an open subset $\U$ of
$n$-dimensional Minkowski spacetime such that $h$ is a causal
isomorphism.}
\end{quote}  
A counterexample of this
assertion is explicitly constructed in \cite{KRONHEIMER-PENROSE}.
Here the set $U_{x}$ can acquire its causal structure either as a causal
subspace of $V$ or as a Lorentzian submanifold of $V$ (observe that for $p\in U_x$, $I^+(p,U_x)\neq I^+(p)\cap U_x$ in general). In the first case the assertion is trivially false, as any vicious spacetime would be an obvious counterexample.
Kronheimer and Penrose argue that the above quoted assertion is also false in the second case, on the grounds that the causal structure of a Lorentzian manifold is
determined by its Lorentzian cone, hence the causal structure of an
arbitrary Lorentzian manifold and that of flat spacetime will differ 
even locally unless the spacetime is locally conformally flat. 
Of course, this is not the case generically, even for the simplest 
examples. To see why this happens these authors present a clarifying 
example: call a set of nine points $p_{ij}$, $i,j=\{1,2,3\}$ in a causal 
space a {\em complete square} if the eighteen conditions    
$$
p_{ij}\rightarrow p_{ij+1},\ p_{ij}\rightarrow p_{i+1j},\ 
p_{ij}\rightarrow p_{ij+2},\ p_{ij}\rightarrow p_{i+2j},
$$
hold. A {\em defective square} is a subset of eight different points satisfying these conditions except for the four conditions involving the `central element' $p_{22}$.
It can be shown that any non-empty open set $\U$ of $n$-dimensional Minkowski spacetime admits an open subset $\U'\subset\U$ such that any defective square 
in $\U'$ can be completed by the addition of a point in $\U'$. 
This property is in general not true for a neighbourhood  
$U_x$ of a (conformally) non-flat Lorentzian manifold leading us to the 
conclusion that a causal isomorphism between a curved Lorentzian 
manifold and flat spacetime cannot in general be established locally.

In the next paragraph we are going to show, however, that an assertion
such as 
``any spacetime is locally equivalent to flat spacetime from the causal 
viewpoint'' does make sense (see theorem \ref{local-isocausal}) 
if we agree to abandon the
concept of causal isomorphism of definition \ref{theta} as the basic 
ingredient to be used for that purpose, and thereby we will see that
the conformal structure is not the appropriate {\em causal structure} 
of a Lorentzian manifold for these matters. 

\subsubsection{Causal relationship. Isocausal Lorentzian manifolds.}
\label{isocausal}
To surmount the mentioned difficulties, the present
authors developed a new approach to the subject in \cite{CAUSAL}.  The
idea is now to consider mappings preserving the causal relations whose
inverses do not necessarily do so. To that end, we apply the 
preservation of the causal properties at the first possible level: the 
algebraic level studied in subsection \ref{causal-tensors}. From this 
the preservation of the rest of the levels (local and global), 
and of the binary causal relations, will follow. 
\begin{defi}
Let $\Phi:V\rightarrow W$ be a global diffeomorphism between
two Lorentzian manifolds. 
We say that $W$ is
{\em causally related} with $V$ by $\Phi$, denoted
$V\prec_{\f}W$, if for every future-directed $\X\in T(V)$,
$\f' \X\in T(W)$ is future directed too.
$W$ is said to be {\em causally related} with $V$,
denoted simply by $V\prec W$, if there exists $\f$ such that
$V\prec_{\f}W$.
Any diffeomorphism $\f$ such that $V\prec_{\f}W$ is called a causal 
mapping\footnote{We used the term ``causal relation'' for these 
mappings in \cite{CAUSAL}, but we have preferred to use here the name 
causal mapping to avoid confusion with the binary causal relations.}.
\label{PREC}
\end{defi}
Of course, a similar definition in which past-directed vectors are mapped
into future-directed ones (anticausal mapping) can also be given.  
All the results described below hold likewise for causal and anticausal
mappings although we only make them explicit for causal mappings. In 
short, the idea behind this definition is that the image by $\Phi$
of the future null cone at every $x\in V$ is contained within the 
future null cone at $\Phi (x)$.

Let $\G$ and $\tilde\G$ be the Lorentzian metrics of $V$ and $W$, 
respectively. The condition imposed in definition \ref{PREC} 
is very easy to check for a fixed diffeomorphism $\Phi: V\rightarrow 
W$, because
$$
V\prec_{\Phi} W\Longleftrightarrow 
\tilde{\G}(\Phi'\X,\Phi'\Y)=\Phi^*\tilde{\G}(\X,\Y)\geq 0,\ \ \
\forall\,\, \mbox{future-directed}\,\,  \X,\Y\in T(V)
$$
which means that $\Phi^*\tilde{\G}$ is a future tensor according to 
definition \ref{CAUS-TENSR}. It can be proven \cite{CAUSAL} that this is
the
characterization of causal and anti-causal mappings. In other words, 
as explained in subsection \ref{causal-tensors}, $\Phi^*\tilde{\G}$
must satisfy the {\em dominant energy
condition} at every point of the manifold $V$.  The algebraic conditions
that make this happen are well known (see e.\ g.\ \cite{FF} for a
presentation
in four dimensions and \cite{CAUSAL-SYMMETRY,thesis} for its
generalization to arbitrary dimension), and several criteria to 
ascertain if a given tensor is future or not can be found in 
\cite{SUP,PI,CAUSAL,CAUSAL-SYMMETRY,thesis}.

In general, all causal objects are preserved, in a precise sense, by 
causal mappings. A summary of these preservations is given next 
\cite{CAUSAL}.
\begin{prop}
If $V\prec_{\f}W$, then
\begin{enumerate}
\item all contravariant (resp.\ covariant) future tensors of $V$ 
(resp.\ $W$) are mapped by $\Phi$ to contravariant (resp.\ covariant)
future tensors on $W$ (resp.\ $V$).
\item all timelike future-directed vectors on $V$ are mapped to 
timelike future-directed vectors. And if the image $\Phi'\X$ of a future
vector 
$\X$ is null, then $\X$ is a null future-directed vector.
\item every continuous future-directed timelike (causal) curve is 
mapped by $\Phi$ to a continuous future-directed timelike (causal) curve.
\item for every set $\zeta\subseteq V$, $\Phi (I^{\pm}(\zeta))\subseteq 
I^{\pm}(\Phi (\zeta ))$, $\Phi (J^{\pm}(\zeta))\subseteq 
J^{\pm}(\Phi (\zeta ))$, and $D^{\pm}(\Phi(\zeta))\subseteq 
\Phi(D^{\pm}(\zeta))$.
\item if a set $S\subset W$ is acausal (achronal), then $\Phi^{-1}(S)$ 
is acausal (achronal).
\item if $\Sigma \subset W$ is a Cauchy hypersurface, then 
$\Phi^{-1}(\Sigma)$ is a Cauchy hypersurface in $V$.
\item $\Phi^{-1}(F)$ is a future set for every future set $F\subset W$; 
and $\Phi^{-1}(\d F)$ is an achronal boundary for every achronal 
boundary $\d F\subset W$.
\end{enumerate}
\label{lista}
\end{prop}
From point {\em (iv)} follows that any causal mapping is in 
particular a chronal-preserving and causal-preserving diffeomorphism 
(see definition \ref{theta}). Notice, however, that the inverse of a 
causal mapping does not need to be a causal mapping. This opens up a 
whole new vista which allows to investigate the causal structures, 
and the causal equivalence, from new perspectives with improved 
results. Of course, the chronal and causal isomorphisms are 
particular cases of causal mappings, and therefore the previous 
results on this subject are included by default under the more 
general causal relationship of definition \ref{PREC}.

Next theorem proven in
\cite{CAUSAL} provides yet more support to the idea that causally
related manifolds share some global causal properties.
\begin{theo}
If $V\prec W$ and $W$ is globally hyperbolic, causally stable, 
strongly causal, distinguishing, causal, 
chronological, or not totally vicious then so is $V$.
\label{preservation}
\end{theo}
Therefore a Lorentzian manifold cannot be causally related to another 
belonging to a different causality class.
 In \cite{CAUSAL} we did indeed show that sometimes two diffeomorphic 
spacetimes are not causally related, namely, there is no diffeomorphism 
$\Phi:V\rightarrow W$ such that $V\prec_{\Phi} W$. 
This is symbolized by $V\not\prec W$. A very simple example is flat 
spacetime which cannot be causally related with 
anti-de Sitter spacetime because the former is globally hyperbolic 
and the latter is not. In view of proposition 
\ref{lista}, theorem \ref{preservation} and many other related results, 
we argued in \cite{CAUSAL} that this impossibility of putting in causal
relation two Lorentzian manifolds is due to the existence of a global
causal property in one of the spacetimes which is absent in the other
one. A more interesting example is provided by de Sitter spacetime and 
the Einstein static universe \cite{FF,MCCALLUM}. As is known,
every inextensible timelike curve $\g$ in de Sitter space
has a non-empty $\d I^+(\g)$ (a ``particle horizon'')
and this is not so for the Einstein static universe \cite{FF,MCCALLUM}. 
One can then prove \cite{CAUSAL}
that this property implies that de Sitter universe is not
causally related with the Einstein static universe. Observe that the 
base manifolds for these two spacetimes are the same ($\r\times S^{n-1}$), 
and more importantly, {\em both of them are globally hyperbolic}. As a 
matter of fact, the Einstein universe {\em is} causally related with 
the de Sitter spacetime.

Given the above comments, an interesting property of causal mappings is 
that they define a binary relation, the relation $\prec$, in the set of
all 
diffeomorphic Lorentzian manifolds. Clearly ``$\prec$'' is reflexive and 
transitive and hence it is a preorder. Nevertheless, as mentioned 
above, there are simple examples (such as de Sitter and Einstein 
static spacetimes) where
a Lorentzian manifold $W$ is causally related with another $V$ but not the
other
way round, namely, $V\prec W$ but $W\not\prec V$. Thus,
the relation ``$\prec$'' is not symmetric. Furthermore, ``$\prec$''
is not antisymmetric either, as there certainly are Lorentzian 
manifolds such that both $V\prec W$ and $W\prec V$ hold \cite{CAUSAL}.
 From all the discussion so far, we deduce that those Lorentzian manifolds
in which causal relations exist in both ways can be expected to share
common causal properties and hence a special name is reserved for
them.
\begin{defi}
Two Lorentzian manifolds $V$ and $W$ are called causally equivalent or
{\em isocausal} if $V\prec W$ and $W\prec V$.  The relation of causal
equivalence is denoted by $V\sim W$.
\label{equivalence}
\end{defi}   
Note that we talk about isocausality as a property between
Lorentzian manifolds, and not as a property of a particular mapping. 
As a matter of fact, there are no explicit particular mappings used here,
and 
more importantly there are {\em two} mappings implicitly used.
Observe that in general the diffeomorphisms setting the relation
$V\prec W$ will be unrelated to those establishing $W\prec V$.  In
this way $V$ and $W$ only need to be diffeomorphic in order to test
their causal equivalence and no geometric conditions such as
conformal equivalence are required or implied, contrary to what 
happened with the approaches
reviewed above.  This is a consequence of working with causal
mappings whose inverse is not necessarily a causal mapping. See 
also theorem \ref{phi-phiminusone} below.

With the definition \ref{equivalence}, which certainly makes sense, 
and using the fundamental proposition 
\ref{local-causal}, we have one of the previously discussed sought
results:
\begin{theo}
Any two Lorentzian manifolds are {\em locally} isocausal.
\label{local-isocausal}
\end{theo}
In other words, the definition given in \cite{KRONHEIMER-PENROSE} and 
quoted in the previous subsection
can be forced to make sense if modified as follows:
\begin{quote}
{\em For each point $x\in V$ we can
find a small open neighbourhood $U_{x}$ of $x$ and an open subset 
$\U$ of flat $n$-dimensional Minkowski spacetime such that $U_{x}$ 
and $\U$ are isocausal: $U_{x}\sim \U$.}
\end{quote}  
Let us stress that isocausality here and in theorem \ref{local-isocausal} 
is used in the sense of definition \ref{equivalence}. Thus, there are 
two chronal-preserving diffeomorphisms: $\Phi : U_{x}\rightarrow 
\U$ and $\Psi : \U\rightarrow U_{x}$, say.

The particular case that a causal mapping has an inverse which is a causal
mapping too can be characterised as a conformal diffeomorphism.
\begin{theo}
For a diffeomorphism $\varphi:(V,\G)\rightarrow(W,\tilde{\G})$ 
the following properties are equivalent
\begin{enumerate}
\item $\varphi^*\tilde{\G}=\l\G$, $\l>0$.
\item $(\varphi^{-1})^*\G=\mu\tilde{\G}$, $\mu>0$.
\item $\varphi$ and $\varphi^{-1}$ are both causal (or anticausal) 
mappings.
\end{enumerate}
\label{phi-phiminusone}
\end{theo}

\subsection{Ordered chains of causal structures}
\label{chains}
The relation ``$\sim$'' is a binary relation on the set of 
diffeomorphic Lorentzian manifolds. This relation is obviously 
reflexive, transitive and symmetric, that is to say, is an 
equivalence relation. Thus we can collect the Lorentzian manifolds in 
equivalence classes and, following standard procedures, partially 
order them. This is the subject of this subsection.

To start with, choose a given a differentiable manifold $M$ 
meeting the conditions of theorem \ref{lorentz-manifold}. Recall that 
the set of Lorentzian metrics which can
be defined on $M$ is designed by Lor($M$). Clearly, for any fixed $M$ we
will find metrics with quite different causal properties. For 
instance, in $M=\r^4$ we can define the flat Minkowskian metric which 
is globally hyperbolic, or the anti de Sitter metric, which is just 
causally simple, or even the G\"odel metric \cite{GODEL,FF,MCCALLUM},
which is 
totally vicious. The nice thing about the relations ``$\sim$'' and 
``$\prec$'' is that all of these metrics can be sorted by means of the
equivalence relation ``$\sim$'', and then orderly classified by the 
appropriate generalization of the preorder ``$\prec$''.
\begin{defi}[Causal structure]
A causal structure on the differentiable manifold $M$ is any element
of the quotient set \mbox{\rm Lor}$(M)$/$\sim$.  Each of these causal
structures
is denoted by coset$(\G)$ where $(M,\G)$ is any
representative of the equivalence class, that is
$$
\mbox{\rm coset}(\G)\equiv \{\tilde\G \in \mbox{\rm Lor}(M): \, 
(M,\tilde\G)\sim (M,\G)\}\, .
$$
\label{causal-structure}
\end{defi}
Obviously the equivalence classes depend on $M$ so that, in case of 
possible confusion one should use the notation coset${}_{M}(\G)$ 
making explicit the base differentiable manifold. Observe that, due to 
theorem \ref{phi-phiminusone}, these causal structures include the 
conformal ones in the sense that all metrics globally conformally related 
to a given $\G$ are elements of coset$(\G)$. However, the causal 
structures are much richer and larger than the conformally related 
metrics, and this is a desirable property which allows to, for 
example, say that all Lorentzian manifolds have the same causal 
structure {\em locally} (theorem \ref{local-isocausal}). Or more 
interestingly, it permits a precise truthful
meaning  to be given to the sentence ``asymptotically flat spacetimes 
\cite{FF,PENROSE,PENROSE-CONFORMAL,PENROSE-DIAGRAM,Wald} (see subsection
\ref{conformal-boundary}) have the 
causal structure of flat spacetime {\em asymptotically}'', see Example 
9 in \cite{CAUSAL}. Concerning this issue, see also \cite{LAU,LOW,RPACN}.

Causal structures are naturally ordered by the binary relation $\preceq$
defined on Lor$(M)$/$\sim$ by
$$
\mbox{coset}(\G_1)\preceq\mbox{coset}(\G_2)\Longleftrightarrow 
(M,\G_1)\prec (M,\G_2).
$$
This is a reflexive, antisymmetric and transitive relation, that is, 
a {\em partial order}. The property measured and classified by this
partial 
order is in some sense the quality of the causal behaviour. This is 
intuitive from theorem \ref{preservation}, which can be re-written 
loosely providing an order of part of 
the hierarchy of standard causality conditions as
$$
\mbox{glob.\ hyp.\ }\preceq\mbox{c.\ stable}\preceq\mbox{strongly c.\ }
\preceq\mbox{disting.\ }
\preceq\mbox{causal}\preceq\mbox{chron.\ }\preceq\mbox{tot.\ 
vicious}\, .
$$
But the order provided by ``$\preceq$'' is indeed finer than this,
because not every pair of globally hyperbolic spacetimes based on the 
same $M$ are causally equivalent (recall the example of de Sitter and 
Einstein spacetimes), and the same happens for pairs of causally 
stable, strongly causal, et cetera, Lorentzian manifolds.  This means 
that each of the subsets defined by the standard hierarchy, such as 
the globally hyperbolic metrics on a given $M$, are themselves also
partially 
ordered by $\preceq$. Hence we can build abstract chains
in the form
$$
\underbrace{\dots\preceq\mbox{coset}(\G_1)\dots\preceq
\mbox{coset}(\tilde\G_1)\dots}_{\mbox{glob. hyp.}}\preceq
\underbrace{\dots\preceq\mbox{coset}(\G_2)\preceq\dots}_{\mbox{causally 
stable}}\preceq
\underbrace{\dots\mbox{coset}(\G_m)\preceq\dots}_{\dots\,\, \dots}
$$ 
Interesting questions are (i) the length of the partial order 
$\preceq$, that is to say, the size of the longest possible chain of 
the above type;
and (ii) the existence of lower or upper bounds for
these chains which could represent Lorentzian manifolds with the
best or worst causal properties, respectively. This would give an 
answer to the question of whether or not it is possible to define the 
best causally behaved metric on a given manifold. One can give
reasons to accept that totally vicious spacetimes are always placed at
the rightmost of a causality chain but it is still unknown what type
of globally hyperbolic Lorentzian manifold (if any) should be at the 
leftmost of these chains, of if the chains continue indefinitely to 
the left. Observe that, from this point of view, the 
coset${}_{\r\times S^{n-1}}$ containing
de Sitter spacetime is ``better'' than that defined by Einstein's
universe.

\section{Causality and topology} 
\label{sec:CandT}
In this section we will discuss the interplay between the causality of
a set in any of the senses presented in previous section and its
topology, if any.  Topology and causality keep a
close relationship in the case of Lorentzian manifolds because the
existence of a Lorentzian metric is not compatible with all topologies
of a differentiable manifold.  Conversely if we put a certain
topology in our manifold certain Lorentzian metrics are not allowed on
it or some causal curves with certain topological properties may not
be present.  Therefore we will start with a brief account of some
classical results settling down these issues for the case of Lorentzian
manifolds and proceed afterwards to more general cases involving
abstract causal spaces.
Other interesting investigations can be found in \cite{CHAMBLIN,MH}. 
An excellent introductory paper, with many examples and questions for 
students and non-experts, is \cite{GERHOR}.

\subsection{Some classical results and their generalizations}
\label{classic}
The simplest question is which manifold topologies are compatible with
the existence of a Lorentzian metric.  This question is answered by the 
next characterization: a differentiable manifold admits a Lorentzian
metric (not necessarily time orientable)
if and only if there exists on $M$ a `line field' \cite{KRIELE}. If there is a
nowhere zero vector field then the existence of the line field is ensured
\cite{KRIELE,ONEILL}, and moreover time orientability does hold. 
Then, the following theorem, whose proof can be
found in
\cite{STEENROD,ONEILL}, follows.
\begin{theo}
If $M$ is a smooth manifold then the following statements are equivalent
\begin{enumerate}
\item $M$ admits a Lorentzian metric.
\item Either $M$ is non-compact or it is compact with zero Euler 
characteristic.
\end{enumerate}
\label{lorentz-manifold}
\end{theo} 
As explained in subsection \ref{hierarchy}, compact manifolds  
always fail to be chronological---they always contain closed timelike 
curves---, so that the restriction put by this theorem is very mild 
in physical terms.

Once we have a Lorentzian manifold the next step is to study the
topological properties of the basic chronological sets $I^{\pm}$,
$J^{\pm}$, etc as well as those of causal curves.  The main results
were presented in section \ref{sec:essentials}, $\S$\ref{J+}. Further 
results will be presented next.

Some of the conditions of the standard hierarchy of definition 
\ref{SER} are in fact topological conditions on relevant sets of
causality theory.  Others admit an alternative formulation in terms of
topological concepts. Recall that a set valued function 
$I:M\rightarrow{\cal P}(M)$ is outer continuous at $x$ if 
for every compact set
$K\subseteq$ext$(I(x))$ we can find a neighbourhood $\U_x$ of $x$ such
that $K\subseteq$ext$(I(y))$, $\forall y\in U_x$.  Similarly we can
define inner continuity at $x$ using the set $I(x)$ instead of
ext$(I(x))$. The set-valued functions defined by $I^{\pm}$ are inner 
continuous \cite{HAWKINGSACHS}. On the other hand, their outer continuity
is 
a property of causally continuous spacetimes:
\begin{prop}
A spacetime is 
\begin{enumerate}
\item Globally hyperbolic if and only if it is strongly causal
and $J^{+}(x)\cap J^{-}(y)$ is compact for all $x,y\in V$.
\item Causally simple if and only if it is distinguishing and
$\d J^{\pm}(x)=E^{\pm}(x)$, for all $x\in V$.
\item Causally continuous if it is distinguishing and
the set-valued functions $I^+$ and $I^-$ are outer continuous.
\end{enumerate}
\label{extension} 
\end{prop}
There are also a number or results concerning the domain of dependence 
\cite{GEROCH-SPLITTING,BEE,FF,PENROSE,SINGULARITY}. 
\begin{theo}
Let $\zeta\subset V$ be a closed achronal set of a Lorentzian 
manifold $(V,\G)$,
\begin{enumerate}
\item if strong causality holds on $\overline{J^+(\zeta)}$, then
$H^+\left[\overline{E^+(\zeta)}\right]$ is non-compact or empty.
\item $(${\em int}$D(\zeta),g)$ is globally hyperbolic.
\end{enumerate}
\label{added}
\end{theo}
The first of these results is crucial in the proof of 
the Hawking-Penrose singularity theorem \cite{HP,FF,SINGULARITY}. The 
second states in particular that if an acausal set without edge 
$\Sigma$ fails to be a Cauchy hypersurface, still its total domain of 
dependence is globally hyperbolic with $\Sigma$ as Cauchy hypersurface. 
This is also of paramount importance for the proof of the singularity 
theorems, due to the following property of globally hyperbolic domains.
In a globally hyperbolic domain of a spacetime, the sets $C(a,b)$ are 
{\em compact} with respect to the $C^0$ topology introduced in
definition \ref{c-0}. 
%% 
 % (This can be generalized to 
 % causal curves with past and future endpoints lying on certain sets $A$ 
 % and $B$, in which case one uses the obvious notation $C(A,B)$ 
 % (see \cite{PENROSE} for further details). 
 %%
This implies that the length functional on curves attains its maximum 
on $C(a,b)$ in globally hyperbolic domains. This argument
together with the fact that, under certain assumptions, the longest
causal curves between points are timelike geodesics with no conjugate
points allows one to prove the incompleteness of causal geodesics
\cite{PENROSE,HP,FF,BEE,Wald,KRIELE,SINGULARITY}, for one can prove 
under physically reasonable assumptions that {\em all} inextensible causal
curves 
must have conjugate points, and this contradicts the global 
hyperbolicity unless the curves are extensible. In this last case the 
reasoning is a little bit more involved and uses essentially point 
{\em (i)} in theorem \ref{added}, see \cite{HP,FF} for details and 
section 6 (p. 707) in \cite{SINGULARITY} for an intuitive description 
of the proofs.

We have already mentioned that many times spacetimes are extensible and that maximal 
(analytical or not) extensions are required. In this regard, it should be mentioned
 here that some maximal
analytic extensions of vacuum solutions to Einstein's equations, such as
the `NUT' extensions of Taub's spherically
symmetric solution through the four distinct Cauchy horizons (see \cite{FF}), may result
 on a non-Hausdorff manifold. Therefore, in some occasions it might be worth 
allowing for these more general type of manifolds. As a matter of fact, 
the causality properties of such non-Hausdorff Lorentzian manifolds were investigated 
by Hajicek \cite{Haj1,Haj2} with the result that they always fail to be strongly causal.

In \cite{WOOLGAR} Sorkin and Woolgar put forward a generalization of
the ordinary causal relation $<$ with the aim of extending the
compactness of $C(a,b)$ on globally hyperbolic
Lorentzian manifolds to the case in which the metric tensor is only
$C^0$.  This new relation is called $K^+$ and it is defined as the
smallest relation containing $I^+$ that is transitive and
topologically closed as a subset of $V\times V$ where
$(V,\G)$ is the Lorentzian manifold.  Using this relation the authors
generalize concepts such as causal curve and global hyperbolicity.

We end this section making some comments about the definition and
use of measures on Lorentzian manifolds. As opposed to the Riemannian
case there is no canonical way to set a measure on a Lorentzian manifold
$V$ and the volume element $n$-form constructed from the metric tensor is 
one among a number of choices. Sometimes one is interested in defining
a measure $\mu$ with the properties that $\mu$ is additive, positive definite, and such that $\mu(V)<\infty$. From such a measure one can 
construct the scalar functions on $V$ \cite{GEROCH-SPLITTING}
$$
t^-(p)=\mu(I^-(p)),\ \ t^+(p)=-\mu(I^+(p)),\ p\in V.
$$ 
These functions appear in the proof of some important results such as 
theorem \ref{geroch-splitting}. In fact the monotonicity 
and differentiability of such functions give alternative characterizations
to some of the conditions of the standard hierarchy
\begin{prop}
A spacetime $V$ is 
\begin{enumerate}
\item chronological if and only if $t^+$ or $t^-$ are strictly increasing
along any future-directed timelike curve \cite{GEROCH-SPLITTING},
\item future (past) distinguishing if and only if $t^+$ ($t^-$) is 
strictly increasing along any future-directed causal curve,
\item causally continuous if both $t^+$ and $t^-$ are continuous 
and strictly increasing along any future-directed causal curve 
\cite{HAWKINGSACHS}.  
\end{enumerate}
\end{prop}
A thorough study of admissible
measures and volume functions on spacetimes can be 
found in \cite{DIECKMANN} (see also \cite{SANCHEZ-BR}).
In \cite{SZABADOS-MEASURE} it is shown how one can construct 
a measure in a space in which only the chronology relation $\less$ 
is present. This might serve as basis of a notion of stable causality 
in abstract causal/etiological spaces. 

\subsection{Splitting theorems for globally hyperbolic spacetimes}
\label{splitting}
Globally hyperbolic spacetimes being the best causally behaved 
Lorentzian manifolds, as well as the arena on which singularity 
theorems were founded upon, have received a great deal of attention 
over the years. Moreover, global hyperbolicity is often assumed as a 
physically realistic restriction, because physicists expect that a 
sensible theory must be capable of defining well-posed initial value
problems, or Cauchy problems, for the physical fields and this is only 
the case in globally hyperbolic spacetimes. Thus, there are many 
results concerning these class of spacetimes, which are 
of a very special type topologically speaking. 

In this section we deal with these topological constraints
on globally hyperbolic Lorentzian manifolds. These are usually known 
as ``splitting theorems'', for the results prove that the manifold 
can be foliated by spacelike hypersurfaces which are, all of them, 
$C^0$ Cauchy hypersurfaces. The pioneering classical result is due to
Geroch \cite{GEROCH-SPLITTING}.
\begin{theo}
Any globally hyperbolic $n$-dimensional Lorentzian manifold $(V,\G)$ is 
homeomorphic to the topological product $\r\times S$ where $S$ is a 
$(n-1)$-dimensional topological manifold. The image of $S$ under 
the homeomorphism is a Cauchy hypersurface on $(V,\G)$.
\label{geroch-splitting}
\end{theo}
Geroch's splitting theorem is the first of a series attempting to
translate the Cheeger-Gromoll splitting theorem in proper Riemannian 
geometry \cite{CG} to the Lorentzian case. This theorems states: ``Let 
$M$ be a complete and connected proper Riemannian manifold with 
non-negative Ricci curvature. If $M$ contains a complete geodesic 
realizing the distance between any two of its points (a ``line'') 
then $M$ splits isometrically as $M'\times \r$''. Partial results
towards this direction were obtained in
\cite{ESCHENBURG,BEEM,GALLOWAY} later generalized
in the next result proven by Galloway \cite{GALLOWAY2}.
\begin{theo}
Let $(V,\G)$ be a connected globally hyperbolic $n$-dimensional spacetime 
whose Ricci tensor $Ric$ satisfies the condition $Ric(\X,\X)\geq 0$ for
all timelike vectors $\X$. If $(V,\G)$ contains a complete timelike geodesic
$\g$
which is maximal between any two of its points, then it is isometric to 
$(\r\times S,dt^2\oplus\, (-\G_{1}))$ where $(S,\G_{1})$ is an
$(n-1)$-dimensional
complete proper Riemannian manifold and $\g$ is represented by the factor 
$(\r,dt^2)$.
\label{galloway-splitting}
\end{theo}
Traditionally, these types of manifolds were called a ``flat 
extension'' of $(S,\G_{1})$, see e.g. \cite{RWW}, so the conclusion 
in the theorem can be 
re-formulated by saying that $(M,\G)$ is a timelike flat extension of 
an $(n-1)$-dimensional complete proper Riemannian manifold. Another 
simpler way of stating the same thing is that there exists a ``global Gaussian 
time coordinate'', see \cite{WTM,PENROSE,SINGULARITY} for the use 
of Gaussian coordinates and its relation with maximal geodesics.
Recently, theorem \ref{galloway-splitting} has been employed in a 
relatively simple proof of the positive mass theorem \cite{CHRUSCIEL}. 

The previous generalizations of the Cheeger-Gromoll splitting theorem 
require, as the original theorem itself, the existence of a 
``line''---an inextensible maximal/minimal geodesic. The conclusion 
is thus stronger than in theorem \ref{geroch-splitting}. However, the 
original Geroch's result can be substantially improved to obtain an 
orthogonal splitting of the type appearing in theorem 
\ref{galloway-splitting}. This has been achieved in recent work by
Bernal and S\'anchez \cite{BERNAL,BERNAL2}, who first of all
were able to prove that the homeomorphism in Geroch's theorem can
be smoothed out to be a diffeomorphism, and the Cauchy hypersurfaces are 
also smooth.
\begin{theo}
Any globally hyperbolic spacetime admits a smooth spacelike Cauchy
hypersurface $S_0$ and is diffeomorphic to $\r\times S_0$.
\label{bernal-sanchez}
\end{theo}
More importantly, these authors \cite{BERNAL2} have managed to prove 
the following fundamental result, probably the most general and 
powerful splitting theorem available so far.
\begin{theo}
Any globally hyperbolic Lorentzian manifold $(V,\G)$ is isometric to the 
smooth product 
spacetime with base manifold $\r\times{\cal S}$ and Lorentzian 
metric
$$
\G=\beta\,\, d{\cal T}\otimes d{\cal T}-\bar{\G}
$$
where ${\cal S}$ is a smooth spacelike Cauchy hypersurface, 
${\cal T}:\r\times{\cal S}\rightarrow\r$ is the natural projection, 
$\beta$ is a smooth positive function on $V$, and $\bar{\G}$ is a 
rank-2 {\em degenerate} symmetric tensor field on $V$, such that 
the following properties are fulfilled
\begin{enumerate}
\item $d{\cal T}$ is timelike and future-directed everywhere (therefore 
${\cal T}$ is a {\em differentiable} time function).
\item All hypersurfaces ${\cal S}_{{\cal T}}:\, \{{\cal T}=const.\}$ are 
mutually diffeomorphic spacelike Cauchy hypersurfaces---with, say, ${\cal 
S}_{0}={\cal S}$.
\item The radical of $\bar{\G}$ is one-dimensional everywhere on $V$ 
and given by {\em Span}$\{\d_{\cal T}\}$.
\end{enumerate}
\label{bernal-splitting}
\end{theo}
Clearly this theorem supersedes theorems \ref{geroch-splitting},
\ref{galloway-splitting} and \ref{bernal-sanchez} and settles a long
history of guesses and hunches
about a possible generalization of Geroch's result, the existence of 
differentiable time functions, and the smoothness of the Cauchy 
hypersurfaces. For an account of this history, consult 
\cite{BERNAL,BERNAL2} and references therein.

\subsection{Topologies on causal spaces and Lorentzian manifolds}
\label{natural}
We now present a brief summary of the topologies which can be defined on a
causal space starting with the results presented in
\cite{KRONHEIMER-PENROSE}. 
These results hold for general causal spaces
and so they carry over to {\em causal} Lorentzian manifolds.  The
Alexandrov
topology $\Ts^*$ already presented in definition \ref{alexandrov} is
the most natural topology for a causal space but there are other
topologies which can be defined on a causal space or a Lorentzian
manifold. Kronheimer and Penrose defined the topology $\Ts^+$ as the
smallest topology in which each set $J^-(x)$ is closed. 
If $X$ is a manifold then its manifold topology is denoted by 
$\Ts^{\mbox{\tiny man}}$. Basic properties of these topologies 
are gathered next.
\begin{prop}
If $(X,\Ts^*)$ is Hausdorff and the causal space $X$ is full then
$(X,\less)$ is 
distinguishing.
\end{prop}
Recall that any Lorentzian manifold is, in particular, a {\em full} 
causal space, and given that for all $x,y\in X$, $I^{+}(x)\cap I^-(y)$ are 
always open in $\Ts^{\mbox{\tiny man}}$, in general $\Ts^*$ is smaller
than 
$\Ts^{\mbox{\tiny man}}$. The previous proposition states that $\Ts^*$ 
is strictly smaller than $\Ts^{\mbox{\tiny man}}$ in 
non-distinguishing spacetimes. As a matter of fact, 
$\Ts^*$ and $\Ts^{\mbox{\tiny man}}$ are
equivalent topologies if and only if the spacetime is strongly causal
\cite{KRONHEIMER-PENROSE,PENROSE}. Other equivalent statements are 
gathered next \cite{KRONHEIMER-PENROSE}.
\begin{theo}
The following statements are equivalent.
\begin{enumerate}
\item $\Ts^*=\Ts^{\mbox{\tiny man}}$.
\item The topological space $(X,\Ts^*)$ is Hausdorff.
\item If $\forall x\in I^-(a)$ and $\forall y\in I^+(b)$ we have $x< y$,
then $b\not < a$ unless $a=b$.
\item If $b\rightarrow a$, and $\forall x\in I^-(a)$ and $\forall y\in 
I^+(b)$ we have that $x\less y$, then $a=b$. 
\item If $X$ is a Lorentzian manifold then it is strongly causal.
\end{enumerate}
\label{tsts}
\end{theo}

Most of the topological properties of the chronological sets discussed
in section \ref{sec:essentials} are carried over to full causal spaces
using the Alexandrov topology. Here, we use $A^{\mbox{\tiny int}*}$,
$A^{\mbox{\tiny cl}*}$ and $A^{\mbox{\tiny bdy}*}$ to mean the interior,
closure, and boundary with respect to the Alexandrov topology, 
respectively.
\begin{theo}
If $X$ is a full causal space then \cite{KRONHEIMER-PENROSE} 
\begin{enumerate}
\item $x\in I^-(x)^{\mbox{\tiny cl}*}\cap I^+(x)^{\mbox{\tiny cl}*}$.
\item $I^+(x)^{\mbox{\tiny cl}*}=\{y\in X: I^+(y)\subset I^+(x)\}$ and 
dually for the past.
\item For any subset $A\subset X$, $J^+(A)\subset I^+(A)^{\mbox{\tiny
cl}*}$ and 
$I^+(A)=J^+(A)^{\mbox{\tiny int}*}$.
\item The following statements are equivalent
\begin{itemize}
\item $X$ is a future-reflecting $\Bf$-space
\item $J^{+}(x)=I^{+}(x)^{\mbox{\tiny cl}*}$ for all $x\in X$
\item If $A$ is compact with respect to $\Ts^+$ then 
$J^+(A)=I^+(A)^{\mbox{\tiny cl}*}$ and 
$J^+(A)^{\mbox{\tiny bdy*}}=E^+(A)$
\item If $A$ is compact with respect to $\Ts^+$ then 
$J^+(A)^{\mbox{\tiny bdy*}}=E^+(A)$
\end{itemize}
\item $\Ts^+$ is smaller than $\Ts^{*}$ if and only if $X$ is 
a past-reflecting $\Bf$-space.
\end{enumerate}
\end{theo}

Other topologies on Lorentzian manifolds can sometimes provide a
deeper insight into certain global causal properties of spacetimes. Of 
particular importance in this sense is the paper by Hawking, King and
McCarthy
\cite{HAWKING} where a new topology for Lorentzian manifolds was
studied. This topology, called the {\em path topology} and denoted by
$\Pp$, is the finest topology such that the induced topology on every
timelike curve agrees with the topology induced from the manifold. 
Intuitively we can think of $\Pp$-homeomorphisms as transformations
mapping bijectively timelike curves to timelike curves. In 
\cite{HAWKING} the following result was proved:
\begin{prop}
For strongly causal spacetimes $\Pp$-homeomorphisms are
homeomorphisms of the manifold mapping null geodesics into null
geodesics.
\label{p-null}
\end{prop} 
This is an intermediate result useful to prove the following theorem
\ref{p-homeo}. To that end a result of Hawking's is invoked
\cite{HAWKING}:
\begin{theo}[Hawking's theorem]
Any homeomorphism of a spacetime which takes
null geodesics into null geodesics is a $C^{\infty}$
diffeomorphism.  
\label{hawking-th}
\end{theo}
From proposition \ref{p-null} and theorem \ref{hawking-th}
the next theorem easily follows.
\begin{theo}
In strongly causal spacetimes, $\Pp$-homeomorphisms are
smooth conformal diffeomorphisms.
\label{p-homeo}
\end{theo}

This result can be generalized in two directions.  The first
generalization
is theorem \ref{malament-result} in which no causality conditions upon
the spacetime are required and the result holds for bijective mappings
with no further topological properties.  The second generalization is
the result of Huang's paper \cite{HUANG} and it only takes into
account null geodesics in its formulation.
\begin{theo}
Let $(V,\G)$ be a strongly causal space-time of dimension greater
than three and let $f: V\rightarrow  V$ be a bijection such that 
both $f$ and $f^{-1}$ take null geodesics into null geodesics.
Then $f$ is a homeomorphism and, by Hawking's theorem 
\ref{hawking-th}, a smooth conformal transformation. 
\label{huang-res}
\end{theo}

\subsection{Topology on {\rm Lor}(V).}
\label{Cr-topo}
Many times it is interesting to introduce topologies in the set
Lor$(V)$ of Lorentzian metrics over the differentiable manifold $M$ in
order to have a definition of ``metric closeness'', which is of
relevance for concepts such as causal stability and related
conditions.  The most studied topologies are the fine $C^r$ topologies
whose definition taken from \cite{BEE} we reproduce here.  Other
related results can be found in \cite{BOMBELLI}.
\begin{defi}[$C^r$ topology on Lor($V$)]
Let $\G_1,\G_2\in \mbox{\rm Lor}(V)$ and define a fixed locally finite 
covering of $V$ by coordinate neighbourhoods whose 
closure lies in a coordinate chart.  Set a continuous function 
$\delta:V\rightarrow
(0,\infty)$.  Then we say that $\G_1$ and $\G_2$ are $\delta$-close in
the $C^r$ topology, written $|\G_1-\G_2|<\delta$, if all the
corresponding components of $\G_1$ and $\G_2$ and of their $j^{th}$
derivatives 
$D^j\G_1$, $D^j\G_2$ ($0\leq j\leq r$) up to order $r$ 
satisfy $\left|D^j\G_1|_p-D^j\G_2|_p\right|<\delta(p)$, $\forall p\in V$.
\label{Crtopology}
\end{defi}
The sets 
$$
\{g_1\in \mbox{\rm Lor}(V):|g_1-g_2|<\delta\}
$$ 
form a basis of the {\em fine $C^r$ topology} of Lor$(V)$ which can be
shown to be independent of the chosen covering. An interpretation of 
these topologies is: metrics close in the fine $C^0$ topology
have light cones which are close; if the metrics are close
in the fine $C^1$ topology then their geodesics are close; if
the metrics are close in the fine $C^2$ topology, their curvature
tensors are close; and so on. In all these cases, closeness of the light
cones,
of geodesics and of curvature tensors are defined in an appropriate 
intuitive sense. A simple application of these ideas is given next
\begin{prop}
The spacetime $(V,\G)$ is causally stable if and only if there is a fine
$C^0$ neighbourhood $U(\G)$ of the metric $\G$ such that for any $\G_1\in
U(\G)$ the spacetime $(V,\G_1)$ is causal.
\label{causa-stable}
\end{prop}
This result tells us that the stable causality condition is nothing but
the stability against fine $C^0$ perturbations of causal spacetimes. 

Actually fine $C^r$ topologies are employed to study the stability of
certain properties of spacetimes such as the existence of complete and
incomplete curves or the causality conditions (cf. chapter 7 of
\cite{BEE}). 
For more information about stability properties in terms of topologies 
defined in tensor field spaces one can consult the standard references 
\cite{Gernew2,Hawnew}.

Definition \ref{Crtopology} has the awkward feature that it is formulated
in terms of local coordinates so the invariance of any concept formulated 
in terms of $C^r$ topologies under coordinate changes
must be checked a posteriori. Therefore it 
is interesting to seek definitions of ``closeness'' between Lorentzian
metrics formulated in a coordinate-free way. In \cite{NOLDUS} Noldus 
discusses three different notions of distance between Lorentzian metrics, all 
formulated with no reference to coordinate systems. They try to 
capture the ideas of when the metrics are causally close, their ``volume'' is close 
or their geodesics are close, respectively.

\section{Causal boundaries}
\label{causal-boundary}
In this section we discuss one of the most studied issues in causality
theory 
with important applications in related fields of General Relativity such
as 
singularity theory or asymptotic properties of fields and gravitation, and 
more recently even in quantum aspects of gravity (through string theory,
supergravity, or conformal field theories) 
via Maldacena's Conjecture \cite{MALDACENA,HOROWITZ}.  
This is certainly a very
remarkable example of the profitability of studying causal boundaries and
causal completions: Maldacena's Conjecture requires the very
concept of causal boundary (at infinity) in its rigorous formulation.  

On mathematical grounds it has always been fruitful to attach a boundary
to 
a topological set $X$ in order to make it ``closed " or ``complete". If
$X$ 
is a metric space then there is a canonical way to accomplish this by
constructing 
its Cauchy completion $\bar{X}$. Hence the boundary $\d X$ of $X$ is defined 
by $\d X\equiv\bar{X}-X$. An important example of this is a proper 
Riemannian manifold which as is well known can be transformed into 
a metric space with the standard notion of geodesic distance so the 
issue of attaching a boundary to a proper Riemannian manifold can be 
addressed with no further complications. Quite often, however, there 
are several inequivalent but sensible ways of attaching a boundary to a 
given set, be it because it has no metric-space structure, or because we 
want to make abstraction of this. This will certainly be the case for 
spacetimes, as they are not metric spaces, so that many boundaries 
are feasible (with good properties) and we will have to decide about 
which particular properties and objects we want to describe or study by 
means of the completion and the boundary. The driving idea in almost all
 of the boundaries conceived so far is to obtain completions which are 
causal sets or Lorentzian manifolds but in such a way that the completion
 is {\em consistent} with the original causal structure. This means that 
causal relations between points of the completed manifold 
must agree with those of the original spacetime. 

The concept of such a ``causal boundary" for spacetimes was first
introduced 
by Penrose more than forty years ago and it turned out to be one of the 
most prolific ideas in General Relativity as we hope to make plain in this 
section. In essence, Penrose's idea is to 
embed the spacetime under study into another bigger Lorentzian manifold 
{\em conformally}, so that the causal properties are trivially kept, and 
obtain properties of the initial spacetime by examining its boundary on 
the encompassing spacetime. After the initial success of this construction
in  
simple but physically relevant examples, the construction was generalized 
in several directions by a number of authors, sometimes keeping the idea 
of the embedding, sometimes not, but usually with the goal of trying to
remove the use of conformal transformations, as they are almost impossible 
to find in generic situations. The most important generalizations are
reviewed 
in this 
section. As a pre-conclusion, unfortunately we must say that in general 
either they are too theoretical, and cumbersome to be tested in explicit
 examples, or they 
present some undesirable features.  However, there have been some recent
 advances and developments \cite{SCOTT,HARRIS1,HARRIS2,CAUSAL,MAROLFROSS}
which might help in the final achievement of a definite and generally
accepted 
definition of causal boundary. These are treated in subsections 
\ref{m-r},  \ref{harris}, \ref{aboundary} and \ref{causal-relationship}.

\subsection{Penrose conformal boundary}
\label{conformal-boundary}
Penrose first introduced his idea of conformal boundary in 
\cite{PENROSE-CONFORMAL} and further developed it 
in \cite{PENROSE-DIAGRAM,PENROSE-RSC}. His aim when defining
 the conformal boundary was to study from another 
point of view questions related with the behaviour and radiative
 properties of spacetimes which are {\em asymptotically 
flat}. In fact the very definition of asymptotic flatness can be
 formulated naturally in the framework of the 
conformal boundary. 

Let $\tilde{\G}$ be the metric tensor of a $n$-dimensional spacetime 
$\tilde{M}$ (in purity, Penrose had General Relativity in mind and 
considered only the case $n=4$, but we will leave the dimension of the
 spacetime free) and suppose that 
we can set a conformal correspondence of this spacetime with a finite
 region $M$ of another $n$-dimensional Lorentzian manifold
with metric $\G$ (this is sometimes called the ``unphysical spacetime").
 This means that at every point of such region we have the relation
$$
\G=\Omega^2\tilde{\G}.
$$       
If $M$ was judiciously chosen (say such that $\tilde{M}$ has
 compact closure in the unphysical spacetime) then the whole of
``infinity'' in 
the physical spacetime can be brought to finite values of the coordinates
of the 
unphysical spacetime. Some other properties of the original spacetime
which 
nevertheless are {\em not} part of it, such as its singularities, can also 
be read off from their guessed places on $M$. Thus, ``infinity" and 
``singularities" are typically parts of the 
boundary $\d\tilde M$ of $\tilde M$ in the unphysical
 spacetime, and thus we can extract properties of the physical spacetime,
 its global structure and its singularities, by 
just analyzing the properties of the boundary $\d\tilde M$. The set
$\d\tilde M$ is called the 
{\em conformal boundary} of $\tilde M$. In some situations, such as 
asymptotically flat spacetimes, the part of the conformal boundary 
describing infinity of $\tilde M$ can be defined as the set of $x\in M$ 
where $\Omega=0$ (and if one wishes to talk about ``null infinity", 
see below, the condition $d\Omega\neq 0$ is added). The part of the
conformal
boundary representing infinity is denoted by $\Is$ and called sometimes 
conformal infinity. 
Therefore, in these cases a substantial part of the boundary are 
$(n-1)$-dimensional hypersurfaces of the unphysical spacetime $ M$.
 However, in general nothing can be said about $\Is$ in this sense, 
and it can be disconnected, or discrete, or $1$-, $2$-dimensional, 
et cetera, as well as a combination of hypersurfaces and all of these, and so on.

The procedure just described is called the {\em conformal
compactification} 
of $\tilde M$ relative to $M$ \cite{FRAUENDIENER}. 
It {\em depends} on $M$ and the chosen 
conformal factor; even if the boundary is {\em complete} 
($\d\tilde M$ has compact closure on $M$) uniqueness 
is not ensured, though it is sometimes claimed that it is fixed exclusively by $\tilde M$. 
The crucial point here, and for the whole construction,
is that the physical spacetime is {\em conformally related} to its image 
on the enlarged manifold, and therefore the causal properties of $\tilde M$
 have been kept. Besides, the boundary acquires causal properties itself
as 
a set of $M$, so that it may be given attributes such as spacelike, 
timelike, or null, or it can have parts to the future/past of $\tilde M$ 
(so that is to the future/past of the entire spacetime!), et cetera.

The conformal compactification can be carried out explicitly in the
case of flat Minkowski spacetime (see \cite{PENROSE-RSC} for details). 
A suitable choice for the unphysical spacetime is the Einstein static 
universe. Many other choices are possible, but this particular one 
has the virtue of making $\d\tilde M$ complete. To see this we write the
flat
Minkowski metric in spherical coordinates 
$$
ds^2=dt^2-dr^2-r^2d\Omega^2,
$$  
where $d\Omega^2$ stands for the standard round metric on the sphere
$S^{n-2}$,
and perform the coordinate transformation (the angular part is simply
omitted for shortness)
\bnr
t=\fr{1}{2}\left(\tan\left(\fr{\bar{t}+\bar{x}}{2}\right)+\tan\left(
\fr{\bar{t}-\bar{x}}{2}\right)\right),\\
r=\fr{1}{2}\left(\tan\left(\fr{\bar{t}+\bar{x}}{2}\right)-\tan\left(
\fr{\bar{t}-\bar{x}}{2}\right)\right),\\
\enr
with coordinate ranges $-\pi<\bar{t}+\bar{x}<\pi$,
$-\pi<\bar{t}-\bar{x}<\pi$, $0<\bar{x}<\pi$. This transformation
brings Minkowski's line element into the form
$$
ds^2=\fr{1}{4\cos^2(\frac{\bar{t}+\bar{x}}{2})\cos^2(\frac{\bar{t}-\bar{x}}{2})
}(d\bar{t}^2-d\bar{x}^2-\sin^2\bar{x}\,
d\Omega^2),
$$
from which we see that it is conformal to a certain region of 
Einstein static universe (which has the line-element in brackets). This is 
depicted in the figure below taken from \cite{PENROSE-DIAGRAM}
 where the Einstein static cylinder is shown
 ($n-2$ spatial dimensions suppressed, so that 
each horizontal circle of the cylinder represents a $S^{n-1}$ sphere)  
and in red the region conformal to Minkowski spacetime.

\parbox{.5\textwidth}{
\includegraphics[width=.4\textwidth]{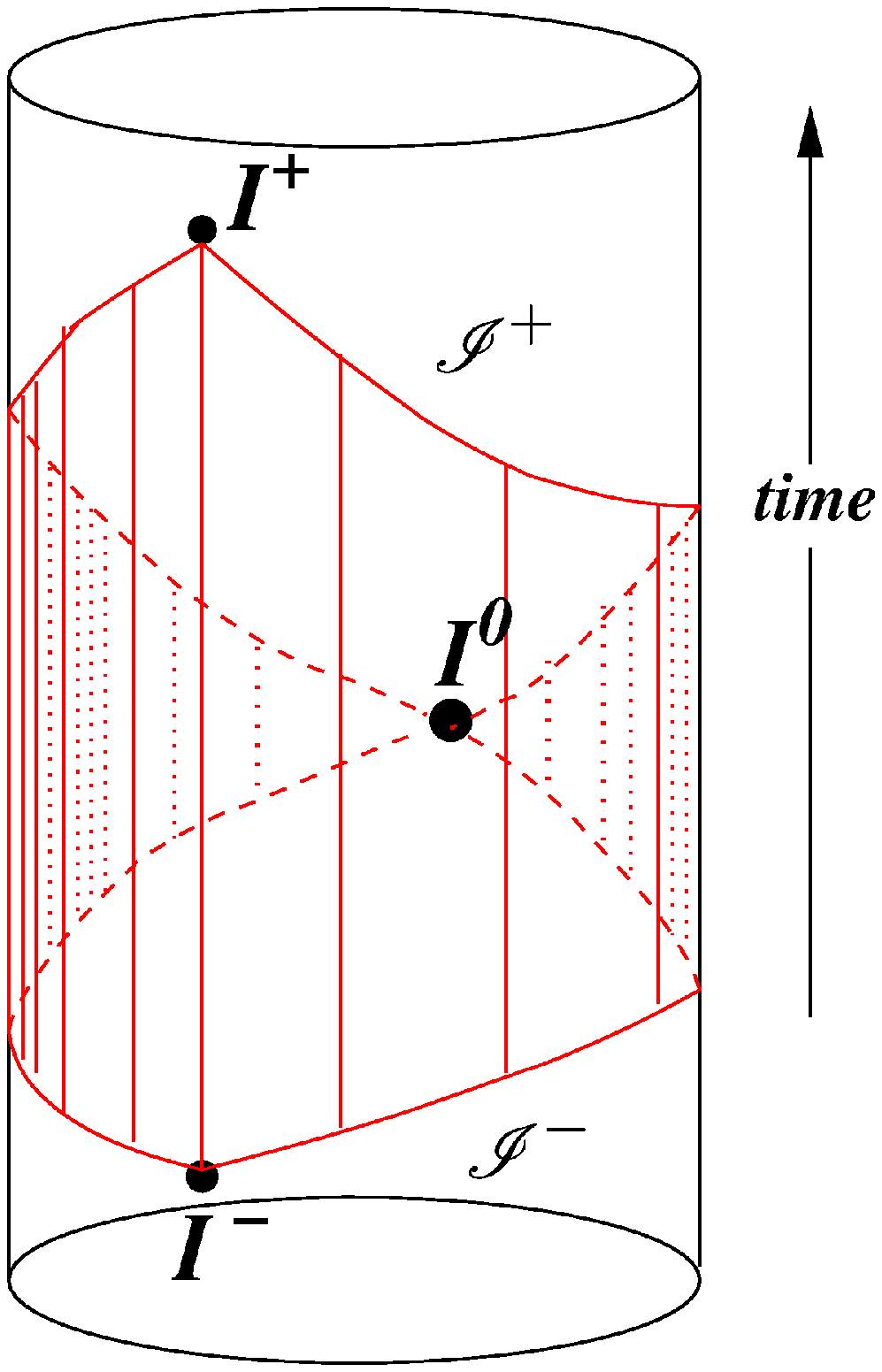}

\medskip
\parbox{5cm}{
\footnotesize Conformal embedding of Minkowski
 spacetime in the Einstein static universe.}}\hspace{-.8cm}
\parbox{.5\textwidth}{In the picture we see that the conformal boundary is
split in different 
regions called by Penrose
$\Is^+$, $\Is^-$, $I^+$, $I^-$ and $I^0$ (many times $i^+$, $i^-$ and
$i^0$ are used nowadays). $i^+$, $i^-$ and $i^0$ are points 
whereas $\Is^{\pm}$ are $(n-1)$-dimensional null hypersurfaces of topology
$S^{n-2}\times\r$ (to make clearer these 
topologies the original Penrose paper \cite{PENROSE-CONFORMAL} presents 
a different picture but the regions names are still the same). The sets
$\Is^+$ ($\Is^-$) are formed by ``endpoints" of the inextensible
future-directed (past-directed) radial null geodesics and $i^{+}$
($i^-$) are the 
``endpoints" of inextensible timelike future-directed
(past-directed) geodesics. Finally $i^0$ is the endpoint of all 
spacelike geodesics, as well as infinity for some spacelike
slices. Several issues related with this conformal compactification are
discussed in Penrose's papers 
\cite{PENROSE-CONFORMAL,PENROSE-RSC}
being some of them seminal ideas for important lines of research developed
in later years, see for a review \cite{FRAUENDIENER}. One of the first
applications was 
the definition of {\em asymptotic flatness} by means of conformal
compactifications. The idea is simply that 
the conformal boundary of any asymptotically flat spacetime must resemble
that of Minkowski spacetime just 
described.} 
\noindent
Penrose suggested that a 4-dimensional spacetime $\tilde{ M}$ is {\em
asymptotically flat} if an unphysical spacetime $ M$ 
exists such that the conformal infinity $\Is$ can be decomposed in two
null three dimensional hypersurfaces 
$\Is^+$ and $\Is^-$ with the topological properties described in the
figure above.
As a matter of fact, it can be proven that, if the vacuum Einstein
equations hold on a neighbourhood of $\Is$, then $\Is$ is a smooth null
hypersurface with two connected components, among other results, see
\cite{FF,FRAUENDIENER,STEWART}. Later this definition was refined and 
the concepts of {\em asymptotic simplicity} and {\em weakly asymptotic
simplicity} \cite{PENROSE-RSC,FF} were 
introduced in the literature. It is worth remarking that these concepts 
only demand that conformal infinity be formed by null hypersurfaces but 
we know that in the Minkowski case the set $\Is$ contains also 
discrete points. A complete correct definition of asymptotic flatness 
conveying the full
structure of $\Is$ has been a difficult issue,
 specially in what refers to spacelike infinity $i^0$. This was studied in
full generality by Ashtekar and Hansen, and an alternative improved treatment 
has been recently performed by Hayward, see 
\cite{AHSTEKAR-HANSEN,HAYWARD, HAYWARD-LETTER} and references therein.
 The problems concerning $i^0$ will re-appear several times in what follows.

It is also possible to study the
gravitational radiation of asymptotically flat systems (isolated
bodies) using these techniques. Indeed, Penrose was able to give explicit
expressions for the gravitational power 
radiated ``at infinity'' by an isolated system as an integral over $\Is^+$ 
\cite{PENROSE-RSC,PENROSE-STRUCTURE} 
(similarly there are expressions for the incoming radiation using
$\Is^-$), see for further details \cite{FRAUENDIENER,STEWART}. 
These results were known previously from the work of 
Bondi, Trautman, Pirani, Sachs and others \cite{SACHS,BONDI} but the
technique of the conformal boundary gave them a fuller, completely
coordinate-independent, 
geometrical significance. 

The conformal compactification can be carried out for other spacetimes
such as de Sitter, anti-de Sitter or some Robertson-Walker geometries
\cite{FF}.  This showed that singularities (such as the big bang) may
be part of the boundary, that the conformal boundary can have new
unexpected properties (an example of this is the non-differentiability
of the metric at $i^0$), and that one can give definite properties to
$\d\tilde M$ as a region of the unphysical spacetime.  In general,
however, the full conformal compactification is very difficult to
achieve, usually impossible.  The good news is that, in certain
particular but relevant cases, it is possible to perform the conformal
compactification of a two-dimensional piece of the metric retaining
the important information.  This is what happens for instance in
spherically symmetric spacetimes where the non-angular part of the
metric (2 dimensions) is conformally flat and thus liable to be
conformally compactified.  In this case we can draw two dimensional
pictures, called {\em Penrose diagrams} \cite{FF,PENROSE-STRUCTURE}, and
define
$\Is^{\pm}$, $i^{\pm}$ for the conformal boundary so obtained if they
exist.  These extremely useful representations of spherically
symmetric spacetimes will be further treated in section
\ref{conformal-diagram}.  Similarly, sometimes one can take a
particularly relevant 2-dimensional surface (ergo conformally flat) of
a given spacetime and draw its Penrose diagram.  This does not give a
full insight into the properties of the spacetime, but it certainly
enlightens our comprehension of some of its features.  A paradigmatic
example of this situation is given by the Penrose diagram of Kerr's
spacetime, which is just the diagram of {\em only} its axis of
symmetry, first found by Carter \cite{CARTER2}, see \cite{FF}.

Despite the fact that the conformal compactification is almost a chimera
in generic spacetimes, Friedrich \cite{FRIEDRICH1,FRIEDRICH2} has been
able to establish a procedure such that it is possible to write down a set
of equations (conformal equations) 
in which $\Omega$ and $\G$ are part of the unknowns. In principle this
could 
help to perform the conformal embedding of a spacetime, but even if not
so, it allows to talk about (and find some of) the properties of the
conformal boundary. Furthermore, the analysis of spacelike infinity, which
is one of the more intricate issues in conformal completions, can also be
recast in a form were $i^0$ is given internal structure so that its
properties become more transparent. For a review about all these matters,
see \cite{FRAUENDIENER}. On this positive side, we should also mention some powerful recent interesting results \cite{ACD,ACD2}, concerning the existence of conformal compactifications with a timelike $\Is$ (such as in anti-de Sitter spacetime, see $\S$\ref{conformal-diagram}) for a large class of globally hyperbolic static line elements. These spaces are shown to be geodesically complete Einstein spaces (with a negative scalar curvature), and no further symmetries apart from staticity. A very remarkable theorem found in \cite{ACD} (for the case $n=4$) states that the manifold $\r\times S^2$ with {\em any} $C^{\infty}$ static Lorentzian metric of non-negative scalar curvature  {\em is the conformal infinity} of a 4-dimensional Lorentzian Einstein space based on $\r^4$ with negative scalar curvature.

Penrose's idea of attaching a ``boundary'' to a spacetime was, and still
is!, very attractive 
and it was pursued by several 
authors over the years with rather assorted techniques and aims. The
question as to why one should be 
interested in enlarging or generalizing the conformal boundary
construction to something else has many answers. First of all, one would
like to avoid objects that are foreign to the spacetime under analysis,
such as $\Omega$ and the fictitious manifold $M$. Thus, a construction of
the boundary using objects of the spacetime exclusively (as is the case of
the Cauchy completions for metric spaces) has been systematically
sought. On the other hand, the conformal compactification is the simplest,
and truly very appealing, way of attaching a boundary to a Lorentzian
manifold, so that sometimes the idea has been to look for ways to avoid
the technical problems in finding the unphysical spacetime and the
conformal factor---which may be very difficult to find in closed form. 
A third reason why the definition of a boundary can be useful is the
analysis of singularities. Singularities 
are not part of the spacetime since they are related with diverging
quantities, or with incompleteness of curves, or with lack of tangent
vectors. 
Another way to look at this is by saying that a singularity is a set of
points ``where" the spacetime itself 
ends, or blows up, so it is sensible to think that a singularity will lie
``at the boundary'' of the spacetime. And this is certainly the case in
the simple examples constructed with Penrose diagrams, and in general in
all examples known so far. Thus, a suitable definition of boundary should
allow to tell apart which of its parts are singularities, and which are at
infinity, among other possibilities, if they arise. If this were achieved,
questions otherwise meaningless 
such as the shape of the singularity or its causal character would make
perfect sense.     

In the forthcoming subsections we review what 
we believe are the most relevant attempts in these directions presented so
far in the literature giving accounts of the motivations lying behind 
each construction.

\subsection{Geroch, Kronheimer and Penrose construction}
\label{GeKrPe}
One of the most famous attempts towards the construction of a causal
boundary was 
performed by Geroch, Kronheimer and Penrose in \cite{GKP} where they put
forward a scheme to attach a ``boundary'' 
to any spacetime fulfilling certain causality restrictions. The method
followed involved advanced topological constructions based only on global
causal objects present in Lorentzian manifolds. They tried to show that a
causal boundary could be associated to certain spacetimes (i) without
invoking ``external" concepts such as the unphysical spacetime or the
conformal factor $\Omega$, and also (ii) by using only causal concepts 
with no attention to the existence of a Lorentzian metric. Only
distinguishing spacetimes were covered by this method (henceforth called
GKP construction or {\em c-boundary}).

Instigating ideas for the GKP boundary were previously published by
Seifert in \cite{SEIFERT}. He proposed a scheme to attach a causal
boundary to space-times by
basically assigning a future and past endpoint to any inextensible causal
curve on a space-time $V$. The set of such points would be the causal
boundary $\d V$.
One can then introduce an ordering $\J$ on the completed space-time
$\bar{V}$ which is an extension of the usual causal relation $<$ on the
space-time $V$. These ideas were much improved in \cite{GKP}.

Before entering into the details of the GKP construction we need to define
certain concepts dealing with 
future and past sets. Recall that future and past sets (definition
\ref{futureset}) are those sets of the manifold $M$ whose chronological 
future (past) is contained in the set itself. Here the authors restrict 
their attention to {\em open} future $F$ or past $P$ sets, which is equivalent to
requiring $I^+(F)=F$ and $I^-(P)=P$. Such a future or past set is
called indecomposable or irreducible (IF or IP resp.) if it is not empty and cannot be
written as the union of two proper
subsets which are themselves future or past sets. Roughly speaking
indecomposable future and past sets can be divided into sets which 
are of the form $I^{\pm}(p)$ for $p\in M$ and those sets which cannot be
written as the chronological future or 
past of any point of the manifold. This distinction gives rise to proper
indecomposable future and past sets 
(abbreviated PIF's and PIP's) and terminal indecomposable future and past
sets (TIF's and TIP's) respectively.  Let us denote the collection of IP's
by $\hat{M}$ and 
 the set of all IF's by $\check{M}$.
For distinguishing spacetimes the manifold $M$ is 
the simplest example of TIP and TIF, but as we show next
terminal indecomposable sets are easily characterised 
(we only formulate the result for the past case) \cite{GKP,FF,KROLAK}.
\begin{theo}
Any IP is of the form $I^-(\g)$ where $\g$ is a future-directed timelike
curve. If $\g$ has a future endpoint $p\in M$ then the IP is the PIP
$I^-(p)$, while if $\g$ is future endless then the IP is a TIP. 
\label{pip}
\end{theo} 
Therefore timelike future-endless curves give rise to TIP's according to
this theorem.
As an example we can take Minkowski spacetime. In this case it is known
that 
TIP's and TIF's are respectively the chronological past and future of
timelike curves with constant acceleration plus the whole manifold itself
(there are no other terminal indecomposable sets in Minkowski spacetime),
and these have endpoints at $\Is$ in the conformal boundary. Actually, one
can set a clear correspondence between 
the TIP's and TIF's and the different regions of the conformal boundary
described in  
section \ref{conformal-boundary}: TIP's and TIF's represent $\Is^+$ and
$\Is^-$ respectively and the TIP or TIF defined by the manifold itself
represents $i^{\pm}$. 
Observe that $i^0$ is missing in this picture. As shown in \cite{GKP}, it
is also possible to determine the set of IP's for asymptotically
simple spacetimes getting a similar structure to that obtained by means of
the conformal compactification. Turning to the general case,
theorem \ref{pip} suggests that we may regard the TIP $I^-(\g)$ as a sort
of future {\em ideal end point} attached 
to the curve $\g$. The attribute ideal means that the point does not
belong to the manifold $M$ but rather to a larger set containing $M$ as a
proper subset. We should therefore try to construct a new manifold
consisting of the points of $M$ plus the ideal points and call it the
completion $\bar{M}$ of $M$. The set $\bar{M}-M$ would then be the {\em
causal boundary} and would contain ideal points only. How to construct
this larger set, and to endow it with a topology, is the next task.

Clearly there is an obvious correspondence between PIP's or PIF's and the 
points of the manifold in a distinguishing spacetime. This correspondence
is set by the injections
$I^+:M\rightarrow \check{M}$, $I^-:M\rightarrow\hat{M}$ so the set of
PIF's is $I^+[M]$ and the set of 
PIP's is $I^-[M]$. Neither $I^+$ nor $I^-$ are bijections, as the TIP's
and TIF's are not included (TIP's and TIF's are {\em ideal points}) so the
sets $\check{M}$ and $\hat{M}$ are ``bigger'' than $M$ which means that
they can provide us with the sought enlargement of $M$.  Consequently,  an
obvious starting point to construct the enlarged manifold is the set
$\check{M}\cup\hat{M}$. Now 
we try to find a natural injection from $M$ into $\check{M}\cup\hat{M}$,
but the problem is that each point $p\in M$ 
naturally corresponds to two elements, $I^+(p)$ and $I^-(p)$, of
$\check{M}\cup\hat{M}$  so it is not 
clear how we can map $M$ into $\check{M}\cup\hat{M}$ in an injective
way. To surmount this difficulty, the proposal in \cite{GKP} was that a
set of {\em identifications} must be carried out among the elements of
$\check{M}\cup\hat{M}$. 
There are some identifications which are obvious (for instance $I^+(p)$
and $I^-(p)$ must be identified) but, unfortunately, sometimes
identifications among the ideal points are also needed. A simple example
of this happens when we try to 
construct the completion for spacetimes whose conformal boundary $\Is$
contains timelike regions. In this case 
any ideal point in one of such regions could stem from a TIP as well and a
TIF, so they should be identified (see figure).

\parbox{.3\textwidth}{
\includegraphics[width=.3\textwidth]{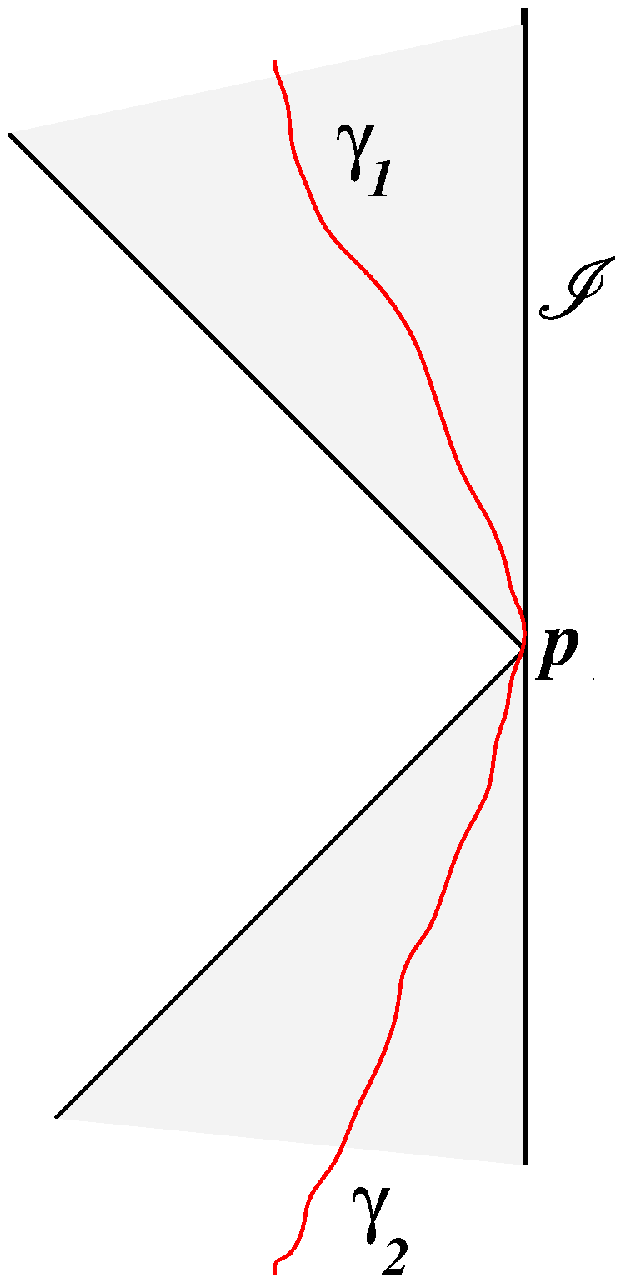}
\parbox{4cm}{\footnotesize In this example the ideal point $p$ stems 
from the TIP $I^-(\g_2)$ and the TIF $I^+(\g_1)$}}\hspace{.5cm} 
\parbox{.6\textwidth}{A further problem which arises in the construction
of the enlarged manifold is the definition 
of a suitable topology and, a posteriori, its differentiable
structure. These two problems, the identification rules and 
the construction of ``good'' topologies and manifold structures, are the
conundrum in the whole GKP construction. 

The answer provided in \cite{GKP} was proven 
later not to be appropriate for some cases and a number of generalizations
were tried
to improve this point as we will review in forthcoming sections. Let us
nevertheless present a summary of the idea in \cite{GKP} as it became the
background on which the generalizations were founded. As explained above
there are obvious identifications, so we define the set $M^{\natural}$ as
$\check{M}\cup\hat{M}$ with the elements $I^+(p)$, $I^-(p)$ identified
$\forall p\in M$.
For any element $P$ of $\hat{M}\cup\check{M}$ we write $P^*$ for the
corresponding element of $M^{\natural}$.
Assuming now the stronger condition of $M$ being strongly causal in order
to ensure 
that  the Alexandrov topology agrees with the manifold topology (theorem
\ref{tsts})
the ``extended Alexandrov topology" is defined on $M^{\natural}$ as the
coarsest topology}
such that for each $A\in\check{M}$, $B\in\hat{M}$ the four sets
 $A^{\mbox{\footnotesize int}}$, $B^{\mbox{\footnotesize ext}}$,
$B^{\mbox{\footnotesize int}}$, $A^{\mbox{\footnotesize ext}}$ are open
sets where
\bnr
A^{\mbox{\footnotesize int}}=\{P^*:P\in\hat{M}\ \mbox{and}\ P\cap
A\neq\varnothing\},\\ 
A^{\mbox{\footnotesize ext}}=\{P^*:P\in\hat{M}\ \mbox{and}\ \forall
S\subset M\ P=I^-(S)\Rightarrow I^+(S)\not\subset A\}.
\enr
The sets $B^{\mbox{\footnotesize int}}$ and $B^{\mbox{\footnotesize ext}}$
have similar definitions with the roles of 
past and future interchanged.
The set $M^{\natural}$ becomes a topological space which in general is not
Hausdorff.
To avoid this awkward feature, an equivalence relation $R_{H}$ is defined
on $M^{\natural}$ by 
the intersection of all the equivalence relations $R\subset
M^{\natural}\times M^{\natural}$ such that
$M^{\natural}/R$ is Hausdorff. This new Hausdorff topological space is
then taken as the desired enlarged manifold $\bar{M}$. If the spacetime
$M$ is strongly causal, it was claimed in \cite{GKP} that the
identifications performed when passing 
from $M^{\natural}$ to $\bar{M}$ will never occur between elements of
$M^{\natural}$ representing original points of $M$. Moreover there 
exists a natural, dense, topological embedding of $M$ into
$\bar{M}$. Stronger related statements will be presented later in
proposition \ref{topology-consistency}. 

The GKP $c$-boundary is very appealing for only simple causal properties
are used in 
its definition and it recovers the structure found with conformal
techniques in relevant cases, such as asymptotically 
simple spacetimes. A general procedure for constructing the $c$-boundary
of static spacetimes can be found in \cite{HARRISNEW}.
However, a number of issues are left open in this
construction aside from the construction of the rule 
$R_{H}$ already commented. First of all the different regions of the
causal boundary
were not studied for general cases (only spacetimes asymptotically simple
are treated) 
and second the construction of causal relations between points at the
boundary was not investigated in \cite{GKP}. These 
features were the subject of subsequent papers published by other authors
who took the GKP scheme as their starting point. Some of these are 
reviewed next.

\subsection{Developments of the Geroch, Kronheimer and Penrose
construction}
\label{others}
In this section we review the different attempts carried out 
to fill in the unfinished steps of the GKP
construction. We must say right from the start that each intended
improvement was sooner or later found to bear undesirable features 
rendering all of them, as well as the original GKP, unsuitable to be
regarded as boundaries for spacetimes in {\em general} situations. Only
the main ideas of each construction 
are given as the details and examples usually result in rather technical
statements. 

\subsubsection{The Budic and Sachs construction.}
\label{budic-sachs}
The paper by Budic and Sachs \cite{BUDICSACHS} develops the GKP
construction for causally continuous spacetimes. In 
this case the authors are able to extend the causal structure and the
topology of the spacetime $M$ to 
 the enlarged spacetime $\bar{M}$ by means of the definition of a suitable
relation $R_{H}$. Before explaining how this 
relation is constructed the following definition is needed
\cite{HAWKINGSACHS}.

\begin{defi}[Common past and future]
For any set $U\subset M$ the chronological common past and the
chronological common future are respectively:
$$
\downarrow U=I^-\{x\in M:x\less p,\ \forall p\in U\},\ 
\uparrow U=I^+\{x\in M:x\greater p,\ \forall p\in U\}. 
$$
\label{common-pf}
\end{defi}
Clearly $\downarrow I^-(p)\subseteq I^-(p)$ and $\uparrow I^+(p)\subseteq
I^+(p)$.  
We denote by ${\cal F}$ and ${\cal P}$ the collection of future and past
sets respectively.  A {\em hull pair} on ${\cal P}\times{\cal F}$ is
any element $(P,F)$ such that $P=\downarrow F$ and $F=\uparrow P$.  A
very important result proved in \cite{BUDICSACHS} is that
$(I^-(p),I^+(p))$ is a hull pair for any $p\in M$ if $M$ is causally
continuous (this is too strong a restriction; it is known that
$(I^-(p),I^+(p))$ is a hull pair if and only if $M$ is just reflecting
\cite{HAWKINGSACHS}).  If we recall that the sets $I^+(p)$ and
$I^-(p)$ must be identified in the GKP construction, and given that
this identification takes place naturally in causally continuous
spacetimes through the relations $\downarrow$ and $\uparrow$, it seems
appropriate to introduce a binary relation on $\hat{M}\times\check{M}$
defined by the elements $(P,F)$ which form a hull pair.  This is an
equivalence relation ``$\sim$" for causally continuous spacetimes and
it is used to construct the quotient $\hat{M}\cup\check{M}/\sim$ which
play the role of the enlarged manifold $\bar{M}$. In order 
to avoid working with a quotient set of $\hat{M}\cup\check{M}$ 
the authors define the set 
$$
{\bf M}=\hat{M}\cup\check{M}- \downarrow(\check{M}\cup\check{{\cal L}}),
$$
which is called the {\em causal completion} of $M$. 
Here set $\check{{\cal L}}$ is the {\em future hull lattice} 
(the past hull lattice is defined dually) and its definition reads
$$
 \check{{\cal L}}\equiv\{X\in{\cal F}: X=\uparrow U,\ U\ \mbox{open}\}.
$$

The causal completion can be endowed with a causal structure by means of 
two binary relations $<$ and 
$\less$ defined on the set ${\cal P}\cup{\cal F}$ making it a causal space 
in the sense of Kronheimer and Penrose, 
definition \ref{causal-space}. 
It is also possible to define a topology $T$ on ${\cal P}\cup{\cal F}$ 
(also called extended Alexandrov topology) as follows:
a set $\C\subset{\cal P}\cup{\cal F}$ is called an enlargement of $M$ if 
it contains either $I^+(x)$ or $I^-(x)$ for all $x\in M$.
Then the extended Alexandrov topology $T$ on the enlargement $\C$ is 
the smallest
topology such that for all $C\in{\C}$ the subsets 
$I^+\{C\},\ I^{-}\{C\},\ C-J^-\{C\},\ C-J^+\{C\}$
are open.
Since ${\bf M}$ is an enlargement of $M$, the causal completion inherits
the 
extended Alexandrov 
topology.
Therefore ${\bf M}$ is both a causal and a topological space.
The causal boundary is now defined as the set $\d{\bf M}={\bf
M}-\check{I}M$
where $\check{I}:M\rightarrow{\bf M}$ is the mapping
$$
\check{I}: x\mapsto\check{I}(x)= I^-(x),\ \forall x\in M.
$$ 
It can be shown that for causally continuous spacetimes $\check{I}$ is a
dense 
embedding of $M$ into ${\bf M}$, see also proposition
\ref{topology-consistency}.

Some other general properties of ${\bf M}$ are studied
\cite{BUDICSACHS}. For example an interesting result is that 
$M$ is globally hyperbolic if and only if for every $\bar{x}\in{\bf M}$,
either 
$\bar{I}^-(\bar{x})$ or $\bar{I}^+(\bar{x})$ 
is the empty set (here $\bar{I}^{\pm}$ are calculated with respect to the
causal relations introduced in ${\bf M}$).

\subsubsection{R\'acz's generalizations} 
\label{rac}  
In \cite{RACZ} R\'acz gave a modification of the topology defined by
Geroch, Kronheimer and Penrose for the set 
$\hat{M}\cup\check{M}$ as the 
coarsest topology $A^*$ in which for each PIF $F$ and PIP $P$ the four
sets $F^{\mbox{\footnotesize int}}$, 
$F^{\mbox{\footnotesize ext}}$, $P^{\mbox{\footnotesize int}}$ and
$P^{\mbox{\footnotesize ext}}$ are open. Here
\bnr
\fl F^{\mbox{\footnotesize int}}=\{A\in\hat{M}\cup\check{M}:A\in\hat{M}\
\mbox{and}\ A\cap F\neq\varnothing\ 
\mbox{or}\ A\in\check{M}\ \mbox{and}\\
 \forall S\subset M: I^+(S)=A\Rightarrow I^-(S)\cap F\neq\varnothing\},\\
\fl F^{\mbox{\footnotesize ext}}=\{A\in\hat{M}\cup\check{M}:A\in\check{M}\
\mbox{and}\
A\not\subset F\ \mbox{or}\\
 A\subset\hat{M}\ \mbox{and}\ \forall S\subset M:A=I^-(S)\Rightarrow
I^+(S)\not\subset F\}.
\enr
The sets $P^{\mbox{\footnotesize int}}$ and $P^{\mbox{\footnotesize ext}}$
have a similar definition. 
Here the topology $A^*$ is set up directly on $\hat{M}\cup\check{M}$
without introducing the intermediate set $M^{\natural}$. Next an
identification rule $R$ is defined on $\hat{M}\cup\check{M}$
yielding the completion $\bar{M}$ and the topology $\bar{A}$ as the
quotients 
$\hat{M}\cup\check{M}/R$ and $A^*/R$, respectively. The minimal
requirement which $R$ must comply with is that 
the sets $I^+(p)$ and $I^-(p)$ be identified. For any such relation $R$
the first point in proposition \ref{topology-consistency} holds, so that the
next goal is finding a relation $R$ such that the topology $\bar{A}$ is
Hausdorff. 
To achieve this, a technical causal condition on $M$ is imposed: for each
$p\in M$ 
there exist $a, b\in M$ such that $a\in I^-(p)$, $b\in I^+(p)$ and there
is no set $S$ satisfying both $I^+(S)\subset I^+(a)$ and $I^-(S)\subset
I^-(b)$ for which $I^+(S)$ is a TIF or $I^-(S)$ is a TIP.

R\'acz also developed the GKP construction for the case of
causally stable spacetimes in \cite{RACZ2}, where an explicit
identification rule and topology were constructed. As an example, in
\cite{RACZ} the singular portion of the causal boundary of Taub
plane-symmetric static vacuum spacetime was shown to be a one-dimensional
set under this construction whereas 
in the original GKP scheme this turned out to be a point. Further details 
about this and other drawbacks of the $c$-boundary construction will be
discussed in $\S$\ref{caveats}. 

\subsubsection{The Szabados construction for strongly causal spacetimes.}
\label{sza}
Almost simultaneously, Szabados tried to address similar questions for
strongly causal 
spacetimes in a couple of important papers \cite{SZABADOS1,SZABADOS2}. The
first of them, which is the work discussed in this 
section, is a truly penetrating study on general identification rules on
the set $M^{\natural}$ defined by Geroch, Kronheimer 
and Penrose, where a new version of the completion $\bar{M}$ together with
a causal structure for $\bar{M}$ was presented.  

Szabados calls future preboundary points to the TIF's (past
preboundary points if they are TIP's) and they are collected in sets
denoted by $\d^+$, $\d^-$ respectively. The remaining elements of
$M^{\natural}$ are the identified pairs $(I^+(p),I^-(p))$, $p\in M$ which
are regarded as the image of the 
injection $i:M\rightarrow M^{\natural}$, $i:p\mapsto i(p)$. Szabados
pointed out that the elements of $\d^{\pm}$ are sometimes termed the
``endpoints" 
of inextensible causal curves in $M$, but this statement needs a
clarification because the concept of endpoint is purely topological (see
definition \ref{endpoint}) so the assertion only makes sense if a topology
$\bar{\Ts}$ has been defined on $\bar{M}$.
More importantly the causal endpoints must be consistent with topological
endpoints which means for instance 
that the TIP $I^-(\g)$ has to be the future endpoint in $\bar{M}$ of the
future inextensible causal curve $\g$. These problems were neatly resolved
in \cite{SZABADOS1}
constructing an appropriate topology $\bar{\Ts}$ on $\bar{M}$ from
the Alexandrov topology $\Ts$ on $M$---which 
coincides with the manifold topology for strongly causal spacetimes {\em
cf.} theorem \ref{tsts}.

The needed topology $\bar{\Ts}$ is constructed as the quotient topology of 
$\Ts^{\natural}$ by a certain equivalence relation $\Rs$ on
$M^{\natural}$. 
The topology $\Ts^{\natural}$ and the equivalence relation $\Rs$ 
are determined at a later stage. A basic consistency requirement
is to impose that 
$i:(M,\Ts)\rightarrow (M^{\natural},\Ts^{\natural})$ be an open dense
embedding 
and that the elements of $M^{\natural}$ given by $P=I^-(\g)$, $F=I^+(\g)$
be
respective past and future endpoints of $i\circ\g$ in the topology
$\Ts^{\natural}$.
Under these assumptions an important conclusion can be drawn.
\begin{prop}
Let $\Rs$ be any equivalence relation on $M^{\natural}$ which is trivial
on the subset $i(M)$ and define the canonical projection
$\pi:M^{\natural}\rightarrow M^{\natural}/\Rs$. If a topology
$\Ts^{\natural}$ on $M^{\natural}$ with the
properties explained above is set then
\begin{enumerate}
\item the mapping 
$\pi\circ i:(M,\Ts)\rightarrow (M^{\natural}/\Rs,\Ts^{\natural}/\Rs)$ is
an open dense embedding.
\item 
$\pi(P)$ and $\pi(F)$ are future and past endpoints of the curve $\pi\circ
i\circ\g$ where as above $P=I^-(\g)$ and 
$F=I^+(\g)$.
\end{enumerate}
\label{topology-consistency}
\end{prop}
From this result we conclude that any equivalence relation acting as the
identity on $i(M)$ 
renders the preboundary points as topological endpoints of inextensible
causal curves. This is a good result but not enough for our purposes
because there are relations $\Rs$ 
such that the endpoints are not unique 
for a single inextensible causal curve. This can be restated as saying
that the quotient topology $\Ts^{\natural}/\Rs$ is not
Hausdorff so in order to get rid of this feature one would have to search
for another equivalence relation $\Rs$ making 
the quotient topology Hausdorff. In fact, the prescription originally
given by Geroch, Kronheimer and Penrose is just ``take the minimal $\Rs$
such that $\Ts^{\natural}/\Rs$ is Hausdorff", but as Szabados shows in
\cite{SZABADOS1} there are explicit examples in which {\em no} such
equivalence relation exists. 

Szabados suggested then to consider quotient topologies with milder
restrictions.
Recall that two points $x_1$, $x_2$ of a topological space are {\em
$T_1$-separated} if there is an open neighbourhood
of $x_1$ which does not contain $x_2$ and there is an open neighbourhood
of $x_2$ not containing $x_1$. 
If this property holds for all pairs $x_1,x_2$ then the set is said to be
$T_1$-separated. $T_2$ separation
is the standard Hausdorff separation property.     
The next result was proven in \cite{SZABADOS1}.
\begin{prop}
Let $\Ts^{\natural}$ be the GKP topology of the set $M^{\natural}$ and
$\Rs$ any equivalence relation defined on $M^{\natural}$. 
Then, each inner point and each boundary point of the topological space
$(M^{\natural}/\Rs,\Ts^{\natural}/\Rs)$ are $T_1$-separated.
\label{tuno}
\end{prop}  
After these important considerations, an explicit identification rule
$\Rs$ was put forward. This rule relies on the concept
of naked TIP's and TIF's introduced in \cite{PENROSE-CLASSIFICATION}. A
TIP $P$ is said to be {\em naked} if for some 
point $p\in M$ we have $P\subset I^-(p)$. The {\em naked counterpart} of
$P$ is a TIF $F$ such that for every $q\in F$ 
the property $P\subset I^-(q)$ holds. If there is no naked counterpart of
$P$ containing $F$, then $F$ is a maximal naked counterpart of $P$. In
general, there are naked TIPs with several maximal naked
counterparts. However, \cite{SZABADOS1}
\begin{prop}
Any naked $P\in\d^+$ possesses a maximal naked counterpart. Moreover,
$\uparrow P=\cup_{\alpha}F_{\alpha}$ where $F_{\alpha}$ are the maximal
naked counterparts of $P$.
\label{counterpart}
\end{prop}
Of course there is a dual formulation in terms of TIFs and their maximal
naked counterparts. Naked TIPs 
and TIFs play the role of points lying in ``timelike'' parts of $\d^{\pm}$
which are in essence the points requiring 
identification. Therefore, a binary relation ``$\sim$'' can be set up as
follows: 
$P\sim F$ if $P$ and $F$ are maximal naked counterparts of each other. If
$F$ is a naked counterpart of the naked 
TIP $P$ then we can always find sets $F_0$ and $P_0$ such that $F\subseteq
F_0$ and $P\subseteq P_0$ where $F_0$ and 
$P_0$ are maximal naked counterparts of each other, so $P_0\sim F_0$. The
relation ``$\sim$'' can then be extended to the whole
$M^{\natural}$ in the following way: $X\sim X$, $\forall X\in
M^{\natural}$ and if $B$, $B'\in\d^+\cup\d^-$, $B\neq B'$ we 
say that $B\sim B'$ if for a finite number of preboundary points
$B_1,\dots,B_r$ the chain of relations 
$B\sim B_1\sim\dots\sim B_r\sim B'$ holds. Interestingly, if $(P,F)$ is a
hull pair then $P\sim F$ which means that Szabados identification 
is a generalization of the Budic and Sachs identification.  

The space $\bar{M}$ can also be endowed with a chronology relation
$\less$ and from this with a causal relation by means of definition
\ref{Bf}, see \cite{SZABADOS1} for details. However, this may not 
agree with the original manifold chronology, see \cite{MAROLFROSS}.
Once the necessary
identifications between TIPs and TIFs has been carried out, still
different TIPs (or TIFs) may need identification.  This is taken care
of in the second paper \cite{SZABADOS2}.  As yet another final
positive result, an asymptotic causality condition is then introduced,
which provides the uniqueness of endpoints of causal curves in the
completed manifold $\bar{M}$.

\subsubsection{The Marolf and Ross recipe.}
\label{m-r}
Recently Marolf and Ross \cite{MAROLFROSS} have proposed a new
identification rule partly based on Szabados work.
They start from the observation that sets which are maximal naked
counterparts of each other can be related in an alternative way as
follows:  
the relation $R_{pf}\subset\hat{M}\times\check{M}$ is the set of pairs
$(P,F)$ such that $F$ is a maximal subset of $\uparrow P$ and 
$P$ a maximal subset of $\downarrow F$. 
%Equivalently, the statements
%$$
%Q^*\subset f(P),\ \nexists R^*\in\check{M}:R^*\neq Q^*,\ Q^*\subsetR^*\subset f(P)
%$$     
%and 
%$$
%P\subset p(Q^*),\ \nexists R\in\hat{M}:R\neq P,\ P\subset R\subsetp(Q^*),
%$$
%hold for $Q^*\in \check{M}$, $P\in \hat{M}$.

The Szabados relation ``$\sim$'' can be obtained as the smallest
equivalence relation containing $R_{pf}$. Observe that ``$R_{pf}$'' is
defined directly on the Cartesian product
 $\hat{M}\times\check{M}$ whereas ``$\sim$'' is
in principle
defined only on $M^{\natural}$. 
Marolf and Ross argue that causal completions need not arise from
equivalence relations and quotient spaces 
because sometimes these relations enforce the identification of ideal
points which should not be identified 
on causal grounds. To avoid this, they suggest characterizing the points
of the causal completion $\bar{M}$ 
by two sets representing their future and their past and, as a key point,
in the case of ideal points one of 
these sets may be empty. Explicitly:
the {\em causal completion} $\bar{M}$ of a spacetime $M$ is the set of
pairs $(P,P^*)$ such that 
\begin{enumerate}
\item $(P,P^*)\in R_{pf}\subset\hat{M}\times\check{M}$ or
\item $P=\varnothing$ and $P^*$ is not an element of any pair in $R_{pf}$
or
\item $P^*=\varnothing$ and $P$ is not an element of any pair in $R_{pf}$.
\end{enumerate}
The elements of $\bar{M}$ are denoted by $\bar{P}\equiv (P,P^*)$.
As discussed in previous cases, a first important property is that the
sets 
$I^{\pm}(p)$ always appear in the pair
$(I^-(p),I^+(p))\in\bar{M}$. Therefore 
there exists a natural embedding
$\Phi:M\rightarrow\bar{M},\Phi(p)=(I^-(p),I^+(p))$, 
which shows the similarity of this construction to the Szabados scheme. 

As shown in \cite{MAROLFROSS}, there is a natural chronology in the
completion $\bar{M}$
\begin{theo}
The binary relation $\less$ defined by
$\bar{P}\less\bar{Q}\Longleftrightarrow P^*\cap Q\neq\varnothing$
is a chronology on $\bar{M}$.
\label{chronology}
\end{theo}
The definition of a causal relation $<$ seems more conflictive and the
authors discuss different approaches, none giving a definite satisfactory
answer adapted to all the examples considered in \cite{MAROLFROSS}. The
construction of a topology upon $\bar{M}$ is also rather technical and, in
fact, more than a single topology is considered.  
The main result regarding these topologies is the 
fact that $M$ is homeomorphic to its image on $\bar{M}$ under the natural
embedding, and its image is dense on $\bar{M}$. Finally, separation
properties of the 
topologies are also considered. 

\subsubsection{Caveats on the GKP-based constructions.}
\label{caveats}
After this account of the attempts based on the Geroch, Kronheimer and
Penrose 
paper it may be a little bit disappointing to learn that the majority of
the GKP-based schemes present unsatisfactory features---see, however,
subsection \ref{harris}. Here we will limit ourselves to give brief
comments about 
the unconvincing aspects referring the interested reader to the literature
for the precise details. Let us remark that the main driving force behind
the successive 
GKP-modifications was in fact to give an answer to the successive
drawbacks found  in certain well-devised examples.

To start with, Penrose \cite{PENROSE-CLASSIFICATION,PENROSE-CENTENARY} 
classified the points of the $c$-boundary as {\em infinity} points 
and {\em finite} points. In the first case the $\infty$-TIP or
$\infty$-TIF representing the 
point is a set of the form $I^-(\g)$ or $I^+(\g)$, respectively, where
$\g$ is a complete 
future (past) inextensible causal curve. The case of finite ideal points
has the same 
definitions but now the curve $\g$ is incomplete. Unfortunately, this
classification does not provide a boundary with the ``right shape'' in
some examples, as we show next.  
The paradigmatic counterexample is the GKP boundary of Taub's
plane-symmetric static vacuum spacetime, whose line-element is   
$$
ds^2=z^{-1/2}(dt^2-dz^2)-z(dx^2+dy^2),\ \ z>0,\ -\infty<t,x,y<\infty.
$$
In \cite{KUANG} the $c$-boundary of this spacetime is explicitly
constructed and it is shown that its singular part 
in the sense of Penrose consists of a single point. However, the Penrose
diagram of the appropriate 2-dimensional subregion spanned by the
coordinates $\{t,z\}$ (see subsection \ref{conformal-diagram}) shows that
the singularity $z=0$ should be a one-dimensional set. Other
counterexamples with analogous behaviours
are also presented in \cite{KUANG}. Thus, a first type of bad behaviour is
that the ``shape'' of the boundary is not the expected one in certain
examples.
The modifications of the $c$-boundary explained in subsections \ref{rac},
\ref{sza} and \ref{m-r} correct this problem for the Taub spacetime,
though.

Unfortunately, there are other type of problems in the GKP-based
constructions dealing with the topology of the completed spacetime.
These problems were illuminatingly pointed out in
\cite{KUANGLIANG1,KUANGLIANG2} where they became apparent through
carefully conceived examples 
in which the $c$-boundaries are explicitly constructed. The failure shown
in all examples is similar in nature:
the topology of the causal boundary $\d M$ using R\'acz or Szabados
versions for the 
$c$-boundary does not match 
the natural topology of the completed spacetime $\bar{M}$. For instance,
many examples are just
two- or three-dimensional Minkowski spacetime with certain regions
removed. The causal completion of the 
excised spacetime is then performed and it is proven that the topology
obtained according to each of the $c$-boundary prescriptions under
surveillance is different from the topology of the closure in Minkowski
spacetime of the excised region. 
This is shown typically using sequences whose limit differ on each
topology.

The authors conclude in \cite{KUANGLIANG2} that the attempt to describe
the singularity
structure via the completion of the spacetime and the definition of a
causal boundary
may be inappropriate. As a matter of fact, all attempts have encountered
undesirable properties so far in certain simple cases, so that one wonders
what could happen in more complicated cases where there is no obvious
``right'' answer to
be obtained or compared to.

For further problems see \cite{RUBE}. On the other hand, for a positive
albeit moderate result regarding the {\em future} $c$-boundary, see next
subsection.

\subsubsection{Harris' approach to the GKP boundary using categories.}
\label{harris}
Harris adopted a different approach to the subject 
 in the companion papers \cite{HARRIS1,HARRIS2}. Apparently, his goal was 
to show that the GKP boundary may be regarded as natural in a categorical
sense under some well-established restrictions. Thus, instead of seeking
yet another construction for the $c$-boundary, the universality of the GKP
boundary construction was settled ``in a categorical manner''. 

As part of the problem resides in that completions of a spacetime may fail
to
be a manifold, in \cite{HARRIS1} only ``chronological sets", which are
similar to the causal spaces of definition \ref{causal-space} but
just having the chronology relation $\less$, are considered. Given a
chronological set $X$, the GKP procedure of attaching a {\em future}
causal boundary to a space-time is carried over to $X$, provided certain
conditions are met. A crucial point here is that only the {\em future}
completion is defined. This is a clever choice, because many of the
problems arising with the GKP construction have their roots in the
intricate problems inherent to the identification of ideal points in the
past boundary $\d^-$ with ideal points in the future boundary
$\d^+$. Independently of this, one can further show that the topology generated for the
Minkowski spacetime does not coincide with that of its conformal
completion of section \ref{conformal-boundary}, see \cite{HARRIS2}. 

The future-completed chronological set is denoted by $X^+$ and the future
chronological boundary by $\d^+(X)$. There are examples with $X^+=X$,
called future-complete chronological sets. 
A map $f:X\rightarrow Y$ between two chronological sets $X$, $Y$ is called
{\em future-continuous} if $f$ is chronal preserving in the sense of
definition \ref{theta} and  if the image $f(x)$ of the limit of a future
chain on $X$ is the future limit of the (necessarily future) image chain
on $Y$.   
Important properties are: (i)  for any future continuous $f$ 
there exists an extension $f^+:X^+\rightarrow Y^+$; (ii) 
the natural inclusion $\iota^+_X:X\rightarrow X^+$ fulfills 
the property $f^+\circ\iota^+_X=\iota^+_Y\circ f$. Using this, the
universality principle is proved: if $Y$ is future-complete, for any
future-continuous map $f:X\rightarrow Y$ $f^+$ is the unique
future-continuous 
extension of $f$ to $X^+$ such that $f^+\circ\iota^+_X=f$. 
This result is very important because it says that any future 
completion of $X$ can be compared with its GKP version preserving the
chronology of $X$, that is to say, the GKP construction is universal (in
the appropriate category, \cite{HARRIS1} states all results in the
language of categories and functors). As remarked before, Harris also
points out that the total $c$-boundary, both future
and past with appropriate points identified, does not seem to have an
analogous universality property. 

A very readable account of this line of research with examples and a
summary of the relevant results can be consulted in \cite{HARRIS4}, by the
same author. Recently, there has appeared a new paper where the relation
between the GKP boundary of a quotient 
spacetime by a set of isometries
and the induced quotient of the GKP boundary is analyzed \cite{HARRIS3}
from this point of view.

\subsection{Other independent definitions}
In this section we summarise other attempts to attach a ``(causal)
boundary'' to any spacetime. As happens with the case of the $c$-boundary,
some of these constructions were found to be unsatisfactory in certain
examples. We start with the cases were no external objects but only the
intrinsic structure of the spacetime are needed (subsections
\ref{gboundary}--\ref{meyer}), and then we consider new approaches where
the idea of ``enlargement" or ``embedding" into larger manifolds has been
reformulated and improved (subsections \ref{aboundary} and
\ref{causal-relationship}).

\subsubsection{$g$-boundary.}
\label{gboundary}
The $g$-boundary was devised by Geroch in \cite{GBOUNDARY} and was
probably one of the first attempts to attach a boundary to Lorentzian
manifolds. The main idea behind the construction was to try and deal with
singularities (understood as inextensible incomplete geodesics), and to
provide them with causal and metric properties. At the time
\cite{GBOUNDARY} was published, there was 
no generally accepted definition of singularity in General Relativity
although, generally speaking, geodesic incompleteness was already agreed
to indicate the presence of singularities (be them removable or essential
singularities, see $\S$\ref{aboundary}).
A (essential) singularity should never be   
considered as part of the spacetime and so it should be placed ``at its
boundary''. Thus, Geroch tried to attach a boundary to each incomplete
inextensible causal geodesic. 

The $g$-boundary is outdated nowadays, since the definition of
singularities requires not only caring about geodesic incompleteness, but
about that of other causal curves as well. This was actually remarked by
Geroch himself in a very interesting paper \cite{GER} where an explicit
future-incomplete inextensible timelike curve with bounded acceleration
was explicitly exhibited in an otherwise geodesically complete spacetime.
However, we have considered interesting to include a summary of the
$g$-boundary here due to its historical interest and for the sake of
completeness.

The $g$-boundary is formed by equivalence classes of incomplete
inextensible geodesics under a certain equivalence 
relation. Essentially the whole idea is to collect in classes geodesics
which stay 
close to each other. Let $(M,\G)$ be an $n$-dimensional spacetime and
denote by $G$ the tangent bundle $T(M)$. As is well known $T(M)$ 
is a differentiable manifold of dimension $2n$. Furthermore each point
$(p,\xiv)\in G$, $p\in M$, $\xiv\in T_{p}(M)$ 
determines one and only one geodesic $\g$ and conversely so we can speak
of the points of $G$ as the geodesics on $M$. Next one defines the
$(2n+1)$-dimensional manifold $H=G\times (0,\infty)$ and the subsets 
\bnr
\fl H_+=\{(p,\xiv,a)\in H:\varphi(p,\xiv)>a\},\ H_0=\{(p,\xiv,a)\in
H:\varphi(p,\xiv)=a\},\\ 
G_I=\{(p,\xiv)\in G:\varphi(p,\xiv)<\infty\},
\enr
where $\varphi(p,\xiv)$ is the affine length of the geodesic $(p,\xiv)$,
which is infinity if the geodesic is complete. We need also the 
mapping $\Psi:H_+\rightarrow M$ defined by the rule
$\Psi(p,\xiv,b)=$ the point $q\in M$ resulting after travelling an affine
distance $b$ 
along the geodesic $(p,\xiv)$. The next step is topologizing $G_I$ as
follows: for any 
open set $O\subset M$ we define the set $S(O)$ as the subset of $G_I$ such
that
$$
\exists\ \mbox{$U\subset H$ open containing $(P,\xiv,\varphi(P,\xiv))\in
H_0$
and $\Psi(U\cap H_+)\subset O$}.
$$
In plain words $S(O)$ consists of the incomplete geodesics entering $O$
and never leaving it. The collection of sets
$S(O)$ with $O$ ranging over all open sets of $M$ serves as a basis for a
topology on $G_I$ called the open set topology. 

We are now ready to define the equivalence relation on $G_I$ leading to
the $g$-boundary. Any two elements 
$\alpha$, $\beta\in G_I$ are related (written $\alpha\approx\beta$) if
every open set in $G_I$
containing $\alpha$ also contains $\beta$ and vice-versa\footnote{Other
equivalence 
relations on $G_I$ are also considered in \cite{GBOUNDARY} but
``$\approx$'' is the weakest in the sense that points of $G_I$ are 
identified if they cannot be distinguished topologically.}.
 The relation ``$\approx$'' is an equivalence relation 
on $G_I$ and the set of equivalence classes, denoted by $\d$, form the
$g$-boundary of $M$. Furthermore the topology 
of $G_I$ induces a topology on the quotient set $\d$ in the standard way.
The completed space-time $\bar{M}$ is then the union of $M$ plus the
singular points $\d$, and can be endowed with a topology whose basis is
formed by pairs of open sets 
$(O,U)\in M\times\d$ with the added property that $U\subset S(O)$. The
restriction of this topology to 
$M$ and $\d$ is consistent with the 
topologies previously introduced.

An interesting feature of the $g$-boundary is the possibility of regarding
$\d$ as if it were a hypersurface, so that concepts such as spacelike or
timelike $\d$ can be defined. Thus, $\d$ is spacelike at $e\in \d$ if
there exists a neighbourhood 
$(O,U)\subset\bar{M}$ such that for every $e'\in U$ there is another
neighbourhood $(O',U')\subset\bar{M}$ of $e'$ with the 
property $[\bar{I}^+(e,(O,U))\cup
\bar{I}^-(e,(O,U))]\cap(O',U')=\varnothing$.
Similarly, $\d$ is timelike at $e$ if for every neighbourhood $(O,U)$ of
$e$ in $\bar{M}$ we can find 
two points $e',e''\in U$ such that $\bar{I}^+(e',(O,U))\cap
\bar{I}^-(e'',(O,U))$ contains an open neighbourhood of $e$ in
$\bar{M}$. Here, $\bar{I}^{\pm}$ are the natural extensions of $I^{\pm}$
to $\bar{M}$.

Some applications of this construction discussed in \cite{GBOUNDARY} deal
with the study of space-time extensions and 
the relationship between the $g$-boundary and the conformal boundary of
Penrose. 
An important limitation of the $g$-boundary 
is that by construction $\d$ only includes singularities and not points
``at infinity''. Furthermore, as remarked above, only {\em some}
singularities are included in $\d$.
These and other unsatisfactory features of the $g$-boundary were
considered in \cite{GEROCHWALD}  for certain examples.

\subsubsection{$b$-boundary}
\label{bboundary}
The $b$-boundary construction invented by Schmidt \cite{SCHMIDT} was also
motivated by the problem of singularity definition in General
Relativity. At the time Schmidt's paper was published, relativists were
already aware that not only inextensible incomplete geodesics were to be
taken into account when constructing the boundary of singular points,
rather all causal (and even non-causal) inextensible curves had to be
considered, \cite{GER}.

To deal with this problem, Schmidt worked with the bundle of frames $L(M)$
constructed from the spacetime $M$ and 
it is actually in this manifold where the completion is carried out
obtaining the set $\overline{L(M)}$ and its boundary 
$\d L(M)\equiv\overline{L(M)}-L(M)$. From this the $b$-boundary of the
manifold is defined as 
$\d M\equiv\pi(\d {L'}(M))$ where $L'(M)$ is one of the connected
components in which $L(M)$ splits for 
any orientable manifold and $\pi$ is a suitable extension of the
projection of the frame bundle onto its base space. 
In order to achieve the completion of the frame bundle 
$L(M)$ the author defines a proper Riemannian metric on this manifold and
perform its standard Cauchy completion. 

Let us summarize next how the Riemannian metric is constructed on $L(M)$
(the reader is assumed familiar with 
fibre bundle theory). A frame bundle is a particular case of a {\em
principal bundle} so one can use in 
this case standard concepts which are specific for them. As is well known,
a connection on a principal bundle 
defines a subspace in the tangent space of any point $u\in L(M)$
(horizontal subspace)
 which is complementary to the vertical space. 
The {\em standard horizontal vector fields} $\{B_i\}_{i=1,\dots n}$ 
are the only horizontal vector fields fulfilling the property
$$
\pi' B_i|_u=X_i,\ \ u=X_1,\dots, X_n,
$$
where $\{X_1,\dots, X_n\}$ is a frame of $L(M)$. Introduce also a set of
1-forms $\{\z^1,\dots,\z^n\}$ dual to the family $\{B_1,\dots, B_n\}$ 
$$
\z^i(B_k)=\delta_k^{\ i}.
$$
The connection can be characterized by means of 
a 1-form $\omega$ with values in $gl(n,\r)$, the Lie algebra of the
structural group $Gl(n,\r)$. In this 
sense horizontal vector fields are characterized by the condition
$$
\omega(X)=0\Longleftrightarrow X\ \mbox{is horizontal}.
$$
The 1-form $\omega$ can be written in components with respect to a basis
of the Lie algebra $gl(n,\r)$, 
$\{E_i^{\ k}\}_{1\leq i,k\leq n^2}$, as
$$
\omega=\omega_k^{\ i}E_i^{\ k},
$$
where the coefficients $\omega_k^{\ i}$ are 1-forms on $L(M)$ called the
connection forms. A proper Riemannian metric $\rmg$ on $L(M)$ is given by 
$$
\rmg(X,Y)=\sum_i\z^i(X)\z^i(Y)+\sum_{i,k}\omega_k^{\ i}(X)\omega_k^{\
i}(Y),
$$ 
for any pair of vector fields $X$, $Y$ on $L(M)$. This is called the {\em
bundle metric}.
Notice that once we get the completion of the frame bundle the projection
$\pi$ must be extended as well 
from $L(M)$ to $\overline{L(M)}$ (we still use the same symbol $\pi$ for
this extended projection). This is done 
through the extension of the right action of the structural group
$Gl(n,\r)$ on the frame bundle.

At first glance this construction may seem rather abstract with no
definite relation
with the points of a boundary for $M$. To start digging the intimate
relationship with incompleteness on $M$, 
we must realize that points of the $b$-boundary are equivalence 
classes of Cauchy sequences in $L(M)$ with respect to the distance defined
by the metric $\rmg$, so it would be interesting to have an interpretation
for the length 
of curves in $L(M)$ with respect to the bundle metric. For curves $u(t)$
such that their tangent vector $\dot{u}$ is horizontal 
(horizontal curves) this length takes the form
\be
L=\int_0^1\left(\sum_{i=1}^{n}\z^i(\dot{u})\z^i(\dot{u})\right)^{1/2}dt \,
. 
\label{b-length}
\ee
As $u(t)$ is horizontal, the frame
$\{X_1(t),\dots,X_n(t)\}$ determined by $u(t)$ is parallel propagated
along the curve $x(t)=\pi(u(t))$ whose tangent 
vector is 
$$
\dot{x}=\pi ' (\dot{u})=\z^i(\dot{u})X_i.
$$ 
From this we conclude that $L$ is nothing but the ``Euclidean length'' of
the curve $x(t)$ measured with respect to a parallel propagated frame. In
particular, the previous reasoning applies to any geodesic because the
lift of any 
geodesic is a horizontal curve. From this it is not difficult to conclude
the following important result 
\begin{theo}
If the bundle metric is complete, then the connection is geodesically
complete.
\label{complete}
\end{theo}    
Therefore geodesically incomplete connections entail incomplete bundle
metrics or in other words any incomplete geodesic
on $M$ determines a point of the $b$-boundary. The advantage of the
$b$-boundary is that the converse of theorem \ref{complete} is not true,
and in fact incomplete curves which are not geodesics also determine
points of the $b$-boundary. 

Equation \ref{b-length} can be seen as an appropriate notion of length 
valid for any $C^1$ curve of $M$ (any $C^1$ curve admits an horizontal
lift). This
new length has some indeterminacies, but if it is finite for a given
choice of frame, then it {\em is} finite for any other choice. So, one can
speak of 
{\em $b$-completeness} of curves without ambiguity, and without resorting
to frame bundle theory in fact.
In this sense a spacetime is said to be singularity-free if it is
$b$-complete, see \cite{FF,SINGULARITY} for details about this.  

Schmidt's paper also discusses some issues related to the topology of the
$b$-boundary. As happened in other cases, 
this is the tricky point and in fact there are examples in the literature
where the $b$-boundary is constructed 
explicitly and shown to have strange topological properties 
\cite{RUSSELL,BOSSHARD}. An interesting alternative to Schmidt's construction
was proposed by Sachs in \cite{SACHS2} where he used the tangent bundle of 
the manifold instead of the principal bundle of orthonormal frames.

\subsubsection{Meyer's metric construction.}
\label{meyer}
In the interesting paper \cite{MEYER}, Meyer describes a boundary
construction based on a definition of metric distance for the spacetime
$(M,\G)$. Firstly, for any set $U\subset M$ its height $d(U)$ is the
supreme of the 
length of all the future-directed causal curves $\g\subset U$. Using the
notation 
$A\triangle B\equiv(A-B)\cup(B-A)$ for the symmetric difference,
the distance between two past sets $A$ and $B$ is 
$$
D(A,B)\equiv d(A\triangle B),
$$
and analogously for future sets. If $p,q\in M$ one introduces the
quantities $D^-(p,q)$ and $D^+(p,q)$ by
$$
D^-(p,q)=D(I^-(p),I^-(q)),\ \ D^+(p,q)=D(I^+(p),I^+(q))
$$ 
(the same notation is used for the distance between future sets and past
sets). Finally the distance between two points 
is simply
$$
D(p,q)=D^+(p,q)+D^-(p,q).
$$
The pair $(M,D)$ is then a metric space but the distance $D(p,q)$ need not
be finite for 
all the pairs $p,q$. To get rid of this and other awkward features, only
spacetimes of finite timelike diameter \cite{BEE} are considered. This is
the main disadvantage of Meyer's construction, as it leaves out many
simple obvious spacetimes. Under 
the mentioned assumption $D(p,q)$ is finite for any pair or points and in
addition to this the function $D$ is continuous on 
$M\times M$. Furthermore the topology induced by this distance agrees with
the manifold topology, and the continuity
of $D$ implies causal continuity of $(M,\G)$. The metric 
space $(M,D)$ can now be completed to $(\bar{M},\bar{D})$ by means of the
Cauchy completion and one gets the boundary 
as $\bar{M}-M$. The causal structure can then be extended to the
completion $\bar{M}$.

The relation between this boundary and the $c$-boundary is analyzed, and
the new boundary is explicitly constructed for some examples, in
\cite{MEYER}.

\subsubsection{$a$-boundary.}
\label{aboundary}
The {\em abstract boundary} or $a$-boundary, first introduced by Scott
and Szekeres in \cite{SCOTT}, was also devised with the study of
singularities in mind.  However, while the $g$- and $b$-boundaries did
only use objects intrinsic to the Lorentzian manifold to be completed,
the $a$-boundary comes back to the original idea on which the
conformal compactification and conformal boundary were founded:
embeddings into larger sets.  Even more, the aim in \cite{SCOTT} was
to put forward a general scheme of how to envelope {\em any}
differentiable manifold, be it Lorentzian or semi-Riemannian or
without metric, or with only a connection..., into a larger one, and
then how to obtain {\em all possible boundaries} that such a manifold
admits.  On a second stage, the abstract construction accommodates
itself very well with a general definition of singularity in the case
of pseudo-Riemannian manifolds, or manifolds with an affine
connection.  Perhaps it could be fair to say that the $a$-boundary put
in rigorous terms some concepts which were used more or less vaguely
many times by relativists.  Further developments of the $a$-boundary
can be found in \cite{ASHLEY}.

Before giving an account of this construction some definitions are in
order. The definition of curve is similar to that 
presented in definition \ref{timelike-curve} with the difference that
half-open intervals $I=[a,b)$, $-\infty <a$ are used. The parameter of the
curve is said to be bounded if $b<\infty$ and unbounded otherwise. A
family of curves $\C$ on 
a manifold $M$ has the bounded parameter property (b.p.p.) if the
following conditions are met
\begin{enumerate}
\item for any point $p\in M$ there is at least one curve $\g$ of the
family passing through $p$.
\item If $\g$ is a curve of the family then so is any subcurve of $\g$.
\item Any pair of curves $\g,\ \g'\subset\C$ related by an allowed
parametrization change have both a bounded parameter 
or an unbounded parameter.
\end{enumerate}
This third point is important at this stage as we do not have a notion of
affine, proper or other well-behaved lengths such as (\ref{b-length}).

Important concepts in the $a$-boundary are the following.
\begin{defi}[Envelopment]
The differentiable manifold $\widehat{M}$ is said to be an envelopment of
the manifold $M$ if there exists a $C^{\infty}$ 
embedding $\phi:M\rightarrow \widehat{M}$ and both $M$, $\widehat{M}$ have
the same dimension. 
Envelopments will be denoted by the triple $(M,\widehat{M},\phi)$.
\label{envelopment}  
\end{defi}
\begin{defi}[Boundary points and sets]
The point $p\in \widehat{M}$ is a boundary point of the envelopment
$(M,\widehat{M},\phi)$ if it belongs to 
the topological boundary of $\phi(M)$ in $\widehat{M}$. A boundary set is
a set consisting of boundary points.
\label{boundary-point}
\end{defi}
One can also say that a curve $\g:I\rightarrow M$ approaches a boundary
set $B$ of the envelopment 
$(M,\widehat{M},\phi)$ if $\phi\circ\g$ has some element of $B$ as an
endpoint.

If different envelopments of the same manifold $M$ are given, one needs to
consider the relation between their boundary points and sets. For example,
let $(M,\widehat{M},\phi)$ and $(M,\widehat{M'},\phi')$ be two different
envelopments 
of the same differentiable manifold $M$ and let $B$, $B'$ be respective
boundary sets.
Then $B$ ``covers" $B'$ if for every open neighbourhood $\U$ of $B$ in
$\widehat{M}$ there exists an open 
neighbourhood $\U'$ of $B'$ in $\widehat{M'}$ such that 
$$
\phi\circ\phi'^{-1}(\U'\cap\phi'(M))\subset\U\, .
$$ 
This definition can be applied to boundary sets consisting of a single 
point too. An equivalence relation $\sim$ between boundary sets on
different 
envelopments can be defined as $B\sim B'$ if $B$ covers $B'$ and $B'$
covers $B$. Any of the equivalence classes of this equivalence 
relation is called an {\em abstract boundary set} denoted by
$[B]$. Clearly single points also give rise to 
equivalence classes, which are called {\em abstract boundary points}.
\begin{defi}[Abstract boundary]
The abstract boundary (or in short $a$-boundary) of a differentiable
manifold $M$ is denoted by ${\cal B}(M)$ and formed by the set of all its 
abstract boundary points. 
\label{abstract-boundary}
\end{defi}

It is also interesting to study the properties of boundary points related
to the curves of $M$. To do this 
one must select a family $\C$ of curves having the b.p.p. seen
before. Then, a
boundary point $p$ of the envelopment $(M,\widehat{M},\phi)$ is a
$\C$-boundary point or $\C-approachable$ 
if it is a limit point of some curve of the family $\C$. Boundary points
which are not $\C$-boundary points 
for any family $\C$ are called {\em unapproachable}.
Explicit examples of unapproachable boundary points are shown in
\cite{SCOTT}. 
This definition can be extended to abstract boundary points and so  
those elements of the abstract boundary which are approachable are called
abstract $\C$-boundary points. 

Up to this point only the differentiable properties of $M$ have been used
to construct the 
$a$-boundary, which can in principle be defined for any differentiable
manifold with no further structures required. If we consider now
pseudo-Riemannian manifolds admitting a $C^k$ metric $\G$ of 
arbitrary signature, then new natural concepts which are in agreement with
traditional ones can be given.
\begin{defi}[Metric extension]
A pseudo-Riemannian manifold $(\widehat{M},\hat{\G})$  is a $C^l$ metrical
extension ($1\leq l\leq k$) of the 
pseudo-Riemannian manifold $(M,\G)$, with $\G$ of class $C^k(M)$, if
$\hat{\G}$ has class $C^l(\widehat{M})$ and 
there exists an envelopment $(M,\widehat{M},\phi)$ such that on $M$
$$
\phi^*(\hat{\G})=\G.
$$
Extensions are denoted by $(M,g,\widehat{M},\hat{g},\phi)$.
\label{cl-extension}
\end{defi}
Boundary points of any envelopment of a pseudo-Riemannian manifold can
then be further classified according to whether they can be forced to
be regular in some metric extension or not.  Thus, a boundary point
$p$ of an envelopment $(M,\widehat{M},\phi)$ is said to be $C^l$ {\em
regular} if there exists a $C^l$ metric extension
$(M,\G,\widehat{M}_1,\bar{\G},\phi)$ (same $\phi$!)  of $(M,\G)$ such
that $\phi(M)\cup\{p\}\subseteq\widehat{M}_1\subseteq\widehat{M}$.  In
other words, the pseudo-Riemannian manifold $(M,\G)$ can be continued
through the boundary point $p$.  Whether or not a boundary point is
regular is independent of any family of b.p.p. curves $\C$.

As a matter of fact, regular points need not be approachable by
certain fixed, previously chosen, families of curves such as
geodesics.  This is shown in \cite{SCOTT} by explicit examples, but it
is already known from the previously mentioned examples of \cite{GER}. 
This is important because boundary points can then be
classified according to whether they are approachable by different
families $\C_1$, $\C_2$ ....  or not, and then depending on their
approachability for each particular family.  This classification, with
some additional details, is spelt out in \cite{SCOTT} resulting in a
very elaborated scheme which shall not be reproduced here.  If
nevertheless one chooses a particular family $\C$ of b.p.p curves,
then a definition of singularity is put forward in \cite{SCOTT}: a
boundary point $p$ of the envelopment $(M,\widehat{M},\phi)$ is $C^l$
singular, also called a $C^l$ singularity, {\em with respect to the
family} $\C$, if $p$ is not a $C^l$ regular boundary point and $p$ is
$\C$-approachable with bounded parameter ---i.e. there exists a curve
in the family $\C$ which approaches $p$ with bounded parameter.

We see that the concept of singularity depends on the envelopment and
on the family of curves.  One may wonder if there are cases in which a
singularity is present for any envelopment of the manifold $(M,\G)$,
given a fixed family $\C$.  This is resolved by calling $C^m$ {\em
removable} singularity to any $\C$-singularity $p$ that can be covered
by a $C^m$ non-singular boundary set of another envelopment. 
Otherwise the point is called an essential singularity with respect to
$\C$.  An important result is that essential singularities have been
proved to be stable \cite{ASHLEY}.  Essential singularities can be
further classified depending on the properties of the curves
approaching them.  One of us \cite{SINGULARITY} put forward a
definition of singularity based on the above by understanding the
inextensibility of b.p.p curves as $b$-completeness.  Thus a concept
of singular extension was also introduced.  Still the subject presents
too many subtleties to be sketched here. In \cite{ASHLEY2} the existence
of {\em curvature singularities} in the $a$-boundary is discussed.

Despite the fact that the $a$-boundary provides a general framework
for the study of completions and boundaries of general manifolds by
using {\em envelopments}, it is too general to provide definite
properties such as shape, causal structure, topology, metric, et
cetera of the boundary.  It can be used, however, as an appropriate
starting point for any try, and as a background for other more
definite constructions, see next subsection.

\subsubsection{Causal relationship and the causal boundary.}
\label{causal-relationship}
After this journey through all the attempts to construct a valid
causal boundary for generic spacetimes, one often reaches the state in
which, on intuitive grounds, one prefers the original plain and elementary
definition of the Penrose conformal compactification.  It is clear,
simple, productive and provides all the required properties: shape,
causal structure, metric, causal character, probably distinction
between infinity and singularities, and so on and so forth.  The only
problem was, in fact, the actual impossibility of finding the
conformal embedding explicitly in general ---recall, however, the powerful encouraging results found in \cite{ACD,ACD2,FRIEDRICH1,FRIEDRICH2}.

This is why we tried to improve the conformal compactification in
\cite{CAUSAL}, where a new definition of causal boundary was
presented.  The idea is very simple: now that we have come to know
that an appropriate definition of isocausal Lorentzian manifolds
(definition \ref{equivalence}) is more general than the conformal one,
and that the causal structure can be defined as in definition
\ref{causal-structure}, we can simply repeat the whole
conformal-boundary construction by replacing any ``conformal" mappings
by the causal mappings of definition \ref{PREC}.  This simple
generalization provides us with the notions of {\em causal extensions,
causal boundaries} and {\em causal compactification} in an obvious
way.  This program has many advantages: on the one hand, it keeps all
the good properties of the conformal scheme, which is in fact
contained as a particular case.  Thus, one can give attributes such as
shape, causal character, dimensionality, connectivity, topology, et
cetera, to the boundaries, and furthermore the traditional Penrose
diagrams can be generalized to get intuitive pictures of complicated
spacetimes, see $\S$\ref{causal-diagram} and \cite{CAUSAL}.  On the
other hand, it avoids to a certain extent the only problem with the
compactification process, which was looking for the conformal factor
and embedding.  Now we only have to look for two mutual causal
mappings, which are certainly easier to find as the restriction to be
a causal mapping is much milder (see the discussion after definition
\ref{PREC}) than that of being a conformal one.

To be precise, let us start with the following definition.  Recall the
equivalence relation ``$\sim$'' of definition \ref{equivalence}.
\begin{defi}[Causal extension]
A causal extension of a Lorentzian manifold $(V,\G)$ is any
envelopment $(V,\tilde{V},\Phi)$ into another Lorentzian manifold
$(\tilde{V},\tilde{\G})$ such that $V$ is isocausal to $\Phi(V)$:
$V\sim\Phi(V)$.
\label{causal-extension}
\end{defi}
Observe that a causal extension for $(V,\G)$ is in fact a causal
extension for the causal structure, that is to say, for the whole
equivalence class coset${}_{V}(\G)\in$ \mbox{\rm Lor}$(V)$/$\sim$
based on $V$ of which $(V,\G)$ is a representative, see definition
\ref{causal-structure}.

Notice that the causal extension is different from the traditional
metric extensions (definition \ref{cl-extension}) in which the {\em
metric} properties of $(V,\G)$ are kept, and also from the conformal
embeddings of subsection \ref{conformal-boundary}, where the conformal
metric properties were maintained.  Here we only wish to keep the
causal structure of $V$, in the sense of definition
\ref{causal-structure}, which is a much more general point of view. 
We would like to remark, however, that obviously all metric extensions
and all conformal embeddings are a particular type of causal
extension, hence they are trivially included in definition
\ref{causal-extension}: any conformal embedding is a causal extension
with the particular choice that the causal equivalence between $V$ and
$\Phi(V)$ is of conformal type.

We arrive at the main definition.
\begin{defi}[Causal boundary]
Let $(V,\tilde{V},\Phi)$ be a causal extension of $(V,\G)$ and $\d V$
the topological boundary of $\Phi(V)$ in $\tilde{V}$.  Then, $\d V$ is
called the causal boundary of $(V,\G)$ with respect to
$(\tilde{V},\tilde{\G})$.  A causal boundary is said to be complete if
$\Phi(V)$ has compact closure in $\tilde V$.
\label{causalbound}
\end{defi}
Note again that all the members in coset${}_{V}(\G)$ have the same
causal boundary {\em with respect to a given causal extension}.  In
principle, however, the causal boundaries of coset${}_{V}(\G)$ may
depend on its causal extensions.  At this stage, this is a problem
more serious here than for conformal completions, as in the latter
case it seems reasonable to conjecture that the main properties of a
causal boundary will be unique {\em if} the boundary is {\em complete}
\cite{FRAUENDIENER}, and this is not the case for the former case,
see the long discussion in Examples 12 and 13 in \cite{CAUSAL}.  We
have persuaded ourselves, though, that this problem can be easily overcome
by
adding the necessary restrictions to definition \ref{causalbound}. 
This is one of the open questions about definition \ref{causalbound}.

With regard to how to distinguish between points at
infinity or singularities at a causal boundary one can use the ideas
of th $a$-boundary in an effective way.  Thus, in \cite{CAUSAL} we
gave the following primary classification
\begin{defi}
Let $\d V$ be the causal boundary of $(V,\G)$ with respect to the causal 
extension $(V,\tilde{V},\Phi)$. A point $p\in\d V$ is said to belong to:
\begin{enumerate}
\item a {\em singularity set} ${\cal S}\subseteq \d V$ if $p$ is the 
endpoint in
$(\tilde V,\tilde\G)$ of a curve which is endless and incomplete within
$(V,\G)$.
\item {\em future infinity} $\Is^+\subseteq \d V$ if $p$ is the 
endpoint
in $(\tilde V,\tilde\G)$ of a causal curve which is complete to the future
in 
 $(V,\G)$. And similarly for the past infinity $\Is^-$.
\item {\em spacelike infinity} $i^0\subseteq \d V$ if $p$ is the 
endpoint
in $(\tilde V,\tilde\G)$ of a spacelike curve which is complete in
$(V,\G)$.
\end{enumerate}
\end{defi}
Let us remark that the traditional $i^{\pm}$ of the conformal
compactification have been included here in the general sets
$\Is^{\pm}$.  We can also collect all past and future infinities in a
set called {\em causal infinity}.  Hitherto, it has not been proved
that all points in a causal boundary belong to one of the
possibilities of the previous definition, nor that the different
possibilities are disjoint in general.  These are interesting
open questions.

To summarize, our scheme provides in principle a simple way to attach
a causal boundary to any Lorentzian manifold.  Its practical
application in specific cases still suffers from the problem of
finding the causal extensions $(V,\tilde{V},\Phi)$ for generic
Lorentzian manifolds $(V,\G)$, but this is certainly easier than
finding the conformal completions.  Besides, it is our opinion that a
systematic way to find out whether or not two Lorentzian manifolds are
isocausal does exist, and so this problem could be fully
resolved.  We thus believe that it is worth exploring this new line
and try to answer the open questions mentioned in this subsection.

For some examples of causal diagrams constructed using causal
extensions see $\S$\ref{causal-diagram} and \cite{CAUSAL}.

\subsection{Examples and applications}
We now briefly present some remarks and comments about the traditional
Penrose diagrams ($\S$\ref{conformal-diagram}), its comparison to the
new causal diagrams ($\S$\ref{causal-diagram}), and some recent
interesting applications and results which have been found in
Lorentzian manifolds of pp-wave type ($\S$\ref{ppwaves}), a subject of
renewed interest for both mathematicians and physicists, specially
with regard to string theory.

\subsubsection{Penrose diagrams.}
\label{conformal-diagram}
Perhaps one of the most severe difficulties faced by the conformal
boundary consists in the high efforts needed to construct the unphysical
spacetime from a given physical metric.  
Friedrich's conformal equations are a first step
towards this direction but they seem to be too complex to allow for an
analytical solution in interesting physical cases such as isolated
bodies.

Despite this difficulty we can still extract useful information out of
the conformal methods in many particular cases.  A paradigmatic
example is the case of spherically symmetric spacetimes: the group
$SO(n-1)$ acts {\em multiply} transitively on the spacetime and the
transitivity
surfaces are spheres $S^{n-2}$.  The most general form of the line
element is
\be
ds^2=A(R,T)dT^2-2B(R,T)dRdT-C(R,T)dR^2-L^2(R,T)d\Omega^2 .
\label{sph}
\ee
The $T-R$ part of the metric can always be brought into an explicitly
conformally flat form by means of a suitable coordinate change
$T=T(t,r)$, $R=R(t,r)$ yielding
\be
ds^2=F(t,r)(dt^2-dr^2)-\Xi^2(t,r)d\Omega^2.
\label{flat-2d}
\ee 
Due to the spherical symmetry the $t-r$ part of the metric contains a
great deal of the relevant information about the global causal
properties of the spacetime and so we can dismiss the angular part in
a first approximation.  This is very useful, since this $t-r$ part can
always be conformally embedded in two-dimensional Minkowski spacetime
in the obvious way.  Thus, if we had a finite conformal diagram for
the two-dimensional Minkowski spacetime, we could represent on them
the conformal boundary for (the $t-r$ part of) the spacetime under
study.  The places where the factor $\Xi$ vanishes or diverges, which
are typically either infinity or (removable or essential)
singularities can also be represented.

The required compactification for two-dimensional Minkowski spacetime
is easily achieved.  Let us start with its canonical line-element
$ds^2_0=dt^2-dx^2$, $-\infty<t<\infty$, $-\infty<x<\infty$, and let us
perform the following coordinate change, bringing infinity to finite
values of the new coordinates,
\bnr
t=\fr{1}{2}\left(\tan\left(\bar{t}+\bar{x}\right)+\tan\left(
\bar{t}-\bar{x}\right)\right),\,\,
x=\fr{1}{2}\left(\tan\left(\bar{t}+\bar{x}\right)-\tan\left(
\bar{t}-\bar{x}\right)\right),
\enr
under which the line-element becomes 
$$
ds^2_0=\fr{1}{4\cos^2(\bar{t}+\bar{x})\cos^2(\bar{t}-\bar{x})}(
d\bar{t}^2-d\bar{x}^2).
$$
The ranges of the new coordinates are $-\pi/2<\bar{t}+\bar{x}<\pi/2$,
$-\pi/2<\bar{t}-\bar{x}<\pi/2$ so the conformal embedding of
two-dimensional Minkowski spacetime in the plane $\r^2$ is depicted as
shown
below.
\begin{center}
\includegraphics[width=.7\textwidth]{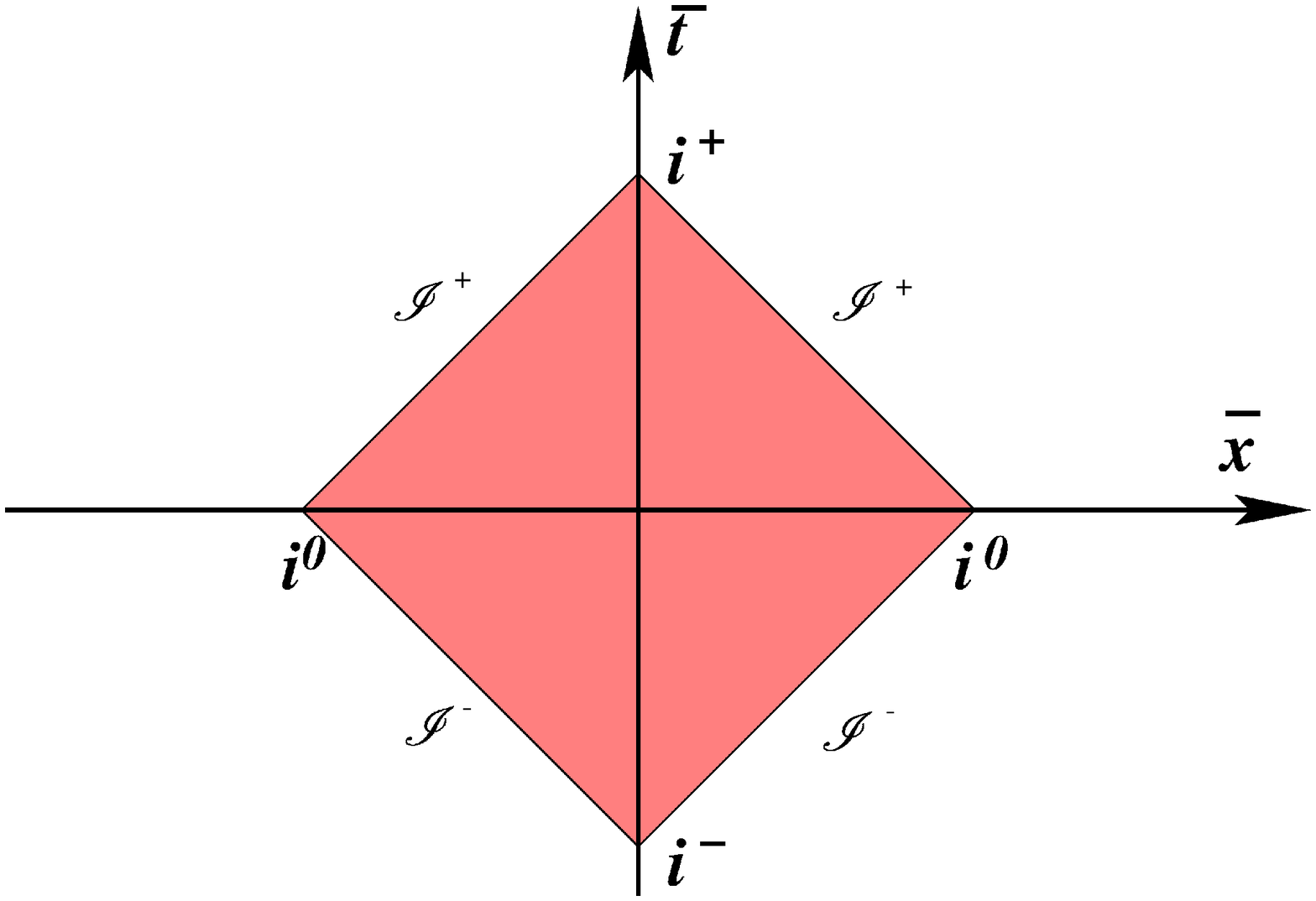}

\parbox{.7\textwidth}{
\footnotesize Conformal compactification of two-dimensional Minkowski
spacetime. The coloured zone corresponds to the 
physical spacetime and the different regions of the conformal boundary are
also marked. Note the disconnected structure 
of spatial infinity $i^0$.} 
\end{center}

Once we have the Penrose diagram for flat spacetime in 2 dimensions,
the ``relevant" $t-r$ part of the line-element given by
(\ref{flat-2d}) for {\em any spherically symmetric} spacetime can also
be represented in two-dimensional pictures like the above figure. 
However, of course, each point of the new picture (with the exception
of the region with $\Xi =0$ if it is not an essential
singularity) represents a
$(n-2)$-sphere.  The diagram and the boundary can adopt
many different shapes, and in some occasions the whole $t-r$ region
cannot be depicted in one single portion of the type of the figure,
see e.g. \cite{CARTER3,FF}.  The pictures just described, called
Penrose diagrams, are very useful, ubiquitous in the relativity
literature, and they were made popular by Carter
\cite{CARTER3,CARTER2} and the textbook \cite{FF}.  Perhaps the
simplest example is the diagram of $n$-dimensional Minkowski spacetime
which, given that $r>0$ for such spacetime,
can be obtained from the above figure by cutting it through the
$\bar{t}$-axis and discarding any of the two symmetric halves.  Most
of the relevant causal information about $n$-dimensional Minkowski
spacetime is kept by its Penrose diagram.  Penrose diagrams of many
relevant spherically symmetric solutions of Einstein equations can be
found in \cite{FF}.  Many others can be found by reading any journal
concerning gravity and relativity.

Perhaps a very instructive example is the case of anti de Sitter 
spacetime, which is also of present relevance concerning the already 
mentioned Maldacena's conjecture (or 
anti-de-Sitter/Conformal-Field-Theory correspondence). The 
inextensible anti de Sitter line-element has the form 
(\ref{sph}) as follows
$$
ds^2=\cosh^2R\ dt^2-dR^2-\sinh^2 R\, d\Omega^2 
$$
where $0<R<\infty$ and $-\infty<t<\infty$. It is very simple to check 
that one cannot {\em conformally} embed this spacetime in a {\em compact 
region} of Minkowski spacetime 
or Einstein static universe. This is why the traditional diagram for 
anti de Sitter spacetime is a non-compact part of Einstein static 
universe with two parallel lines as infinity, and one adds artificially 
two points $i^{\pm}$ to the picture, see e. g. \cite{FF}. However, one 
can draw a true Penrose diagram following the comments of the previous 
paragraph. One forgets about the angular part of the metric and the 
following change of coordinate $r=\arctan (\sinh R)$
brings the line-element to its form (\ref{flat-2d})
$$
ds^2=\frac{1}{\cos^2 r}(dt^2-dr^2)-\tan^2 r\, d\Omega^2
$$
where now $0<r<\pi/2$. Therefore, a finite Penrose diagram can be 
drawn now by just taking the part of 2-dimensional flat spacetime 
defined by $0<x<\pi/2$, compare to \cite{TIPLER}. This proves that
infinity is timelike 
everywhere. A 3-dimensional representation of anti de Sitter 
spacetime can then be given as its Penrose diagram; this is shown in the 
following figure taken from \cite{CAUSAL}.

\parbox{.2\textwidth}{\includegraphics[width=.3\textwidth]{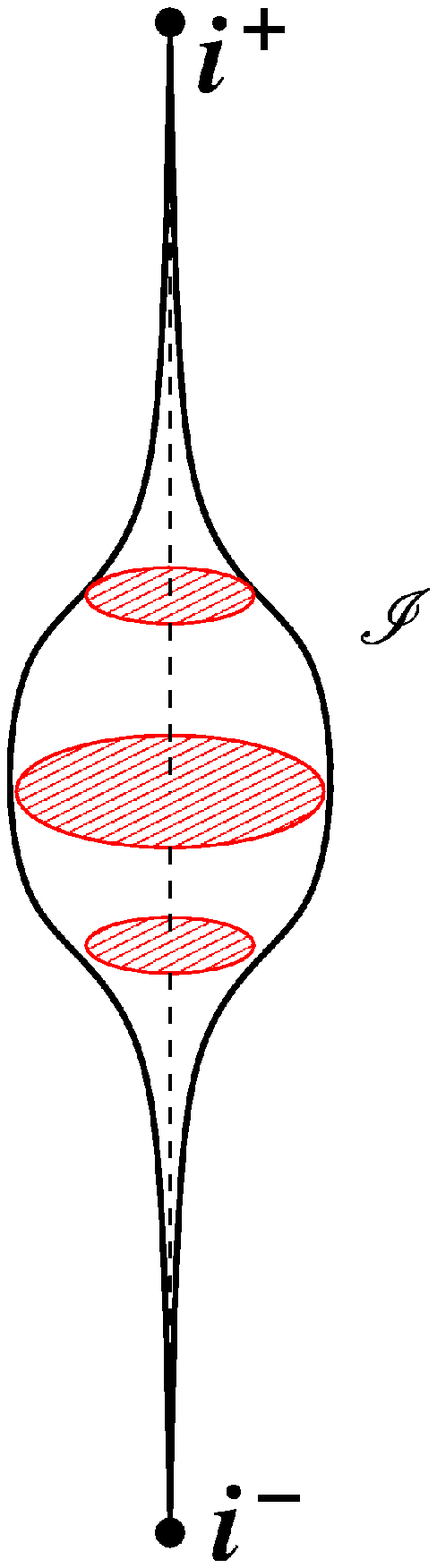}}
\hspace{1.5cm}\parbox{.6\textwidth}
{\footnotesize In this picture we show a Penrose-like diagram 
for anti-de Sitter spacetime. In this case we have
preferred to draw a 3-dimensional diagram to get a clearer picture of
the causal infinity. Every $t=$const. slice (in red)
has been reduced to an open horizontal disk, so that
every point in the diagram represents a $(n-3)$-sphere except for the
vertical axis which is the origin of coordinates. 
The boundary of the picture
represents the conformal infinity $\Is$ of anti-de Sitter. It is
remarkable that this boundary has precisely the shape of the
Einstein universe. Thus, one is tempted
to say that the causal boundary of $n$-dimensional anti-de Sitter
spacetime is the $(n-1)$-dimensional Einstein universe. 
Notice also the timelike
character of $\Is$ and the non-global hyperbolicity of the anti-de Sitter
spacetime.}

\medskip
\noindent
The above discussion is essentially based on the study of the 
causal properties of the $t-r$ part of the metric (\ref{flat-2d}) 
but it is possible to consider other forms for such 2-dimensional metric.
For instance any 2-dimensional metric with a Killing 
vector field can be written in local coordinates as
$$
ds^2=2dx^0dx^1+h(x^0)(dx^1)^2.
$$
In \cite{STROBL} constructions of Penrose diagrams for 2-dimensional 
spacetimes with the above line element are pursued and practical 
recipes to build the diagrams are provided. 

\subsubsection{Causal completions and causal diagrams.}
\label{causal-diagram}
In $\S$\ref{causal-relationship} we introduced the generalization of
conformal embeddings called causal extensions, and defined the causal
boundary.  Given that the Penrose diagrams are based on the conformal
compactification, we can generalize these diagrams by using causal
mappings.  In this way, as with Penrose diagrams, we can try to
analyze and understand the properties of complicated spacetimes by
studying the simpler ones to which they are isocausal, and by drawing
pictures which will generally include the causal boundary---showing
some of its properties.

To draw a causal diagram for $(V,\G)$, we must look for a causal
extension $(V,\tilde V,\Phi)$ (definition \ref{causal-extension}). 
There is no general rule to find an appropriate $\Phi$, but the fact
that we are dealing with causal mappings instead of conformal
relations may be many times an advantage.  This is so because less
restrictive conditions are needed for the isocausality condition
$\tilde\G|_{\Phi(V)}\in \mbox{coset}_V(\G)$, than for a conformal
relation $\Phi^*\tilde\G|_{V}=\Omega^2\G$.

In \cite{CAUSAL} we presented different explicit examples for which
the causal extension was performed explicitly and the causal diagrams
drawn.  These diagrams provide a shape and a {\em causal character}
for the boundary.  A remarkable example which shows how this scheme
works and its applicability is given for instance by the Kasner-type
spacetime
$$
ds^2=dt^2-\sum_{j=1}^{n-1}t^{2p_j}(dx^j)^2,\ \ 0<t<\infty,\
-\infty<x^j<\infty
$$
where $p_j$ are constants.  If $p_j<1$ then it is shown in
\cite{CAUSAL} that the above spacetime is isocausal to
$$
ds^2=dt^2-(B+e^{-kt})^2\sum_{j=1}^{n-1}(dx^j)^2,\ k=\mbox{max}\{1-p_j\},\
B>1,\ 0<t<\infty. 
$$
But this a Robertson-Walker line-element!  Thus this can be easily
written in explicitly conformally flat form by means of the simple
coordinate transformation $\tau=\log(1\!+\!Be^{kt})/(kB)$ getting
$$
ds^2=\fr{B^2e^{2kB\tau}}{(e^{kB\tau}-1)^2}\left(d\tau^2-\sum_{j=1}^{n-1}(dx^j)
^2\right),\
0<\tau<\infty.
$$
From here we see that this spacetime can be conformally embedded in
(the upper half of) Minkowski spacetime and this very conformal
embedding is a causal extension for our original Kasner-type model. 
Performing the conformal compactification of Minkowski spacetime if
desired we can obtain a {\em complete} causal boundary.  The picture
of this conformal embedding is called a {\em causal diagram} of the
original Kasner-like spacetime and it conveys most of the causal
information about this spacetime despite that Kasner and FRW models
are not conformally related.

The case with one of the Kasner exponents $p_1=1$ and the rest as
before was also treated, and an interesting diagram found in
\cite{CAUSAL}.  Other examples of causal diagrams can be found in
\cite{CAUSAL}.

\subsubsection{Causal properties of the Brinkmann spacetimes: pp-waves.}
\label{ppwaves}
Plane fronted waves with parallel rays (pp-waves) have attracted a lot
of interest in recent years specially within the string theory
community.  The main reason is that all the scalar curvature
invariants are zero in these spacetimes and so they are exact
solutions (solutions ``at any order in the expansion of the string
action") of the classical theory, providing backgrounds upon
which string theorists can explicitly try to study quantum phenomena
involving gravity.  Besides this modern interest, the relevance of
these spacetimes is clear as they describe under certain circumstances
the simplest (though highly idealized) models of exact gravitational
waves.  A general study of the global causal properties and the causal 
completions of these and
some related spacetimes has been systematically carried out
only recently.  The main
results are the subject of this subsection.

The spacetimes with all scalar curvature invariants vanishing have
been explicitly found in \cite{CMPP} for arbitrary dimension, see also
references therein and \cite{VSI}.  They include in particular all
spacetimes characterized by the existence of a null vector field $\k$
which is parallel (also called covariantly constant).  The most
general local line-element for such a spacetime was discovered by
Brinkmann \cite{Br} by studying the Einstein spaces which can be
mapped conformally to each other: there exists a local coordinate
chart in which the line element takes the form \cite{Br}
\be
\fl ds^2=2dudv+H(x,u)du^2-A_i(x,u)dx^idu-g_{ij}(x,u)dx^idx^j,\
x=\{x^1,\dots,x^{n-2}\},
\label{brink}
\ee   
where the functions $H$, $A_i$ and $g_{ij}=g_{ji}$ (det$(g_{ij})\neq
0$) are independent of $v$, otherwise arbitrary, and the parallel null
vector field is given by $\k=\d/\d v$.  This null vector field has
then vanishing shear, rotation and expansion so the hypersurfaces
orthogonal to $k^a$ (wave fronts) are locally planes.  However, they
are not planes!

The Brinkmann spacetimes include to the so-called pp-waves
\cite{EK,MCCALLUM}, 
and nowadays it has become commonplace to 
consider the existence of a covariantly constant null vector field as 
the ``definition of pp-waves". Thus, many times the whole family in 
(\ref{brink}) are called pp-waves. However, this is a clear 
misunderstanding. The term pp-wave arises as a shorthand for ``plane 
fronted" gravitational waves with ``parallel" rays, 
\cite{EK,MCCALLUM}, and it should be reserved to those cases contained 
in (\ref{brink}) for which $g_{ij}(x,u)$ can be reduced to 
$g_{ij}=\delta_{ij}$ for all (fixed) $u$ simultaneously. 
In other words, for the cases where 
$(M,g_{ij})$ (with $g_{ij}$ at fixed $u$) is a flat Riemannian
manifold, 
$M$ being the manifold with coordinates $\{x^i\}$.
The reason is that the Brinkmann metrics 
certainly contain parallel rays (the null geodesics defined by 
$\k$), and these rays are orthogonal to 
$(n-2)$-surfaces defined locally by $u=$const. However, these 
surfaces, whose first fundamental form is determined by the $g_{ij}$ 
at the fixed $u$, are {\em not} planes nor flat in any sense in 
general. As a matter of fact, they can have any possible geometry, see
e.g. \cite{CANDELA}. As a simple example, consider the obvious 
metric build on $\r^2 \times S^{n-2}$. This misunderstanding arises because, in 
General Relativity (that is, for $n=4$ exclusively), all solutions of 
the Einstein vacuum or Einstein-Maxwell equations have in appropriate 
coordinates $g_{ij}=\delta_{ij}$ \cite{EK,MCCALLUM}---and, as a matter 
of fact, $A_i=0$ too---. In other 
words, {\it all vacuum or Einstein-Maxwell solutions in General 
Relativity with a parallel null vector field are pp-waves}. This is 
not so in general, though, and it does not hold in $n=4$ for metrics 
with other type of Ricci tensor, or for other values of $n$. If, 
nonetheless, one wishes to keep a semantic relation to pp-waves, a 
more adequate term would be ``Mp-waves", or something similar, 
indicating by the `M' the Riemannian manifold(s) defined by the $g_{ij}$
at 
constant $u$. 

Keeping these remarks in mind, important particular cases of the
line-element (\ref{brink}) are given by: (i) as already mentioned,
line elements with $A_i=0$ and a flat metric $g_{ij}(x)$ (ergo
reducible to $\delta_{ij}$) which are called pp-waves; (ii) pp-waves
with the additional restriction that $H(x,u)=f_{ij}(u)x^ix^j$ are
called {\em plane waves}; and (iii) if the plane wave has also $\d
f_{ij}/\d u=0$, so that the $f_{ij}$ are
constants, are {\em locally symmetric} spacetimes in the classical
sense \cite{ONEILL} that the curvature tensor is parallel (that is to
say, covariantly constant).  These locally symmetric plane waves have
sometimes been termed as ``homogeneous plane waves", but this could be
misleading as there is a more general class of plane waves with a
group of symmetries acting transitively on the spacetime which are
thus traditionally called homogeneous, see \cite{MCCALLUM} and
references therein.

In the classical paper \cite{PENROSE-WAVE} Penrose proved that plane
waves are in general not globally hyperbolic.  These results have been
enlarged recently by a number of authors with many results of
interest.  Marolf and Ross \cite{MAROLF-ROSS} performed a general
study of the causal boundary for locally symmetric plane waves.  After
a rotation in the coordinates $\{x^i\}$ the line element for these
waves can be reduced to
$$
ds^2=2dudv -du^2 \sum_{i=1}^{n-2}\epsilon_i
\mu^2_ix^2_i-\delta_{ij}dx^idx^j,
$$  
where $\epsilon_i$ are signs given by $\epsilon_i=1$ for
$i\in\{1,\dots , j\}$ and $\epsilon_i=-1$ for $ i\in\{j+1,\dots
,n-2\}$, and the $\mu_i$ are constants.  A straightforward calculation
shows that the Weyl curvature tensor vanishes if and only if $j=n$ or
$j=0$, that is to say, if all the $\epsilon_i$ have the same sign. 
These are the only cases in which finding a conformal embedding into
Minkowski or other conformally flat spacetime will not be too hard a
task.

\hspace{-1cm}\parbox{.26\textwidth}{
\includegraphics[width=.25\textwidth]{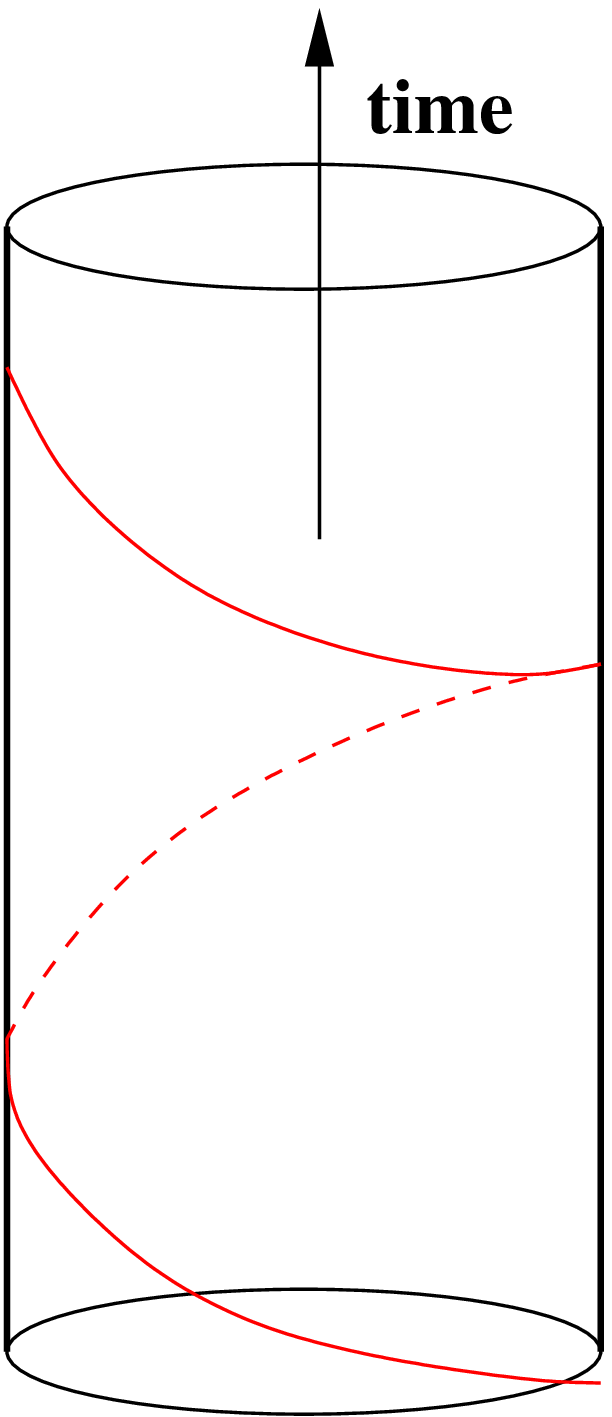}

\footnotesize Conformal embedding of a homogeneous plane wave with
positive 
eigenvalues into Einstein static universe. The red one-dimensional 
line represents $\Is$ (taken from \cite{BERENSTEIN}).}
\parbox{.75\textwidth}{
Remarkably these embeddings are known and the result depends on the
sign of all the $\epsilon_i$.  The case with $\epsilon_i=1$ for all
$i=1,\dots n$ was addressed in \cite{BERENSTEIN} for $n=10$ where the
conformal embedding of this particular plane-wave spacetime into
$10$-dimensional Einstein static universe was constructed.  The
conformal boundary $\Is$ consists of a one-dimensional null line which
winds around the compact dimensions.  In a three-dimensional picture
this line would look like an helix contained on the Einstein cylinder,
see the figure.  The case of negative $\epsilon_i$ for arbitrary $n$
can be conformally embedded in $n$-dimensional Minkowski spacetime
resulting in a sandwich region limited by two parallel null planes.

These results were generalized by the same authors in
\cite{MAROLF-ROSS,MAROLF-ROSS2} for all locally symmetric plane waves
with at least one positive $\epsilon_i$, and in \cite{HUBENY} for
general plane waves and some pp-waves, but now instead of the
conformal boundary, the GKP boundary is constructed
explicitly.  A very important remark was noted in \cite{MAROLF-ROSS2}:
as these plane waves have a non-vanishing but constant Weyl curvature
tensor, the conformal completion is actually impossible. 
This is an explicit case where a {\em complete} conformal boundary
cannot be defined (observe that the causal completion and a {\em
complete} causal boundary in the sense of
$\S$\ref{causal-relationship} and $\S$\ref{causal-diagram} can
certainly be looked for).  This and other results in
\cite{MAROLF-ROSS2,HUBENY,HR} 
are very interesting and can be summarized as
follows:
} 

\begin{enumerate}
\item The $c$-boundary for locally symmetric plane waves with at least one
$\epsilon_1=1$ is again a one-dimensional null set plus two points
(identified as $i^{\pm}$) \cite{MAROLF-ROSS}.

\item However, ``spatial" infinity ---or the part of the boundary
approached
by spacelike curves--- cannot be included in the total boundary as the
completion $\bar{M}$ is non-compact in the topologies introduced in
\cite{MAROLFROSS}, see $\S$\ref{m-r}.  A new topology is introduced
and an analysis of the points that should be added to the $c$-boundary
to include spatial infinity is performed in \cite{MAROLF-ROSS2} with
no definite conclusion.

\item For the case of plane waves the behaviour at infinity is determined
by
the properties of the functions $f_{ij}(u)$.  The case with
tr$f_{ij}\geq 0$ and $f_{ij}(u)$ regular $\forall u\in\r$ was treated
in \cite{HUBENY} with the outcome that if any of these $f_{ij}(u)$
approaches zero for large values of $u$ {\em fast enough} then the GKP
boundary is of dimension higher than one.  This case includes for
instance Minkowski spacetime in which all the $f_{ij}(u)$ are zero. 
On the contrary, if these functions exhibit a polynomial,
trigonometric or hyperbolic behaviour, such that the geodesics exhibit
an oscillatory regime for large values of $u$, or even if the
$f_{ij}\rightarrow 0$ for large $u$ as a rational function of $u$,
then it can be shown that the boundary is again a one-dimensional null
line.

\item In \cite{HR}, by using techniques of geodesic connectivity, the
result
that every point of any general plane-wave spacetime can be joined to
infinity (not increasing the value of $u$ too much) is proved
regardless of the properties of the functions $f_{ij}(u)$.  In fact,
tighter results regarding geodesic connectivity can be proved, 
see \cite{CANDELA,SANCHEZ}.  This was
interpreted in \cite{HR} by saying that plane waves cannot have
``event horizons".  A stronger and more precise version of this result
can be found in \cite{VSI}, where the complete absence of closed
trapped surfaces in the general spacetime with vanishing curvature
invariants---including in particular all Brinkmann metrics
(\ref{brink})--- was demonstrated.

\item The $c$-boundary for pp-waves was also investigated in
\cite{HUBENY}. 
Now, we have a more diverse set of behaviours since the function
$H(x,u)$ may be singular in both the coordinate $u$ and the transverse
coordinates $x^i$.  Some conditions under which the GKP boundary is
again one-dimensional (and null) are presented in \cite{HUBENY},
usually under the assumptions that the spacetime is geodesically
complete and distinguishing.
\end{enumerate}

Linking to this last comment, the question of the degree of causal
``virtue" of general pp-waves is of great interest.  After the seminal
paper \cite{PENROSE-WAVE} mentioned before, a recent important advance
in the study of the causal properties of ``Mp-waves" with $A_i(x,u)=0$
(see above) follows from the work of Flores and S\'anchez
\cite{SANCHEZ}.  Thus, a classification according to the standard
hierarchy of causality conditions (definition \ref{hierarchy}) was
found for these spacetimes.  Assuming that $H(x,u)$ is smooth, the
classification depends on the behaviour of $H(x,u)$ at large values of
the transverse coordinates $x^i$, and in this way the notions of
subquadratic, at most quadratic or superquadratic behaviour at spatial
infinity are introduced: $H(x,u)$ behaves {\em subquadratically} at
spatial infinity if there exists $\bar{x}\in M$ (recall that
$(M,g_{ij})$ is the transverse Riemannian manifold with constant $u$)
and continuous functions $R_1(u),R_2(u)\geq 0$, $p(u)<2$ such that
$$
H(x,u)\leq R_1(u)d^{p(u)}(x,\bar{x})+R_2(u),\ \forall\ (x,u)\in
M_0\times\r,
$$
where $d(x,y)$ is the canonical distance function on $(M,g_{ij})$.  If
$p(u)=2$ then $H(x,u)$ is said to behave {\em at most quadratically} at
spatial infinity.  Finally $H(x,u)$ is {\em superquadratic} if there
exists a sequence $\{y_n\}\subset M$ and a point $\bar{x}\in M$ such
that $d(\bar{x},y_n)\rightarrow\infty$, when $n\rightarrow\infty$ and
$$
H(y_n,u)\geq R_1d^{2+\epsilon}(y_n,\bar{x})+R_2,\ \forall u\in\r,
$$   
for some quantities $\epsilon$, $R_1$, $R_2\in\r$ with $\epsilon$, $R_1$
$>0$.
The results in \cite{SANCHEZ} can be summarized in the next theorem.
\begin{theo}
All general Mp-wave spacetimes with $A_i(x,u)=0$ are causal. If in
addition
\begin{enumerate}
\item the proper Riemannian manifold $(M,g_{ij})$ is complete, $H(x,u)\geq
0$ and $H(x,u)$ 
behaves superquadratically, then they are non-distinguishing.
\item $H(x,u)$ behaves at most quadratically at spatial infinity then they
are strongly causal.
\item $H(x,u)$ behaves subquadratically at spatial infinity and the
Riemannian distance on $M$ is 
complete, then they are globally hyperbolic.
\end{enumerate}   
\label{pfw}
\end{theo}
Notice that the cases presented in this theorem are not mutually 
exclusive and other behaviours of $H(x,u)$ not covered by this 
result may result.

In \cite{CANDELA}, a thorough analysis of the geodesic properties and
geodesic connectivity 
of Mp-waves was performed, and again the sub- or super-quadratic 
properties of $H(x,u)$ revealed themselves as essential to classify the
different 
possibilities. More related results were found in \cite{VERONIKA} for
cases 
with an explicitly non-smooth $H(x,u)$, but only for pp-waves: with a flat
 $(M,g_{ij})$. If $H(x,u)$ satisfies the inequality 
$H(x,u)\leq A_{ij}(u)x^ix^j$, $\forall x^j$ 
(in the above terminology $H(x,u)$ would be at most
quadratic)
 for certain functions $A_{ij}(u)$ which may have singularities, 
then the pp-waves are causally stable. All 
these results and theorem \ref{pfw} are thus very interesting 
and have physical implications, because the {\em critical} at most
quadratic 
behaviour is in fact the one usually relevant in General Relativity, 
for this is {\em the} behaviour of the plane waves which are exact
 solutions of the vacuum or Einstein-Maxwell equations.

\section{Causal transformations and causal symmetries} 
\label{symmetry}
Transformations preserving the ``causal structure'' or the causal
relations have been already described in this review (section
\ref{causal-preservation}). As remarked there, it seems that the 
right concept of causal structure is that of definition 
\ref{causal-structure}, arising from the idea of isocausality 
(definition \ref{equivalence}) and causal mappings (definition
\ref{PREC}), 
and that this structure is more general than the conformal 
one: the causal structure is defined by a metric {\em up 
to causal mappings}. This idea can be pursued further and in
such way we can try to generalize the group of conformal transformations,
and 
the conformal Killing vector fields \cite{MCCALLUM}, by defining and
studying 
sets of transformations
whose elements are mappings preserving the causal relations (or the
causal properties of Lorentzian manifolds). In general however, the
algebraic structures stemming from these transformations are no longer
groups, but {\em monoids}, as the inverse of a {\em
causal-preserving map} does not need to be a causal preserving map.  A
monoid is a set $G$ endowed with an internal associative operation
``$\cdot$'' admitting a neutral element $e$.  If there is no such
neutral element the pair $(G,\cdot)$ is called a {\em
semigroup}.  Semigroups
(specially Lie subsemigroups) and (sub-)monoids have been largely studied
in mathematics (standard references are
\cite{LJAPIN,SEMIGROUP,HILGERT}) due to their independent interest. 
Along this section we will try to convey the idea that these algebraic
structures are truly the ones needed to describe the
``submonoids of causal-preserving mappings''. Furthermore, we will 
define the set of causal transformations, obtain its algebraic 
structure, derive the infinitesimal generators of {\em one-parameter} 
submonoids, and find their characterization in terms of the Lie 
derivative of the metric.

\subsection{Causal symmetries}
\label{causal-symmetry}
In this section we are interested in finding the structure and 
properties of the set of
transformations preserving the Lorentzian cones of a Lorentzian
manifold.  The starting point for this was given in
sections \ref{causal-preservation} and \ref{causal-tensors}
where transformations preserving causal
relations and the Lorentzian cone on a manifold were introduced (this
was made explicit in definitions \ref{theta} and \ref{PREC}).
\begin{defi}[Causal symmetries]
Causal mappings for which both the domain and the target spacetimes 
are the same differentiable Lorentzian manifold $(V,\G)$ are called 
{\em causal transformations} or {\em causal symmetries}. The set of 
all causal transformations is denoted by $\C(V,\G)$.
\label{causal-transformation}
\end{defi}
As we saw in section \ref{causal-preservation}, a transformation 
$\phi$ is a causal symmetry if and only if $\phi^{*}\G$ is a future 
tensor. From the properties of causal mappings it is
clear that the composition of transformations is an internal operation in 
$\C(V,\G)$ which is associative with identity element (the identity 
transformation). Therefore,
\begin{prop}
The set $C(V,\G)$ is a submonoid of the group of
diffeomorphisms of the manifold $V$.
\end{prop}
However, $\C(V,\G)$ is not a
subgroup because the inverse of a causal symmetry is not in general a
causal symmetry. As a matter of fact, only conformal transformations of 
$(V,\G)$ will have such an inverse, see theorem 
\ref{phi-phiminusone}. Recall that if $S\subset G$ is a {\it proper
submonoid} of a group $G$, its {\em group of units} $H(S)$ is given by
$S\cap
S^{-1}$. The group of units is the maximal subgroup contained in $S$ in 
the sense that there is no other bigger subgroup of $G$ contained in $S$ 
possessing $H(S)$ as a proper subgroup. See \cite{SEMIGROUP} for the proof
of
this and other properties of monoids and semigroups.
Denoting by Conf$(V,\G)$ the set of all 
conformal transformations of $(V,\G)$ we have then
\begin{prop}
The maximal subgroup of $\C(V,\G)$ is the group of conformal 
transformations, that is to say, the group of units of $\C(V,\G)$ is
$$
\C(V,\G)\cap\C(V,\G)^{-1}=\mbox{\rm Conf}(V,\G).
$$
\end{prop}
The causal symmetries which are not conformal transformations are 
given the name of {\em proper} causal symmetries.  
Obviously, $\C(V,\G)$ depends on the background metric $\G$, but the 
set $\C(V,\G)$ is conformally invariant, a desirable property.
\begin{prop}
$\C(V,\G)=\C(V,\sigma\G)$ for any positive smooth function $\sigma$ on
$V$.
\label{CONFINVARIANT}
\end{prop}
The set ``$\C[$coset${}_V(\G)]$'' is not well-defined, however, so that 
we cannot say that the causal symmetries are the same for a given 
causal structure in the sense of definition \ref{causal-structure}. 
Nevertheless, given a causal structure, the causal symmetries of any 
of its metric representatives are in bijective correspondence 
\cite{CAUSAL-SYMMETRY}:
\begin{prop}
For any $\G_{1},\G_{2}\in$ {\em coset}$(\G)$ there is a one-to-one
correspondence 
between the sets $\C(V,\G_{1})$ and $\C(V,\G_{2})$.    
\end{prop}

The most extensive account dealing with causal symmetries are the
papers \cite{LETTER,CAUSAL-SYMMETRY} published recently
(the nomenclature and notation followed here is taken from that papers)
although similar ideas under different terminology can also be found
in the literature \cite{HARRIS-METHOD,HARRIS-SHAPE,PATRICOT}. 

\subsection{Causal-preserving vector fields}
\label{CPVF}
Our aim now is to obtain infinitesimal
generators for one-parameter submonoids of causal transformations and try
to find 
out the differential conditions fulfilled by these generators in much the
same
way as it has been done with isometries, homotecies or conformal
symmetries
\cite{YANO,MCCALLUM}. Despite causal
transformations not forming a group, infinitesimal generators for them
can still be defined.  This is
accomplished by considering local one-parameter submonoids of causal
symmetries defined by the condition
\be
\{\varphi_s\}_{s\in I\cap\r^+}\subset\C(V,\G),
\label{condition-rmas}
\ee 
where $\{\varphi_s :V\rightarrow V\}$, $s\in\r$ is a one-parameter group 
of global diffeomorphisms and 
$I\subset \r$ is a connected interval of the real line containing zero. 
The most interesting 
case occurs when $I\cap\r^+=\r^+$ in which case 
we say that $\{\varphi_s\}_{s\in\r^+}$ is a global one-parameter submonoid 
of causal symmetries. It can be easily seen 
\cite{LETTER,CAUSAL-SYMMETRY} that $\varphi_{0}=$Id is the only conformal 
transformation contained in $\{\varphi_{s}\}$ unless the submonoid is 
in fact a subgroup of conformal transformations. Therefore, 
there cannot be realizations of $S^1$ as a one-parameter submonoid of 
{\em proper} causal symmetries.

We can give our main definition in this section.
\begin{defi}[Causally preserving vector fields]
A smooth vector field $\xiv$ defined on an entire Lorentzian manifold
is said to be causal preserving if the local one-parameter group
generated by $\xiv$ complies with (\ref{condition-rmas}) for some
interval $I$.
\label{cpvf}
\end{defi}
These causal-preserving vector fields are a strict generalization of
conformal 
Killing vectors, which are particular cases of them.

The next step is to derive the necessary and sufficient conditions 
for a vector field $\xiv$ to be causal preserving. To that end, we need to
classify such vectors in terms of the set of null directions which are
kept invariant under the mappings of the one-parameter submonoid. 
These are called {\em canonical null directions} of the submonoid, 
or in short, of $\xiv$.  The set of
canonical null directions, denoted by $\mu_{\xiv}$, only depends on
the specific submonoid if the metric tensor $\G$ is analytic on $V$
and indeed we can calculate it from the vector field $\xiv$ (this is
the reason for the chosen notation).  Whenever
$\mu_{\xiv}\neq\varnothing$, these null vectors are the part of the
null cone preserved by the submonoid so we can regard causal
transformations in this case as partly conformal transformations.  The
canonical null directions can be calculated explicitly by means of the
condition
$$
\mu_{\xiv}=\{\k\ \mbox{null}: \pounds_{\xiv}\G(\k,\k)=0\}.
$$
Then we have the main result in \cite{LETTER,CAUSAL-SYMMETRY}.
\begin{theo}
Let $\xiv$ be a smooth complete vector field and suppose there
exists a function $\alpha$ such that $\pounds_{\xiv}\G-\alpha\G$ is a 
future tensor field with the same algebraic type at every point of the
manifold. 
Then, if
\begin{enumerate}
\item $\mu_{\xiv}=\varnothing$, $\xiv$ is a causal preserving vector 
field (with no canonical null direction).
\item $\mu_{\xiv}\neq\varnothing$ and $\lie{\bf\Omega}\propto{\bf\Omega}$ 
where ${\bf\Omega}$ is a $p$-form constructed as the wedge product of a
maximal 
set of linearly independent elements of $\mu_{\xiv}$, $\xiv$ is a causal 
preserving vector field with $\mu_{\xiv}$ as the set of its canonical null 
directions.  
\end{enumerate} 
\label{conditions-cpvf}
\end{theo}
Observe that if $p=n$, then 
$\mu_{\xiv}$ contains all possible null directions and the causal 
preserving vector field is a conformal Killing vector. Note also that, 
for the case {\em (i)} of this theorem, 
the condition that $\pounds_{\xiv}\G-\alpha\G$ be a future tensor can 
be replaced by $\pounds_{\xiv}\G(\k,\k)>0$ for all null vectors $\k$, 
making no mention of the function $\alpha$. There is an analogous 
statement for the case {\em (ii)} of the theorem which is more 
involved, see \cite{CAUSAL-SYMMETRY}.

The physical relevance of causal preserving vector fields is
still unclear. Tentative interpretations can be found in
\cite{PATRICOT,PITTS}, and a generalization shedding some light as to 
their applicability and geometrical properties in \cite{BICONFORMAL}. 
Let us simply remark here that sometimes 
conserved quantities and constants of motion can be found, 
see \cite{CAUSAL-SYMMETRY,LETTER} for details. For example, 
for general affinely parametrized null geodesics whose tangent vector 
is $\v$ it follows that $\G(\xiv,\v)$ is monotonically non-decreasing to
the 
future along the geodesic. Moreover, if $\v$ is tangent to a canonical 
null direction for all $x$ on the curve,
then $\G(\xiv,\v)$ is constant along this null geodesic. Hence, the null 
geodesics along the canonical null directions of a causal motion 
$\xiv$ have a {\em constant} component along $\xiv$. Furthermore,
the construction of conserved currents (divergence-free vector fields)
is also possible using causal-preserving vector fields as shown in
\cite{CAUSAL-SYMMETRY,LETTER}, where particular examples for the 
electromagnetic field and the Bel-Robinson tensor can be found.

Further interesting properties of causal preserving vector fields can
be found in a paper by Harris and Low \cite{HARRIS-SHAPE} (these authors
employ the terminology ``causal decreasing vector fields'' for causal
preserving vector fields).  In the forthcoming results we adapt the
language of these authors to the notation followed here.
\begin{prop}
If $\xiv$ is a complete timelike future-directed causal preserving vector
field 
with $\mu_{\xiv}=\varnothing$ then for any integral curve $\g$ of $\xiv$
we have
that $I^-(\g)=V$.
\label{half-omniscient}
\end{prop}
This allows us to prove a splitting theorem.
\begin{prop}
Under the assumptions of the previous proposition,
the space $Q$ of integral curves of $\xiv$ 
is a manifold and we can write $V=Q\times\r$.
\label{factorization}
\end{prop}

\subsection{Lie subsemigroups}
\label{Lie-subsemigroup}
Suppose that we are given a set of causal preserving vector fields 
with the same canonical null directions. 
This set cannot form a Lie algebra, and in fact they only have the 
structure of a cone, that is, we can only form linear combinations 
with {\em positive coefficients}. This is due to the fact that, 
as explained in the previous subsection, causal symmetries do not form
a group. This means that
if we are to describe them in terms of actions upon manifolds we
cannot use full groups as it is done with ordinary symmetries.
Can we nevertheless obtain the corresponding generated subset of 
causal symmetries? The answer is yes, for it is possible to define 
actions of submonoids and subsemigroups on
Lorentzian manifolds.  

Submonoids have received attention in
Mathematics in the framework of the theory of Lie subsemigroups. 
Standard references about this subject are \cite{SEMIGROUP,HILGERT}
(see \cite{LAWSON} for a good review).  Here we will limit ourselves
to a brief account mainly to bring together pieces of information
which ordinarily lie in journals and books only read by mathematicians.
We will thus use concepts and notation which are standard in Lie groups
theory.  A Lie subsemigroup $S$ is a subset of a Lie group $G$ which
is a semigroup.  For these objects we can define an analog to the Lie
algebra as
$$
L(S)=\{x\in\mathfrak{g}:\exp(\r^+x)\in S\},
$$ 
where $\mathfrak{g}$ is the Lie algebra of $G$.  $L(S)$ is a {\em
cone} in the vector space $\mathfrak{g}$.  Recall that a cone in a
vector space $L$ is a subset $C$ such that $\forall v_1, v_2\in C$,
$\l_1v_1+\l_2v_2\in C$ where $\l_1,\l_2$ are any pair of positive (or
negative) scalars and ${\bf 0}\in C$.  From the cone $L(S)$ we may
define the vector space $H(L(S))\equiv L(S)\cap-L(S)$ called the edge
of the cone.  If such edge satisfies the property
$$
e^{ad[h]}L(S)=L(S),\ \ \forall h\in H(L(S)),
$$
then $L(S)$ is called a Lie wedge.  The set $L(S)$ can be mapped to
the remainder of the Lie group $G$ by means of the differential map of
the right action $R_g:G\rightarrow G$.  In this way the Lie group
becomes a conal manifold where the cone $C(g)$ at each $g\in G$ is
defined by
$$
C(g)=\{\vec{w}\in T_{g}(G):\exists x\in\mathfrak{g}\ 
\mbox{with}\ \vec{w}=R'_g|_{e}(x)\},
$$
where $e$ is the neutral element of $G$. 
Note that $C(g)$ can in principle be constructed from any cone on
$\mathfrak{g}$, be it the Lie wedge of a Lie subsemigroup or not.  An
interesting question now is to find out the conditions that a cone on
$\mathfrak{g}$ (and by extension a cone field on $G$) be the Lie wedge
of a Lie subsemigroup $S\subseteq G$.  The next result proven in
\cite{NEEB} addresses this matter.
\begin{theo}
Let $G$ be an analytic group, $\mathfrak{g}$ its Lie algebra and
$W\subset\mathfrak{g}$ a cone such that $\mathfrak{g}$ is the smallest
Lie algebra containing $W$.  Then $W$ generates a Lie subsemigroup
$S\subseteq G$ if and only if the subgroup of $G$ generated by the
subalgebra $H(W)$ is closed and there exists a function $f\in
C^{\infty}(G)$ with the property
$$
df|_{g}\langle R'_g|_e(x)\rangle>0,\ \ \ \ \forall x\in W-H(W),
\ \ \ \ \ \forall g\in G.
$$  
\label{lie-wedge}
\end{theo}
\noindent
{\bf Remark.} This theorem takes the form of the stable causality
condition for the manifold $G$ with respect to the causality induced
by the cone $C(g)$.

If $V$ is a differentiable manifold let us consider the action
$\varphi:G\times V\rightarrow V$ of a group $G$ complying with the
assumptions of theorem \ref{lie-wedge} plus the property
\be
Ad(g)[x]\in L(S),\ \ \forall x\in L(S),\ \forall g\in G,
\label{adjoint-invariance}
\ee
where $Ad$ is the adjoint representation of the lie group $G$ on its
Lie algebra $\mathfrak{g}$.  In this case one can show that there is a
natural cone field $C(x)$, $x\in V$ which is invariant under this
action, namely,
$$
\varphi'_g|_x(C(x))=C(\varphi_g(x)),
$$
where $\{\varphi_g\}_{g\in G}$ is the group of transformations of $V$
defined from the action $\varphi$.  
The explicit relation between the cones $C(g)$ of the Lie
group and $C(x)$ of the differentiable manifold is
$$
C(x)=\{\vec{Y}\in T_x(V):\exists \vec{w}\in C(g): 
\Psi'_y(\vec{w})|_g=\vec{Y}\},\ \ \Psi_y(g)\equiv\varphi(g,y)=x,
$$
where a simple calculation shows that $C(x)$ does not depend on $y$ 
(full details can be found in \cite{HILGERT}).

If one wishes to study actions in which the invariance of the cone is
replaced by $\varphi'_g|_x(C(x))\subseteq C(\varphi_g(x))$ (this is
the case of causal symmetries with $C(x)$ the Lorentzian cone) then
this last condition will only be true for elements $g$ belonging to
the semigroup $S\subset G$ if we demand that condition
(\ref{adjoint-invariance}) hold only for $g\in S$.  Note however, that
causal symmetries are more general than this as they are ruled by
actions of infinite-dimensional groups so the theory just presented
could be useful to study {\em finite-dimensional} submonoids of causal
symmetries. As a matter of fact, one can sometimes consider a set of 
causal preserving vector fields as a cone in a subgroup of {\em 
biconformal vector fields} \cite{BICONFORMAL}, and the previous 
construction applies automatically.

Further details can be consulted in 

\cite{LETTER,CAUSAL-SYMMETRY,BICONFORMAL,HH,HILGERT2,HO,LEVICHEV,
LEVICHEV2,MITT,PANEITZ,PANEITZ2}.

\section{Future perspectives}
\label{future}
It is time now to recapitulate drawing conclusions from our review and
suggesting some possible fruitful lines of research. 
Lots of topics have been discussed, some of them classical,
well-established and known 
among relativists, some others mainly known in mathematics circles and
also others which probably have not yet reached a widespread knowledge due
to their novelty. 
It is precisely this last category which we would like to emphasize in 
these conclusions.

To start with, the definition of future and past tensors shown in
subsection \ref{causal-tensors} has many potential applications. Some have
been already studied and commented in this review, specially in connection
with the preservation of the null cone, the existence of causal
symmetries, or the setting of causal mappings; others are still to be
fully explored, specially the generalized null-cone algebraic structure
which induces on the whole bundle $T^r_s(V)$, and the classification and
possible decompositions that tensor fields inherit from this. This is a
subject characteristic and exclusive of Lorentzian geometry, as it
requires essentially the existence of the Lorentzian metric.

Concerning section \ref{abstract}, we believe there is not much to do
related to the definition and characterization of abstract
causal/etiological spaces, with an important exception: quantum causality
and the theory of ``causal sets" outlined in subsection
\ref{quantum}. This is a very important line of research, with a strong
vitality, and it is worth devoting some efforts in that direction. The
main sought but so far unreached result is how to connect the discrete
causal set or spin network with the smooth spacetime we believe to see.

One of the main results presented in this review is the possibility of 
sorting Lorentzian manifolds in abstract causality classes or {\em causal
structures} in 
the sense of definition \ref{causal-structure}. In our opinion, this
settles the 
issue of what the concept of ``causal structure'' of a Lorentzian manifold
really means.
More importantly, it has allowed a long-standing question to be resolved by
means of the theorem \ref{local-isocausal}, giving a precise meaning to
the {\em local} causal equivalence of Lorentzian manifolds. This also
permits us to talk about the causal asymptotic equivalence of asymptotically
flat spacetimes, which was only possible if a full conformal completion
was previously achieved. Similar local or asymptotic equivalences can be
further investigated, and one can easily produce a definition of
``asymptotical causal equivalence to $(M,\G)$", where $(M,\G)$ is any
particular or preferred spacetime, say, de Sitter, or anti de Sitter, or a
particular Robertson-Walker geometry, or a plane wave, and so on. This
route is yet to be examined. At the same time, causal mappings (definition
\ref{PREC})
enable us to refine the standard hierarchy of causality conditions, and to
construct causal chains of causal structures where, on a fixed manifold,
we can compare Lorentzian metrics from the causal point of view, and
classify them according to their ``goodness" in relation to causality. 
There are important open problems concerning this line of research
which we want to bring to the attention of the mathematical relativity
community.
For instance, an even more practical procedure to decide if two Lorentzian
manifolds are causally related, and then if they are isocausal, would be
very convenient.
Another question is how many different causal
structures can be defined on a given background differential manifold, and
whether 
there exists upper and lower bounds for them in terms of the partial order
introduced
in subsection \ref{chains}. 
As an additional remark to the previous comments, let us stress that the
definition of causal structure carries over to abstract etiological spaces
(ergo also to causal spaces)
using the concepts put forward in definition \ref{theta}. Thus
the results and questions of the above paragraph can be formulated as well
for these more abstract spaces being possible to define and classify
causal structures for them, and causal completions and boundaries too.

Another big issue in this review is the definition of the causal boundary
of Lorentzian manifolds. This problem has deserved renewed interest
recently in the wake of Maldacena's conjecture, and may become a
fundamental issue of the subject termed ``holography" (see
\cite{HOROWITZ}): the probable correspondence between string theory on
spacetime backgrounds and a conformal field theory {\em on the boundary at
infinity} of that background. Prior to this, researchers tried all sort of
recipes and methods in order to find a kind of boundary of {\em universal
} application to all possible spacetimes. Apart from the inspiring, clear
and very fruitful ideas of Penrose's
concerning the conformal boundary, subsection  \ref{conformal-boundary},
the main breakthrough here was Geroch, Kronheimer and Penrose's paper
described in subsection \ref{GeKrPe}, but further research
proved that strong though this procedure undoubtedly is, it has its own
limitations
that one should bear in mind. All the subsequent amendments suggested over
the 
years and described in subsection \ref{others}
were found either too theoretical, requiring daunting efforts to be
applied in practical cases, or not
mending the drawbacks they were supposed to mend. Several
other alternative 
constructions were devised with this aim but it is fair to say that they
have remained
as a set of rather too complicated definitions with no possible
translation into explicit relevant examples.
These difficulties to actually construct the causal boundary by whatever
procedure
have hampered its study in physically relevant cases. There are fresh good
news, however: recently new interesting and encouraging results have been
obtained, insufflating renewed air to this subject and arousing again the
interest of the mathematical and physics communities. We are referring
specially to 
\begin{itemize}
\item the startling result that the GKP boundary of certain classes of
pp-wave spacetimes consists of a one dimensional null line, see
$\S$\ref{ppwaves}. It is quite surprising that this result has been
obtained only recently and not, say, twenty years ago.   
In this sense it would be interesting trying to find 
the full GKP boundary in other pertinent cases instead of paying a deeper
attention to 
the wrong topological properties of the boundary in certain spacetimes
with 
strange or unphysical causal behaviours. 
\item Harris' ideas of caring about just the future boundary, and his
results ($\S$\ref{harris}) showing the universality of the future GKP
boundary using category theory.
\item the generalization of Penrose's original ideas by using causal
mappings instead of conformal ones. As is known the conformal and the GKP
boundaries (in any of its improved versions) are very difficult to
construct and we believe that the causal 
boundary based on the concept of isocausality (cf. $\S$
\ref{causal-relationship}) could come in aid. To this end, the technical
difficulties concerning causal mappings pointed out above should be
addressed and, if these problems were resolved satisfactorily, then the
causal boundary in the sense of $\S$\ref{causal-relationship}, as well as
the causal diagrams and causal completions ($\S$\ref{causal-diagram}),
could be constructed and analyzed explicitly in a large variety of cases
very easily. This would allow us to recover the simple and very powerful
applications of the conformal boundaries of spacetimes by keeping their
enormous virtues and intelligibility but avoiding on the one hand the
traditional problem of the impossibility of finding {\em conformal}
completions explicitly, and on the other hand the extreme difficulties to
build the other types of causal boundaries, which are also not exempt from
inconsistencies, as we have just explained.
\end{itemize}

Causality conditions impose restrictions on the topology and the metric
tensor 
of a spacetime as we explained in section \ref{sec:CandT}. Here 
we wish to stress theorem \ref{bernal-splitting} which may have
interesting applications. It is not necessary to stress the importance of 
globally hyperbolic spacetimes in Physics because they are present in any
model 
where a well-posed formulation of the Cauchy problem for Einstein field
equations 
is required. Therefore this new result could help improving well-posedness
results involving
globally hyperbolic spacetimes. Moreover we could adapt formulations of
the field 
equations such as the classical ADM to the hypotheses of theorem
\ref{bernal-splitting} and thereby obtain relevant
simplifications. Finally numerical simulations always assume globally
hyperbolic spacetimes
so one could set the equations modelling the system with a metric tensor
as given by theorem
\ref{bernal-splitting} with no loss of generality.

Last but not least, yet another issue we feel worth researching by
relativists and mathematicians alike
is the subject of section \ref{symmetry}, namely, causal
transformations/symmetries and their infinitesimal generators,
causally-preserving vector fields. There is a mathematical 
sub-branch lying behind these type of transformations (Lie subsemigroups
and their actions on manifolds, cf. $\S$\ref{Lie-subsemigroup}) which has
its own independent interest. The interpretation we have provided for
these transformations and vector fields may help improving or suggesting
the mathematical advances apart from bringing to light important physical
applications. A physical interpretation for 
these new transformations or their generators is also an open, seemingly
solvable, important question. It is interesting to note that
as opposed to other classical symmetries, 
causal symmetries might be present in virtually all spacetimes with
reasonable
causality conditions. Therefore they could allow us to formulate
rigorously results 
involving {\em approximate} or {\em asymptotic} isometries or conformal 
transformations. Furthermore, they may provide new conserved, or
monotonically increasing, quantities. Finally, they have a direct
application to important splitting theorems of spacetimes
\cite{CAUSAL-SYMMETRY,BICONFORMAL}, of a more general nature than those
presented in subsection \ref{splitting}, which are also worth mentioning
here as they give characterizations of spacetimes splittable in two
orthogonal distributions of any $p$ and $q$ dimensions ($p+q=n$). Observe
in this sense that causal-preserving vector fields may leave a set of,
say, $p$ null directions invariant, and therefore they act as conformal
Killing vectors in the distributions spanned by these null directions. One
can then try to construct tensors which are invariant under such general
splittings, or characterize spacetimes decomposable in two conformally
flat pieces, and so on. In summary, causal symmetries and
causal-preserving vector fields can always be considered as {\em partly
conformal} transformations and {\em partly conformal Killing vectors},
respectively, and the many implications and possibilities deriving from
this fact which immediately spring to mind are certainly worth
investigating.

\section*{Acknowledgements}
We thank  M. S\'anchez for a careful reading of the manuscript and
many improvements and suggestions. Financial support from the Wenner-Gren
Foundations 
(JMMS) and the Applied Mathematics Department at Link\"oping University
(AGP), 
Sweden, are gratefully acknowledged. We thank the mentioned department,
where 
this work was partly discussed, for hospitality. Financial support under
grants BFM2000-0018 and FIS2004-01626 of the Spanish CICyT and 
no. 9/UPV 00172.310-14456/2002 of the University of the Basque 
Country is also acknowledged. 

Finally we thank the valuable comments and corrections on a previous version 
received from V Manko, S Hayward, I R\'acz, S Ross, L B Szabados, S B Edgar, 
T Strobl, W Simon, E Ruiz, S G Harris, and two anonymous referees.

\end{document}